\documentclass[twocolumn,eqsecnum,aps,pra,superscriptaddress,10pt,showpacs]{revtex4-1}
\usepackage[usenames,dvipsnames]{color}
\usepackage[english]{babel}
\usepackage{epsfig}
\usepackage{graphicx,subfigure}
\usepackage{tikz}
\usepackage{pgfplots}
\pgfplotsset{compat=1.12}
\usetikzlibrary{calc}
\usetikzlibrary{decorations.markings}
\usetikzlibrary{decorations.pathmorphing}
\usetikzlibrary{positioning,shapes,arrows,decorations.markings,calc}
\usepackage{longtable}
\usepackage{multirow}
\usepackage{arydshln}
\usepackage{hyperref}
\usepackage{amsmath,bbm,bm,amssymb}
\usepackage[mathcal,mathscr]{eucal}
\DeclareFontFamily{OT1}{pzc}{}
\DeclareFontShape{OT1}{pzc}{m}{it}{<-> s * [1.10] pzcmi7t}{}
\DeclareMathAlphabet{\mathpzc}{OT1}{pzc}{m}{it}
\usepackage[latin1]{inputenc}
\usepackage[T1]{fontenc}
\usepackage{times}
\usepackage[Euler]{upgreek}
\usepackage{enumerate}

\usepackage{bookmark}

\usepackage{textcomp}

\setcitestyle{numbers,square,citesep={,\kern-.24em}}

\hypersetup{
  colorlinks=false,
  pdfborder={0 0 0},
}

\tikzset{midarrow/.style={decoration={
  markings,
  mark=at position #1 with {\arrow{angle 45}}},postaction={decorate}}
}
\tikzset{waved/.style={decorate,decoration=snake}}

	\definecolor{aaltoBlack}{RGB}{0,0,0}%
	\definecolor{aaltoGray}{RGB}{146,139,129}%
	\definecolor{aaltoRed}{RGB}{237,41,57}%
	\definecolor{aaltoBlue}{RGB}{0,101,189}%
	\definecolor{aaltoYellow}{RGB}{254,203,0}%
	\definecolor{aaltoPurple}{RGB}{102,57,183}%
	\definecolor{aaltoTurquoise}{RGB}{0,168,180}%
	\definecolor{aaltoGreen}{RGB}{0,155,58}%
	\definecolor{aaltoLightGreen}{RGB}{105,190,40}%
	\definecolor{aaltoOrange}{RGB}{255,121,0}%
	\definecolor{aaltoFuchsia}{RGB}{177,5,157}%
\usepackage{soul}

\begin{document}

\title{Multi-scale model for disordered hybrid perovskites: the concept of organic cation pair modes}

\author{Jingrui Li}
\email{jingrui.li@aalto.fi}
\affiliation{Centre of Excellence in Computational Nanoscience (COMP) and Department of Applied Physics, Aalto University, P.O.Box 11100, FI-00076 AALTO, Finland}
\author{Jari J\"arvi}
\affiliation{Centre of Excellence in Computational Nanoscience (COMP) and Department of Applied Physics, Aalto University, P.O.Box 11100, FI-00076 AALTO, Finland}
\affiliation{Department of Physics, University of Helsinki, P.O.Box 64, FI-00014 University of Helsinki, Finland}
\author{Patrick Rinke}
\affiliation{Centre of Excellence in Computational Nanoscience (COMP) and Department of Applied Physics, Aalto University, P.O.Box 11100, FI-00076 AALTO, Finland}

\begin{abstract}
We have studied the properties of the prototype hybrid organic-inorganic perovskite $\text{CH}_3^{}\text{NH}_3^{}\text{PbI}_3^{}$ using relativistic density functional theory (DFT). For our analysis we introduce the concept of $\text{CH}_3^{}\text{NH}_3^+$ ``pair modes'', that is, characteristic relative orientations of two neighboring $\text{CH}_3^{}\text{NH}_3^+$ cations. In our previous work [Phys. Rev. B \textbf{94}, 045201 (2016)] we identified two preferential orientations that a single $\text{CH}_3^{}\text{NH}_3^+$ cation adopts in a unit cell. The total number of relevant pairs can be reduced from the resulting 196 combinations to only 25 by applying symmetry operations. DFT results of several $2\!\times\!2\!\times\!2$ supercell models reveal the dependence of the total energy, band gap and band structure on the distribution of $\text{CH}_3^{}\text{NH}_3^+$ cations and the pair modes. We have then analyzed the pair-mode distribution of a series of $4\!\times\!4\!\times\!4$ supercell models with disordered $\text{CH}_3^{}\text{NH}_3^+$ cations. Our results show that diagonally-oriented $\text{CH}_3^{}\text{NH}_3^+$ cations are rare in optimized $\text{CH}_3^{}\text{NH}_3^{}\text{PbI}_3^{}$ supercell structures. In the prevailing pair modes, the $\text{C--N}$ bonds of the two neighboring $\text{CH}_3^{}\text{NH}_3^+$ cations are aligned approximately vertically. Furthermore, we fit the coefficients of a pair-mode expansion to our supercell DFT reference structures. The pair-mode model can then be used to quickly estimate the energies of disordered perovskite structures. Our pair-mode concept provides combined atomistic-statistical insight into disordered structures in bulk hybrid perovskite materials.
\end{abstract}


\maketitle
\thispagestyle{empty}


\section{Introduction}

Hybrid perovskite photovoltaic (HPPV) technology \cite{Snaith13,Green14} is the most recent rising star in the emerging solar-cell community due to its record increase in power-conversion efficiency (PCE) during the last five years \cite{NRELchart}. The current state-of-the-art HPPV architecture was proposed in 2012 achieving $\sim\!\!10\%$ PCE \cite{KimHS12,LeeMM12}. Now the PCE of HPPV cells has already reached $22\%$, overtaking the best-performing inorganic-based single-junction thin-film cells such as $\text{CdTe}$ and copper-indium-gallium-selenide (CIGS) cells \cite{NRELchart,Jackson16}. The most common photoactive material in HPPV cells is methylammonium ($\text{MA}\!\equiv\!\text{CH}_3^{}\text{NH}_3^{}$) lead triiodide ($\text{CH}_3^{}\text{NH}_3^{}\text{PbI}_3^{}$, shortened as $\text{MAPbI}_3^{}$ hereafter). This hybrid perovskite material exhibits several advantageous features for photovoltaic applications, such as a band gap close to the optimal value for single-junction solar-cell absorbers \cite{Stoumpos13}, excellent absorption strength in the visible part of the solar spectrum \cite{deWolf14}, and high mobilities for both electron and hole transport \cite{Stranks13,Xing13}. It can be synthesized in solution at low temperature from common starting materials that have limited harm to the environment. Therefore HPPV cells are considered as promising candidates that can offer clean, affordable and sustainable energy.

Apart from the PCE improvement, recent experimental and theoretical studies in HPPV technology have focused on the origin of the high mobility and low recombination rate \cite{Stranks13,Xing13,PonsecaJr14,Johnston16}, the observed current-voltage hysteresis \cite{Stoumpos13,Snaith14a,KimHS15,ChenB16}, and the stability of hybrid perovskites materials \cite{Noh13,Niu14,Niu15}. To resolve open questions  in hybrid perovskites it is imperative to develop a comprehensive understanding of their atomic structure, which is both fundamental and challenging due to the structure's complexity. Taking the prototype hybrid perovskite $\text{MAPbI}_3^{}$ as an example, the central cation $\text{MA}^+$ is not spherical (as, e.g., $\text{Cs}^+$ in the conventional perovskite $\text{CsSnI}_3^{}$) but exhibits polarity and an orientational preference in the lattice. At low temperatures, $\text{MAPbI}_3^{}$ assumes a minimal-energy structure with regularly-aligned $\text{MA}^+$ cations and thereby a regularly-deformed inorganic $\text{PbI}_3^-$ matrix, resulting in an orthorhombic phase. Conversely, at room temperature or above, the $\text{MA}^+$ cations are thought to be randomly oriented due to thermal fluctuations, forming (dynamically-)disordered structures \cite{Poglitsch87,Stoumpos13,Weller15,Egger16}.

The detailed mechanism leading to disorder is not yet fully understood. Wasylishen \textit{et al.} \cite{Wasylishen85} claimed that the change of an $\text{MA}^+$ ion's $\text{C--N}$ bond direction (called reorientation hereafter) occurs on a sub-picosecond time scale in the cubic phase of $\text{MA}$-based perovskites based on $^{14}\text{N}$-NMR measurements. Conversely, Poglitsch and Weber (using millimeter-wave spectroscopy) \cite{Poglitsch87}, Bakulin \textit{et al.} (two-dimensional infrared vibrational spectroscopy) \cite{Bakulin15}, and Chen \textit{et al.} (quasi-elastic neutron scattering) \cite{ChenT15} reported characteristic times for $\text{MA}$-reorientation on a picosecond time scale. For the activation energy of the $\text{C--N}$ bond rotation, recent first-principles molecular dynamics (MD) simulations give a value of $42~\text{meV}$ \cite{Meloni16}, whereas other theoretical \cite{Mosconi14b,LeeJH15,LiJ18a,LiJ18b} and also experimental \cite{OnodaYamamuro92,Mosconi14b} studies suggested that it is of the order of $100~\text{meV}$. Such a discrepancy can lead to significantly different scenarios. A small reorientation energy implies that the $\text{MA}^+$ cations are loosely attached to the inorganic cage and can rotate almost freely within the lattice at room temperature. This would lead to dynamical three-dimensional isotropy on a length scale of one single cell ($\sim\!\!6~\text{\AA}$). In contrast, if the activation energy is $\sim\!\!100~\text{meV}$ (corresponding to $\sim\!\!1200~\text{K}$), the probability for an $\text{MA}^+$ to overcome such a barrier is very low  at room temperature. The $\text{MA}^+$ cations will then remain bound to the inorganic framework via hydrogen bonds \cite{Egger14,LiJ16,LiJ18a} for a relatively long time. In this scenario, localized $\text{MA}$-patterns \cite{Frost14a,Frost14b,Leguy15} would form on short length scales. On  large length scales, $\text{MAPbI}_3^{}$ appears effectively cubic.

Modeling the orientational disorder of $\text{MA}^+$ ions in $\text{MAPbI}_3^{}$ is a challenging task. Quantum mechanical first-principles techniques are required to correctly describe the hydrogen bonding of $\text{MA}^+$ ions to the inorganic cage and the corresponding distortions of the cage. However, even density-functional theory (DFT), which in local or semi-local approximations is currently the most computationally efficient first-principles technique, cannot scale up to the required length scales or the large number of candidate structures. The simple primitive-cell model is not representative of $\text{MAPbI}_3^{}$'s atomic structure, as it effectively describes a system of infinitely many aligned polar $\text{MA}^+$ ions. The dipole moment introduced by each $\text{MA}^+$ unit can be canceled by compensating alignments of $\text{MA}^+$ ions in an appropriately chosen supercell model. For such compensated models, the atomic and electronic structure, especially in the low-temperature orthorhombic and tetragonal phases, can then be calculated at the DFT or beyond level by means of small supercell models such as $\,\sqrt[]{2}\!\times\sqrt[]{2}\!\times\!2$ and $2\!\times\!2\!\times\!2$ \cite{Umari14,MenendezProupin14,Mosconi14b,Yin15,Brivio15}. $2\!\times\!2\!\times\!2$ supercell models have also been adopted to study the distribution of $\text{MA}^+$ orientations at a finite temperature using \textit{ab initio} MD \cite{Frost14b,Deretzis16}. However, to really model disorder we would need to know the structure of $\text{MAPbI}_3^{}$ on a length scale of a few to a few tens of single (primitive) cells. $2\!\times\!2\!\times\!2$ supercell models do not suffice for this purpose, because of the periodic boundary conditions, while DFT calculations for larger supercell models become computationally very demanding.

Only recently three studies employed large supercell models to approach the structural complexity \cite{Meloni16,Lahnsteiner16}. Meloni \textit{et al.} \cite{Meloni16} used \textit{ab initio} MD to study the time-dependent autocorrelation function of $\text{C--N}$ bond directions at different temperatures in a $2\,\sqrt[]{2}\!\times\!2\,\sqrt[]{2}\!\times\!4$ supercell model, and claimed an activation energy of $42~\text{meV}$ for $\text{MA}$-reorientation. Lahnsteiner \textit{et al.} \cite{Lahnsteiner16} constructed a series of $n\!\times\!n\!\times\!n$ supercell models with $n=2,4,6$ to model $\text{MAPbI}_3^{}$ at different temperatures. Their results indicate that the $\text{C--N}$ bonds are very rarely oriented along the diagonal direction within a single cell, and the angles between the $\text{C--N}$ bonds of two $\text{MA}^+$ cations follow certain static and dynamical correlation in the cubic phase. Lahnsteiner \textit{et al.} provide an important reference for our study, especially for the analysis of $\text{C--N}$ bond-direction and the alignment of $\text{MA}^+$ ions in $\text{MAPbI}_3^{}$. In addition, they reported that the upper bound for the reorientation of an $\text{MA}^+$ is $7~\text{ps}$ at room temperature. For even larger supercells ($n=8,12$), Ma and Wang \cite{Ma15} studied the electronic structure of $\text{MAPbI}_3^{}$ using the \textit{ab initio} three-dimensional fragment method. In their model systems, the $\text{C--N}$ bonds were randomly oriented along the diagonal directions of a single cell. Their results indicate that the orientational disorder induce a charge-density localization of both valence-band-maximum and conduction-band-minimum on small length scales.

In our previous work \cite{LiJ16}, we have comprehensively analyzed the atomic structure of hybrid perovskites using the primitive-cell model. We found several stable locations of $\text{MA}^+$ in the lattice. Moreover, our analysis revealed that the stability of hybrid perovskites is closely related to the deformation of the inorganic cage, which acts synergetically with the organic ions analogous to a chicken-and-egg paradox. In this work, we performed DFT calculations for a number of different $\text{MAPbI}_3^{}$ supercell models and focus on the pairs of neighboring $\text{MA}^+$ ions. We devised a pair-mode description that reduces each $\text{MA}^+$ ion to a dipole \cite{NOTEsupercell} with discrete orientations that were adopted from our previous primitive-cell results \cite{LiJ16}. We then defined the relative geometry of two nearest individual dipoles as a ``pair mode''. With the pair-mode concept, we were able to relate the dependence of certain $\text{MAPbI}_3^{}$ properties (e.g., total energy, band structure) to the distribution of $\text{MA}$-orientations and to the dipoles' alignment. This was done by studying a series of $2\!\times\!2\!\times\!2$ supercell models. We further investigated $4\!\times\!4\!\times\!4$ supercell models focusing on the distribution of pair modes. This distribution tells us for a given dipole which dipole-orientations are preferred in its surrounding, thus providing knowledge of the local structure beyond a single $\text{MAPbI}_3^{}$ unit cell. 

The remainder of this paper is organized as follows. In Sec.~\ref{comp}, we briefly describe the model systems and the computational details of our DFT calculations. Section~\ref{results} outlines the concept of pair modes, and uses this concept to discuss the results of the supercell models. Finally, Sec.~\ref{concl} concludes with a summary.

\section{Computational details}\label{comp}

The supercell models considered in this paper were constructed based on single (primitive) cells. In each single cell, the $\text{MA}^+$ is located close to the centre of the cell, $\text{Pb}^{2+}$ at the corners, and $\text{I}^-$ at the edge-centers. We considered a series of $2\!\times\!2\!\times\!2$ and $4\!\times\!4\!\times\!4$ supercell models for different purposes. In $2\!\times\!2\!\times\!2$ supercell models the total dipole moment can be easily canceled with regular alignments of $\text{MA}^+$ ions. Although a $2\!\times\!2\!\times\!2$ supercell model is larger than a $\,\sqrt[]{2}\!\times\sqrt[]{2}\!\times\!1$ model, it has the advantage that it does not introduce artificial differences between the three lattice directions \textit{a priori}. Compared with the $2\!\times\!2\!\times\!2$ counterparts, the larger $4\!\times\!4\!\times\!4$ supercell models contain 64 $\text{MA}^+$ cations, thus providing appropriate model systems to mimic the disordered structures by introducing randomly oriented $\text{MA}^+$ ions. The considered supercell models were fully-randomly initialized with different $\text{MA}^+$ alignments.

The choice of $4\!\times\!4\!\times\!4$ supercell models (containing $768$ atoms) is based on the following considerations: (a) The smaller $3\!\times\!3\!\times\!3$ supercell models cannot properly host the octahedron-tilting of the perovskite structure due to the odd number of single cells along each lattice vector \cite{Even16}. Nonetheless, we have also studied a series of $3\!\times\!3\!\times\!3$ supercells and the results are provided in Sec.~S4 of Ref.~\cite{SMsupercell} for comparison. (b) Some smaller supercell models, such as  $2\,\sqrt[]{2}\!\times\!2\,\sqrt[]{2}\!\times\!4$, may introduce artifacts, as they limit the number of possibilities of $\text{MA}$-alignments along the two shorter lattice vectors so that the results would be very sensitive to the random initial geometry of $\text{MA}^+$ ions \cite{Lahnsteiner16}. (c) Even larger supercell models contain $>\!1~000$ atoms and are thus too computationally expensive for conventional DFT modeling.

In our previous study \cite{LiJ16} we demonstrated that the ``PBE+vdW'' exchange-correlation functional produces the lattice constants of hybrid perovskite systems in good agreement with experiment, and can properly describe the interaction between the organic cations and the inorganic framework. Thus we adopted this functional for all DFT calculations in this work. Specifically, the Perdew-Burke-Ernzerhof (PBE) generalized gradient approximation \cite{Perdew96} was used as the exchange-correlation functional, and the long-range van der Waals (vdW) interactions were described by employing the Tkatchenko-Scheffler method based on the Hirshfeld partitioning of the electron density \cite{Tkatchenko09}. In addition, scalar relativistic effects were included via the zero-order regular approximation (ZORA) \cite{vanLenthe93}. Although spin-orbit coupling \cite{Even12,Even13,Brivio14,Katan15} and exact exchange or many-body corrections \cite{Chiarella08,Brivio13,Brivio14} have significant impact on the band structure of $\text{MAPbI}_3^{}$, we did not include them in the band-structure calculations (for the $2\!\times\!2\!\times\!2$ supercell models), since they are computationally very demanding. For the relative differences of band gaps between different supercell models PBE+vdW+ZORA is sufficient.

All calculations were carried out using the all-electron numeric-atom-centered orbital code \textsc{fhi-aims} \cite{Blum09,HavuV09,Levchenko15}. For the $2\!\times\!2\!\times\!2$ supercell models, we used a $\Gamma$-centered $4\!\times\!4\!\times\!4$ $k$-point mesh and tier~2 basis sets for both structure relaxation and band-structure calculations. We performed direct lattice-vector optimization with the analytical stress tensor implemented in \textsc{fhi-aims} \cite{Knuth15}. For the larger $4\!\times\!4\!\times\!4$ supercells, we have reduced the computational expense by employing a $2\!\times\!2\!\times\!2$ $k$-point mesh and tier~1 basis sets. The geometries were optimized for a fixed size of the cubic unit cell, for which the lattice parameter $a=25.25~\text{\AA}$ was adopted based on the experimental value $a_0^{}=6.31~\text{\AA}$ of the primitive cell \cite{Stoumpos13}. The results of all relevant calculations of this work are available from the Novel Materials Discovery (NoMaD) repository \cite{NoMaDsupercell}.

\section{Results and discussions}\label{results}

\subsection{Definition of \texorpdfstring{$\text{CH}_3^{}\text{NH}_3^+$}{} ``pair modes''}

The essential difference between conventional perovskites such as $\text{CsPbI}_3^{}$ and hybrid perovskites is the monovalent central cation. Metal cations such as $\text{Cs}^+$ are spherically symmetric, while the polar organic $\text{CH}_3^{}\text{NH}_3^+$ cation has a permanent dipole moment pointing from the $\text{C}$-end (the methyl group) to the $\text{N}$-end (the ammonium cation group). Thus, in a primitive-cell model all $\text{MA}^+$ ions will be aligned parallel. This would result in a large dipole moment in the bulk material, which is not observed experimentally. Supercell models allow us to cancel the total dipole moment in the supercell. In this paper, we calculated the total dipole moment within a supercell by the vector sum of individual $\text{MA}^+$ dipole moments (or its average per $\text{MAPbI}_3^{}$ unit) and represent it in terms of $p_0^{}$, the permanent dipole moment of an isolate $\text{MA}^+$ in vacuum. Our PBE+vdW/tier~2 result of $p_0^{}$ is $2.2~\text{D}$, very close to the B3LYP/6-31G* result of $2.3~\text{D}$ \cite{Frost14a}.

\begin{figure}[!ht]
\leftline{(a)~Two stable orientations of $\text{MA}^+$ in the primitive-cell model}
\begin{tabular}{ccc}
\includegraphics[clip=true,trim=3.6in 9.65in 3.6in 1.in,page=1,scale=1.]{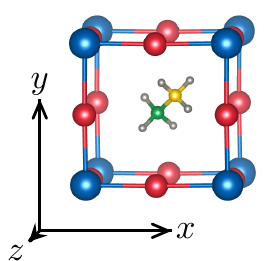} & $\quad$ &
\includegraphics[clip=true,trim=3.6in 9.65in 3.6in 1.in,page=2,scale=1.]{./Fig01.pdf} \\
Diagonal $\text{MA}^+$ & & Face-to-face $\text{MA}^+$ \\
$\quad$
\end{tabular}

\leftline{(b)~Possible directions of a diagonal $\text{MA}$-dipole}
\includegraphics[clip=true,trim=3.8in 10in 3.8in 1.in,page=3,scale=1.]{./Fig01.pdf} $\quad$
\includegraphics[clip=true,trim=3.8in 10in 3.8in 1.in,page=4,scale=1.]{./Fig01.pdf} $\quad$
\includegraphics[clip=true,trim=3.8in 10in 3.8in 1.in,page=5,scale=1.]{./Fig01.pdf} $\quad$
\includegraphics[clip=true,trim=3.8in 10in 3.8in 1.in,page=6,scale=1.]{./Fig01.pdf} \\
\includegraphics[clip=true,trim=3.8in 10in 3.8in 1.in,page=7,scale=1.]{./Fig01.pdf} $\quad$
\includegraphics[clip=true,trim=3.8in 10in 3.8in 1.in,page=8,scale=1.]{./Fig01.pdf} $\quad$
\includegraphics[clip=true,trim=3.8in 10in 3.8in 1.in,page=9,scale=1.]{./Fig01.pdf} $\quad$
\includegraphics[clip=true,trim=3.8in 10in 3.8in 1.in,page=10,scale=1.]{./Fig01.pdf} \\
$\quad$

\leftline{(c)~Possible directions of a face-to-face $\text{MA}$-dipole}
\includegraphics[clip=true,trim=3.8in 10in 3.8in 1.in,page=11,scale=1.]{./Fig01.pdf} $\quad$
\includegraphics[clip=true,trim=3.8in 10in 3.8in 1.in,page=12,scale=1.]{./Fig01.pdf} $\quad$
\includegraphics[clip=true,trim=3.8in 10in 3.8in 1.in,page=13,scale=1.]{./Fig01.pdf} \\
\includegraphics[clip=true,trim=3.8in 10in 3.8in 1.in,page=14,scale=1.]{./Fig01.pdf} $\quad$
\includegraphics[clip=true,trim=3.8in 10in 3.8in 1.in,page=15,scale=1.]{./Fig01.pdf} $\quad$
\includegraphics[clip=true,trim=3.8in 10in 3.8in 1.in,page=16,scale=1.]{./Fig01.pdf} \\
\caption{Symbolic representation of a $\text{CH}_3^{}\text{NH}_3^+$ ion by an arrow pointing from $\text{C}$ to $\text{N}$. (a) Two stable structures (left: with diagonally-oriented $\text{MA}^+$, and right: with $\text{MA}^+$ oriented approximately along face-to-face) obtained from primitive-cell calculations (data taken from Ref.~\cite{LiJ16}). $\text{C}$, $\text{N}$, $\text{H}$, $\text{Pb}$ and $\text{I}$ atoms are colored in green, yellow, gray, blue and red, respectively. (b,c) Dipole representation for $\text{MA}^+$ along (b) diagonal and (c) face-to-face orientations. $\odot$ and $\otimes$ indicate dipoles that are perpendicular to the plane of the paper and point out of and into it, respectively. Likewise, thick solid arrows in panel (b) point out of the plane of the paper, while dashed arrows point into the paper.}\label{dipoles}
\end{figure}

Our previous DFT calculations \cite{LiJ16} revealed two stable structures of the cubic primitive-cell model, as shown in Fig.~\ref{dipoles}(a). Specifically, in the left structure of Fig.~\ref{dipoles}(a), the $\text{C--N}$ bond is oriented along the diagonal ($[111]$ or equivalent) direction of the single unit cell, while in the right structure it is oriented along the face-to-face ($[100]$ or equivalent) direction with a small deviation. The face-to-face $\text{MA}^+$ structure is $21~\text{meV}$ more stable than the diagonal structure. We attribute this stability to the considerably larger deformation of the inorganic-framework in the face-to-face structure. As a side note, the internal atomic geometry of $\text{MA}^+$ in $\text{MAPbI}_3^{}$ is nearly independent of its location in the unit cell.

To simplify our notation, we abstract each $\text{MA}^+$ in the optimized $\text{MAPbI}_3^{}$ structure by an arrow along its $\text{C--N}$ bond that represents its dipole moment. Corresponding to the $8$ diagonal and $6$ face-to-face directions in the cubic single cell, there are altogether $14$ possible directions for such a dipole. They are illustrated in Figs.~\ref{dipoles}(b) and (c), respectively.

\begin{figure}[!ht]
\includegraphics[clip=true,trim=3.6in 9.8in 3.6in 1.in,page=1,scale=1.]{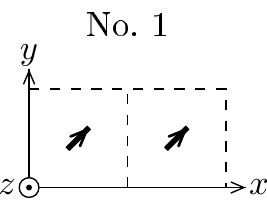} $\quad$
\includegraphics[clip=true,trim=3.6in 9.8in 3.6in 1.in,page=2,scale=1.]{./Fig02.pdf} $\quad$
\includegraphics[clip=true,trim=3.6in 9.8in 3.6in 1.in,page=3,scale=1.]{./Fig02.pdf} $\quad$
\includegraphics[clip=true,trim=3.6in 9.8in 3.6in 1.in,page=4,scale=1.]{./Fig02.pdf} $\quad$
\includegraphics[clip=true,trim=3.6in 9.8in 3.6in 1.in,page=5,scale=1.]{./Fig02.pdf} \\
\includegraphics[clip=true,trim=3.6in 9.8in 3.6in 1.in,page=6,scale=1.]{./Fig02.pdf} $\quad$
\includegraphics[clip=true,trim=3.6in 9.8in 3.6in 1.in,page=7,scale=1.]{./Fig02.pdf} $\quad$
\includegraphics[clip=true,trim=3.6in 9.8in 3.6in 1.in,page=8,scale=1.]{./Fig02.pdf} $\quad$
\includegraphics[clip=true,trim=3.6in 9.8in 3.6in 1.in,page=9,scale=1.]{./Fig02.pdf} $\quad$
\includegraphics[clip=true,trim=3.6in 9.8in 3.6in 1.in,page=10,scale=1.]{./Fig02.pdf}\\
\includegraphics[clip=true,trim=3.6in 9.8in 3.6in 1.in,page=11,scale=1.]{./Fig02.pdf} $\quad$
\includegraphics[clip=true,trim=3.6in 9.8in 3.6in 1.in,page=12,scale=1.]{./Fig02.pdf} $\quad$
\includegraphics[clip=true,trim=3.6in 9.8in 3.6in 1.in,page=13,scale=1.]{./Fig02.pdf} $\quad$
\includegraphics[clip=true,trim=3.6in 9.8in 3.6in 1.in,page=14,scale=1.]{./Fig02.pdf} $\quad$
\includegraphics[clip=true,trim=3.6in 9.8in 3.6in 1.in,page=15,scale=1.]{./Fig02.pdf} \\
\includegraphics[clip=true,trim=3.6in 9.8in 3.6in 1.in,page=16,scale=1.]{./Fig02.pdf} $\quad$
\includegraphics[clip=true,trim=3.6in 9.8in 3.6in 1.in,page=17,scale=1.]{./Fig02.pdf} $\quad$
\includegraphics[clip=true,trim=3.6in 9.8in 3.6in 1.in,page=18,scale=1.]{./Fig02.pdf} $\quad$
\includegraphics[clip=true,trim=3.6in 9.8in 3.6in 1.in,page=19,scale=1.]{./Fig02.pdf} $\quad$
\includegraphics[clip=true,trim=3.6in 9.8in 3.6in 1.in,page=20,scale=1.]{./Fig02.pdf} \\
\includegraphics[clip=true,trim=3.6in 9.9in 3.6in 1.in,page=21,scale=1.]{./Fig02.pdf} $\quad$
\includegraphics[clip=true,trim=3.6in 9.9in 3.6in 1.in,page=22,scale=1.]{./Fig02.pdf} $\quad$
\includegraphics[clip=true,trim=3.6in 9.9in 3.6in 1.in,page=23,scale=1.]{./Fig02.pdf} $\quad$
\includegraphics[clip=true,trim=3.6in 9.9in 3.6in 1.in,page=24,scale=1.]{./Fig02.pdf} $\quad$
\includegraphics[clip=true,trim=3.6in 9.9in 3.6in 1.in,page=25,scale=1.]{./Fig02.pdf}
\caption{The 25 pair modes of neighboring $\text{MA}^+$ ions considered in this paper. Each $\text{MA}^+$ ion is represented by arrows indicating the direction from $\text{C}$ to $\text{N}$, that is, the direction of the $\text{CH}_3^{}\text{NH}_3^+$ dipole.}\label{pairmodes}
\end{figure}

\begin{figure}[!ht]
\includegraphics[clip=true,trim=2.6in 9.6in 2.6in 1.in,page=1,scale=1.]{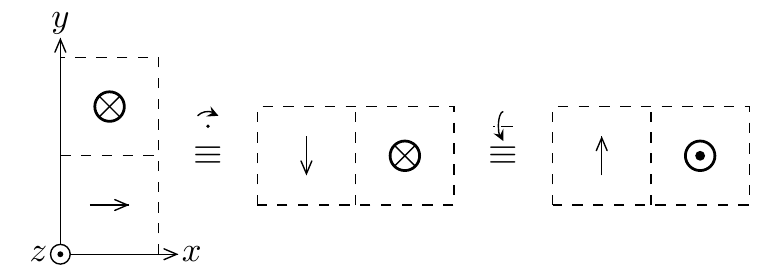}
\caption{Conversion of an arbitrary pair mode (left) into mode No.~25 (right) via a series of symmetry operations: rotation around $-z$ for $90\text{\textdegree}$ then rotation around $x$ for $180\text{\textdegree}$.}\label{equivalence}
\end{figure}

In our definition, a ``pair mode'' is the alignment of a pair of neighboring $\text{MA}^+$ ions. From the $14$ dipole directions shown in Fig.~\ref{dipoles} we can derive $14^2=196$ pair modes. However, this number can be significantly reduced to $25$ by considering only symmetry inequivalent modes, since many modes can be transformed into each other. The $25$ inequivalent modes are listed in Fig.~\ref{pairmodes}. Figure~\ref{equivalence} shows an example of transformations among several equivalent modes. We will discuss our supercell models in terms of the pair modes listed in Fig.~\ref{pairmodes} hereafter. Specifically, modes No.~1--6 and No.~11--13 are constructed by two diagonal dipoles, modes No.~7--10 and No.~14--17 by one diagonal and one face-to-face dipole, and modes No.~18--25 by two face-to-face dipoles. In this paper we will only discuss dipoles with ``strict'' face-to-face orientations. Pair modes resulting from dipoles that deviate from the face-to-face line by the angle found in our previous work \cite{LiJ16} are presented in Sec.~S1 of Ref.~\cite{SMsupercell}.

\onecolumngrid

\twocolumngrid

\subsection{Properties of optimized \texorpdfstring{$2\!\times\!2\!\times\!2$}{} supercell models}

\subsubsection{General properties analysis based on dipole-direction distribution}

In our previous primitive-cell study we found only two stable structures \cite{LiJ16}, as alluded to before. For $2\!\times\!2\!\times\!2$ supercell models, the situation changes dramatically. There are many possible alignments of $\text{MA}$-dipoles in the initial structures. The optimization of them using the aforementioned DFT approach results in different atomic geometries corresponding to different local total-energy minima. Here we first select from the many local minima and structures that we found the nonpolar structures, in which the vector sum of $\text{MA}$-dipole moments (approximately) vanishes.

\begin{figure}[!ht]
\includegraphics[clip=true,trim=1.4in 4.9in 1.4in 1.3in,page=1,scale=.23]{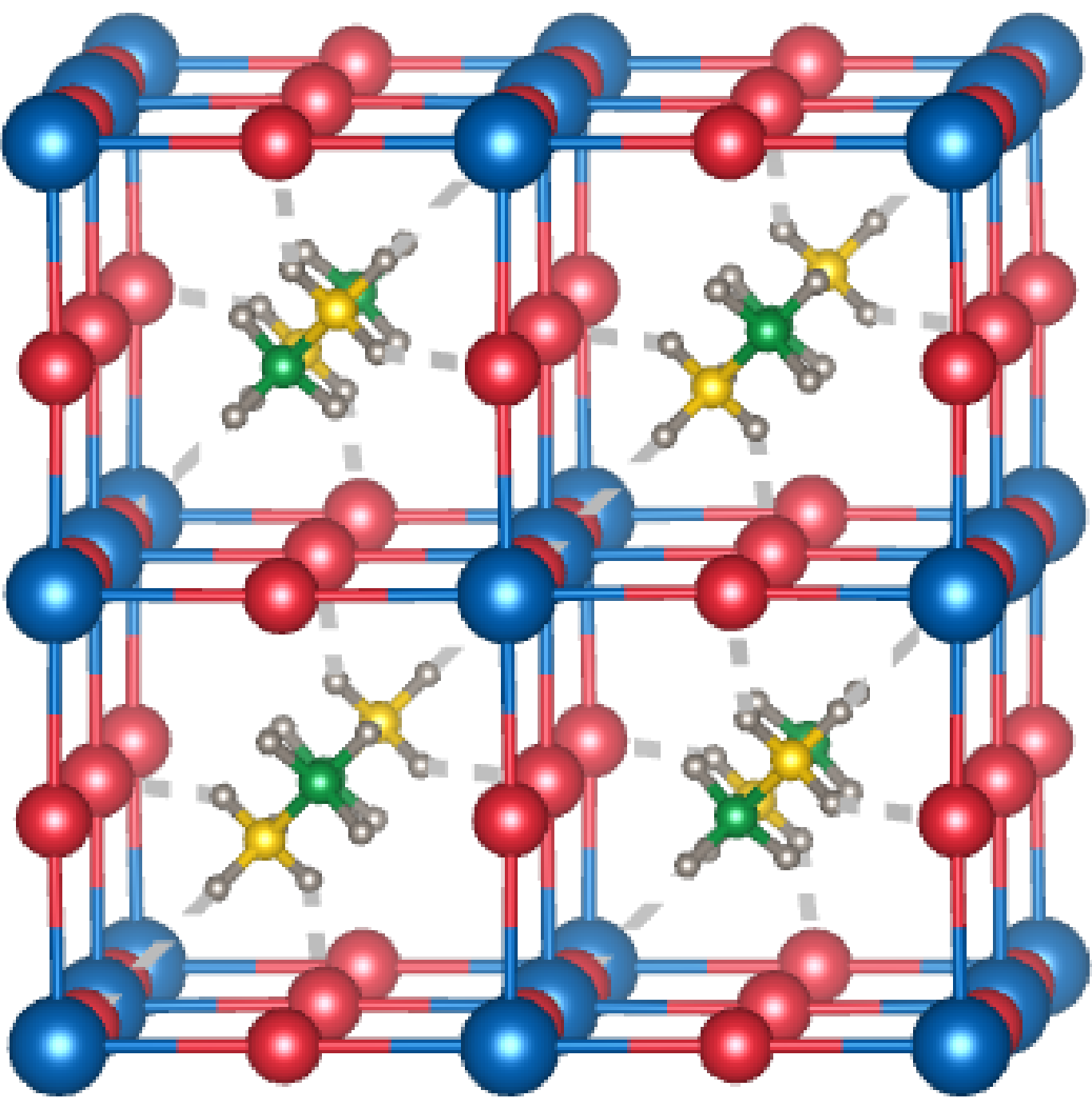}
\caption{An optimized $2\!\times\!2\!\times\!2$ supercell structure in which all $\text{MA}^+$ cations are oriented diagonally. The hydrogen bonds between $\text{I}^-$ anions and $\text{H}$ atoms in the $\text{-NH}_3^+$ group are highlighted by gray dashed lines.}\label{diag}
\end{figure}

Figure~\ref{diag} shows a geometry optimized $2\!\times\!2\!\times\!2$ supercell in which all $\text{MA}^+$ ions are oriented diagonally. The $\text{C--N}$ bonds are aligned parallel and alternatingly point in opposite directions. This results in a nearly-vanishing net dipole moment in the supercell: the three components of the average dipole-moment vector are $0.001$, $0.004$ and $0.003~p_0^{}$.

\begin{figure*}[!ht]
\begin{tabular}{p{.23\textwidth}p{.23\textwidth}cp{.23\textwidth}p{.23\textwidth}}
\multicolumn{2}{c}{(a) Structure I} & $\quad$ & \multicolumn{2}{c}{(b) Structure II} \\
\multicolumn{2}{c}{\includegraphics[clip=true,trim=1.1in 4.1in 1.1in 1.1in,page=1,scale=.23]{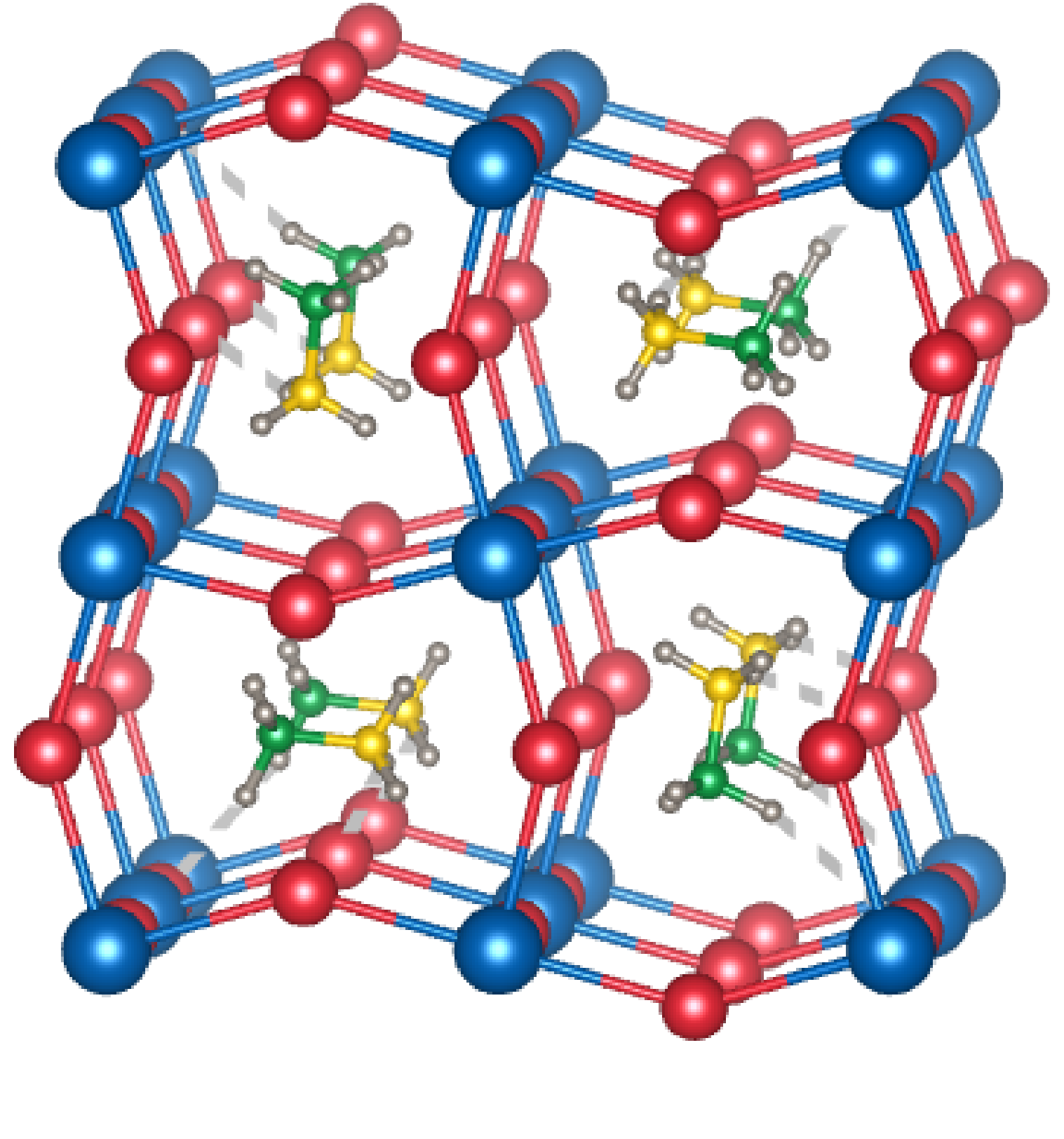}
\includegraphics[clip=true,trim=2.5in 8.in 2.5in 1.in,page=1,scale=.6]{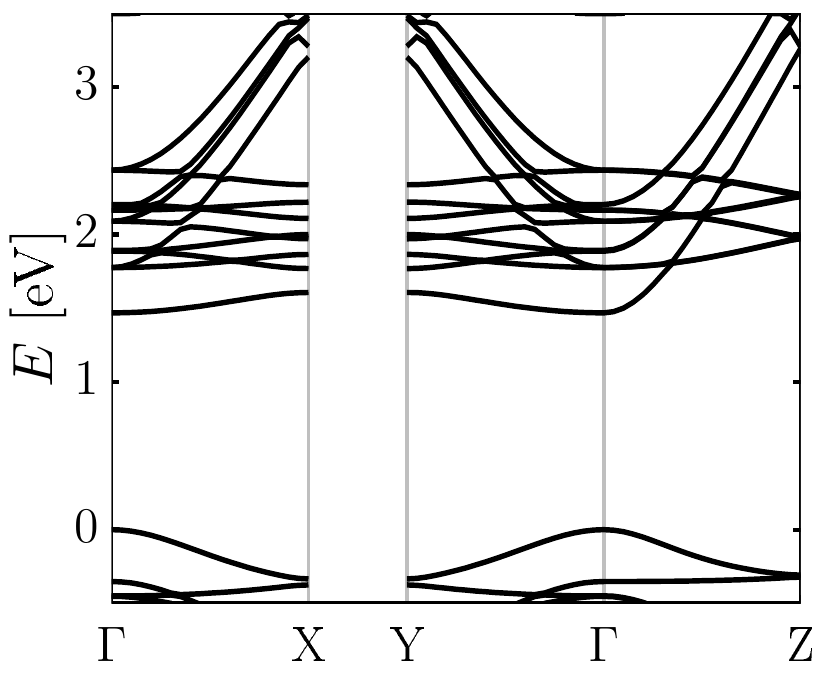}} & &
\multicolumn{2}{c}{\includegraphics[clip=true,trim=1.1in 4.1in 1.1in 1.1in,page=4,scale=.23]{./Fig05_1.pdf}
\includegraphics[clip=true,trim=2.5in 8.in 2.5in 1.in,page=2,scale=.6]{./Fig05_2.pdf}} \\
\multicolumn{2}{c}{$\text{MA}$-dipole pattern} & & \multicolumn{2}{c}{$\text{MA}$-dipole pattern} \\
\multicolumn{1}{c}{Bottom layer} & \multicolumn{1}{c}{Top layer} & & \multicolumn{1}{c}{Bottom layer} & \multicolumn{1}{c}{Top layer} \\
\multicolumn{1}{c}{\includegraphics[clip=true,trim=3.5in 9.6in 3.5in 1.in,page=2,scale=1.]{./Fig05_1.pdf}} &
\multicolumn{1}{c}{\includegraphics[clip=true,trim=3.5in 9.6in 3.5in 1.in,page=3,scale=1.]{./Fig05_1.pdf}} & &
\multicolumn{1}{c}{\includegraphics[clip=true,trim=3.5in 9.6in 3.5in 1.in,page=5,scale=1.]{./Fig05_1.pdf}} &
\multicolumn{1}{c}{\includegraphics[clip=true,trim=3.5in 9.6in 3.5in 1.in,page=6,scale=1.]{./Fig05_1.pdf}} \\
\\
\\
\multicolumn{2}{c}{(c) Structure III} & & \multicolumn{2}{c}{(d) Structure IV} \\
\multicolumn{2}{c}{\includegraphics[clip=true,trim=1.1in 4.1in 1.1in 1.1in,page=7,scale=.23]{./Fig05_1.pdf}
\includegraphics[clip=true,trim=2.5in 8.in 2.5in 1.in,page=3,scale=.6]{./Fig05_2.pdf}} & &
\multicolumn{2}{c}{\includegraphics[clip=true,trim=1.1in 4.1in 1.1in 1.1in,page=10,scale=.23]{./Fig05_1.pdf}
\includegraphics[clip=true,trim=2.5in 8.in 2.5in 1.in,page=4,scale=.6]{./Fig05_2.pdf}} \\
\multicolumn{2}{c}{$\text{MA}$-dipole pattern} & & \multicolumn{2}{c}{$\text{MA}$-dipole pattern} \\
\multicolumn{1}{c}{Bottom layer} & \multicolumn{1}{c}{Top layer} & & \multicolumn{1}{c}{Bottom layer} & \multicolumn{1}{c}{Top layer} \\
\multicolumn{1}{c}{\includegraphics[clip=true,trim=3.5in 9.6in 3.5in 1.in,page=8,scale=1.]{./Fig05_1.pdf}} &
\multicolumn{1}{c}{\includegraphics[clip=true,trim=3.5in 9.6in 3.5in 1.in,page=9,scale=1.]{./Fig05_1.pdf}} & &
\multicolumn{1}{c}{\includegraphics[clip=true,trim=3.5in 9.6in 3.5in 1.in,page=11,scale=1.]{./Fig05_1.pdf}} &
\multicolumn{1}{c}{\includegraphics[clip=true,trim=3.5in 9.6in 3.5in 1.in,page=12,scale=1.]{./Fig05_1.pdf}}
\end{tabular}
\caption{Four optimized $2\!\times\!2\!\times\!2$ supercell structures I, II, III and IV (in the upper-left panels) in which all $\text{MA}^+$ cations are oriented approximately face-to-face. For each system, the band structure (in the upper-right panel) as well as the patterns of $\text{MA}$-dipoles (in the lower panel) in the bottom ($0\!<\!z\!<\!0.5$) and top ($0.5\!<\!z\!<\!1$) layers of the unit cell (bottom panels) are also shown.}\label{f2f222}
\end{figure*}

Figure~\ref{f2f222} shows four optimized $2\!\times\!2\!\times\!2$ supercells, in which all $\text{MA}^+$ cations are oriented face-to-face. Also shown are their band structures along three high-symmetry lines $\Gamma$--$\text{X}$, $\Gamma$--$\text{Y}$ and $\Gamma$--$\text{Z}$ around the band gap (for band structures along more high-symmetry lines we refer to Sec.~S2 of Ref.~\cite{SMsupercell}). Corresponding key parameters of these structures (e.g., relative stability, band gap) are listed in Table~\ref{222}. Structure~III [Fig.~\ref{f2f222}(c)] is the most stable one as it corresponds to the lowest total energy among them. We set its total energy to $0$ hereafter. The total energy of structures~I [Fig.~\ref{f2f222}(a)], II [Fig.~\ref{f2f222}(b)] and IV [Fig.~\ref{f2f222}(d)] are $898$, $182$ and $500~\text{meV}$ per unit cell, or $112$, $23$ and $63~\text{meV}$ per $\text{MAPbI}_3^{}$, respectively. The total energy of the structure shown in Fig.~\ref{diag} is $170~\text{meV}$ per $\text{MAPbI}_3^{}$, much higher than structures~I--IV. Following the analysis of our previous work \cite{LiJ16}, this can be rationalized by the occurrence of diagonally-oriented $\text{MA}^+$ ions. These diagonally oriented dipoles prevent  the inorganic framework from releasing energy by deforming the inorganic cage, which leads to a significantly higher total energy. It is thus unlikely that many diagonally oriented dipoles occur in $\text{MAPbI}_3^{}$, which is also confirmed by our supercell calculations. Therefore we only focus on structures I--IV hereafter, in which only face-to-face $\text{MA}^+$ ions are involved. Some structural parameters and the band structure of the systems shown in Fig.~\ref{diag} are given in Sec.~S3 of Ref.~\cite{SMsupercell}.

For each optimized supercell structure, we illustrate the $\text{MA}$-alignment ``pattern'' in both Fig.~\ref{f2f222} and Table~\ref{222}. These patterns show that in structures~I--III the $\text{MA}^+$ cations are (approximately) located within the $xy$ plane and regularly aligned. For the properties of each of these system, we therefore observe an equivalence between the $x$ and $y$ directions, whereas the $z$-direction exhibits differences. For example, the lattice parameters $a$ and $b$ are approximately equal, whereas $c$ differs (cf. Table~\ref{222}). This equivalence is also reflected in the band structures [see Figs.~\ref{f2f222}(a)--(c), upper-right panels]: for each system, the band structures in the $\Gamma$--$\text{X}$ and $\Gamma$--$\text{Y}$ directions are identical, while $\Gamma$--$\text{Z}$ shows a different band dispersion.

\begin{table*}[!ht]
\caption{Geometry parameters and properties of PBE+vdW optimized $2\!\times\!2\!\times\!2$ supercell structures I, II, III and IV. Listed are lattice parameters (in $\text{\AA}$), average $\text{MA}^+$ dipole moment (in $p_0^{}$), patterns of $\text{MA}$-dipoles in the unit cell, distribution of $\text{MA}$-dipole directions, pair-mode distribution, relative total energy (in $\text{meV}$ per unit $\text{MAPbI}_3^{}$) and band gap (in $\text{eV}$).}\label{222}
\begin{tabular}{cccccllcc} \hline\hline
\multirow{2}{*}{Structure} & Lattice      & Average $\text{MA}^+$ & Bottom-layer & Top-layer & \multicolumn{1}{c}{Dipole-direction} & \multicolumn{1}{c}{Pair-mode}
& Relative       & \multirow{2}{*}{Band gap} \\
                           & ~parameters~ & dipole moment         & pattern      & pattern   & \multicolumn{1}{c}{distribution}     & \multicolumn{1}{c}{distribution}
& ~total energy~ & \\ \hline
I   & $a=12.28$ & $|p_x^{}|=0.003$ & \multirow{6}{*}{\includegraphics[clip=true,trim=3.5in 9.6in 3.5in 1.in,page=2,scale=.7]{./Fig05_1.pdf}}
                                   & \multirow{6}{*}{\includegraphics[clip=true,trim=3.5in 9.6in 3.5in 1.in,page=3,scale=.7]{./Fig05_1.pdf}}
                                   & 2~$+x$, 2~$-x$, & ~8 modes No.~20,~ & $112$ & $1.469$ \\
& $b=12.28$ & $|p_y^{}|=0.000$ & & & 2~$+y$, 2~$-y$  & ~8 modes No.~22,  &       & \\
& $c=12.68$ & $|p_z^{}|=0.195$ & & &                 & ~8 modes No.~23   &       & \\
& & & & & & & & \\
& & & & & & & & \\
II  & $a=12.26$ & $|p_x^{}|=0.001$ & \multirow{6}{*}{\includegraphics[clip=true,trim=3.5in 9.6in 3.5in 1.in,page=5,scale=.7]{./Fig05_1.pdf}}
                                   & \multirow{6}{*}{\includegraphics[clip=true,trim=3.5in 9.6in 3.5in 1.in,page=6,scale=.7]{./Fig05_1.pdf}}
                                   & 2~$+x$, 2~$-x$, & ~8 modes No.~20,~ & ~~$23$ & $1.654$ \\
& $b=12.24$ & $|p_y^{}|=0.002$ & & & 2~$+y$, 2~$-y$  & ~8 modes No.~22,  &         & \\
& $c=12.71$ & $|p_z^{}|=0.008$ & & &                 & ~8 modes No.~24   &         & \\
& & & & & & & & \\
& & & & & & & & \\
III & $a=12.24$ & $|p_x^{}|=0.001$ & \multirow{6}{*}{\includegraphics[clip=true,trim=3.5in 9.6in 3.5in 1.in,page=8,scale=.7]{./Fig05_1.pdf}}
                                   & \multirow{6}{*}{\includegraphics[clip=true,trim=3.5in 9.6in 3.5in 1.in,page=9,scale=.7]{./Fig05_1.pdf}}
                                   & 2~$+x$, 2~$-x$, & ~8 modes No.~20,~ & ~~~~$0$ & $1.762$ \\
& $b=12.24$ & $|p_y^{}|=0.000$ & & & 2~$+y$, 2~$-y$  & ~8 modes No.~22,  &         & \\
& $c=12.62$ & $|p_z^{}|=0.008$ & & &                 & ~8 modes No.~24   &         & \\
& & & & & & & & \\
& & & & & & & & \\
IV  & $a=12.55$ & $|p_x^{}|=0.003$ & \multirow{6}{*}{\includegraphics[clip=true,trim=3.5in 9.6in 3.5in 1.in,page=11,scale=.7]{./Fig05_1.pdf}}
                                   & \multirow{6}{*}{\includegraphics[clip=true,trim=3.5in 9.6in 3.5in 1.in,page=12,scale=.7]{./Fig05_1.pdf}}
                                   & 1~$+x$, 1~$-x$, & ~8 modes No.~20,~ & ~~$63$ & $1.745$ \\
& $b=12.24$ & $|p_y^{}|=0.017$ & & & 1~$+y$, 1~$-y$, & ~8 modes No.~22,  &       & \\
& $c=12.38$ & $|p_z^{}|=0.012$ & & & 2~$+z$, 2~$-z$  & ~4 modes No.~24,  &       & \\
&           &                  & & &                 & ~4 modes No.~25   & & \\
& & & & & & & & \\ \hline\hline
\end{tabular}
\end{table*}

Conversely, in structure~IV the $\text{MA}$-dipoles are oriented along the six different face-to-face directions $\pm x$, $\pm y$ and $\pm z$. The $\text{MA}^+$ alignment exhibits a quasi-random character and no equivalence between any two directions can be observed. This results in different lattice parameters $a$, $b$ and $c$. However, the root-mean-square deviation of $\{a,b,c\}$ of structure~IV is $0.13~\text{\AA}$, clearly smaller than structure~I ($0.19~\text{\AA}$), II ($0.22~\text{\AA}$) and III ($0.18~\text{\AA}$). In addition, Fig.~\ref{f2f222}(d) shows that the band structures of structure~IV along $\Gamma$--$\text{X}$, $\Gamma$--$\text{Y}$ and $\Gamma$--$\text{Z}$ are generally similar.

\subsubsection{Pair-mode analysis}

In the previous section, we discussed the lattice parameters and band structures of four characteristic $2\!\times\!2\!\times\!2$ supercell models in terms of their dipole distribution. Now we analyze their total energies and band gaps in terms of pair modes. Figure~\ref{f2f222} shows the dipole pattern for each geometry in the notation established in Fig.~\ref{dipoles} (also listed in the fourth and fifth columns of Table~\ref{222}).

For a $2\!\times\!2\!\times\!2$ supercell model, each $\text{MA}^+$ has 6 nearest neighbors, thus there are altogether $\frac{1}{2}\cdot6\cdot2^3=24$ pair modes after eliminating double counting. Since all dipoles in these four systems are oriented along face-to-face directions, only modes No.~18--25 will contribute. Specifically, in modes No.~18, 19 and 21, the two dipoles are (approximately) linearly aligned; modes No.~20 and 22 include two vertical dipoles; modes No.~23 and 24 two parallel dipoles; and the two dipoles in mode No.~25 are orthogonal to each other and not located in the same plane. These geometric characteristics are summarized in Table~\ref{f2fmodes}.

\begin{table}[!ht]
\caption{Relative geometry of dipoles in pair modes No.~18--25 defined in Fig.~\ref{pairmodes}.}\label{f2fmodes}
\begin{tabular}{ccc} \hline\hline
Mode No. & $\quad$ & Dipole-dipole Geometry \\ \hline
18 & & linear, extending \\
19 & & linear, head-to-head \\
20 & & vertical, in-plane \\
21 & & linear, tail-to-tail \\
22 & & vertical, in-plane \\
23 & & parallel \\
24 & & antiparallel \\
25 & & vertical, out-of-plane \\ \hline\hline
\end{tabular}
\end{table}

In structure~I, II or III, the orientation of dipoles within a layer (i.e., with similar $z$ coordinate) alternates between $x$ and $y$. This results in 8 pair modes of type~20 and 8 modes of type~22 (these two different ``vertical'' modes, in which the two dipoles are vertical and approximately in-plane, appear in pairs due to the periodic boundary conditions for $2\!\times\!2\!\times\!2$ supercell models). As a result, these three systems exhibit similar geometric properties as alluded to in the previous section.

However, both the total energy and band gap of structure~I are significantly different from those of structures~II and III, which we attribute to the difference in the remaining pair modes. In structures~II and III, the dipoles in the ``bottom'' and the ``top'' layers are oriented in opposite directions, introducing 8 antiparallel modes (No.~24). This similarity leads to similar band gaps (difference $\sim\!\!0.1~\text{eV}$) and relatively close total energies (difference $\sim\!20~\text{meV}$ per unit $\text{MAPbI}_3^{}$). In contrast, in structure~I, the identical alignment in these two layers results in 8 parallel modes (No.~23). The significantly higher total energy ($>\!100~\text{meV}$ per unit) suggests that mode No.~23, in which the two dipoles are aligned parallel, is less favorable than mode No.~24 that contains two antiparallel dipoles. In addition, the occurrence of the parallel mode also gives rise to a smaller band gap (by $\sim\!\!0.3~\text{eV}$ compared to structure~III). Our results for structures~I--III suggest a correlation between the stability of hybrid perovskites and the size of the band gap, that is, the higher the stability the larger the band gap. This observation agrees well with the trend reported by a recent experimental-theoretical study \cite{Baikie13}.

Despite the apparently different dipole-direction distribution, structure~IV has a similar although not identical pair-mode distribution compared to structure~III. The three-dimensional dipole-network results in 8 pair modes of type~20, 8 modes of type~22, 4 modes of type~24 and 4 modes of type~25. Hence, the major difference arises from the presence of mode No.~25, in which the two dipoles are approximately vertical and not located within the same plane. Our DFT results give a moderately higher total energy for structure~IV ($\sim\!\!60~\text{meV}$ per unit) and a band gap that is very close to that of structure~III. This suggests that mode No.~25 does not significantly reduce the overall stability of the system.

Recapping the pair-mode distribution in $2\!\times\!2\!\times\!2$ supercell models: The in-plane vertical modes, No.~20 and 22, are abundant. The antiparallel mode (No.~24) can significantly stabilize the system, whereas mode No.~25 decreases stability. The parallel mode (No.~23) is energetically unfavorable. The linear modes (No.~18, 19 and 21) are not present in structures~I--IV.

\subsection{Pair-mode distribution in optimized \texorpdfstring{$4\!\times\!4\!\times\!4$}{} supercell structures}

\subsubsection{Reference structures and dipole distributions}

In this section we discuss the DFT (PBE+vdW) results for a series of $4\!\times\!4\!\times\!4$ $\text{MAPbI}_3^{}$ supercell models. Our objective is to generate snapshots of the cubic (high-temperature) $\text{MAPbI}_3^{}$ phase, in which the $\text{MA}^+$ ions are disordered. We therefore fixed the lattice parameters to $a=b=c=25.25~\text{\AA}$ for the $4\!\times\!4\!\times\!4$ supercells. This value corresponds to four times $6.31~\text{\AA}$ \---- the experimental lattice parameter of the cubic-phase \cite{Stoumpos13}.

\begin{figure}[!ht]
\includegraphics[clip=true,trim=1.7in 5.0in 1.7in 1.in,scale=.6]{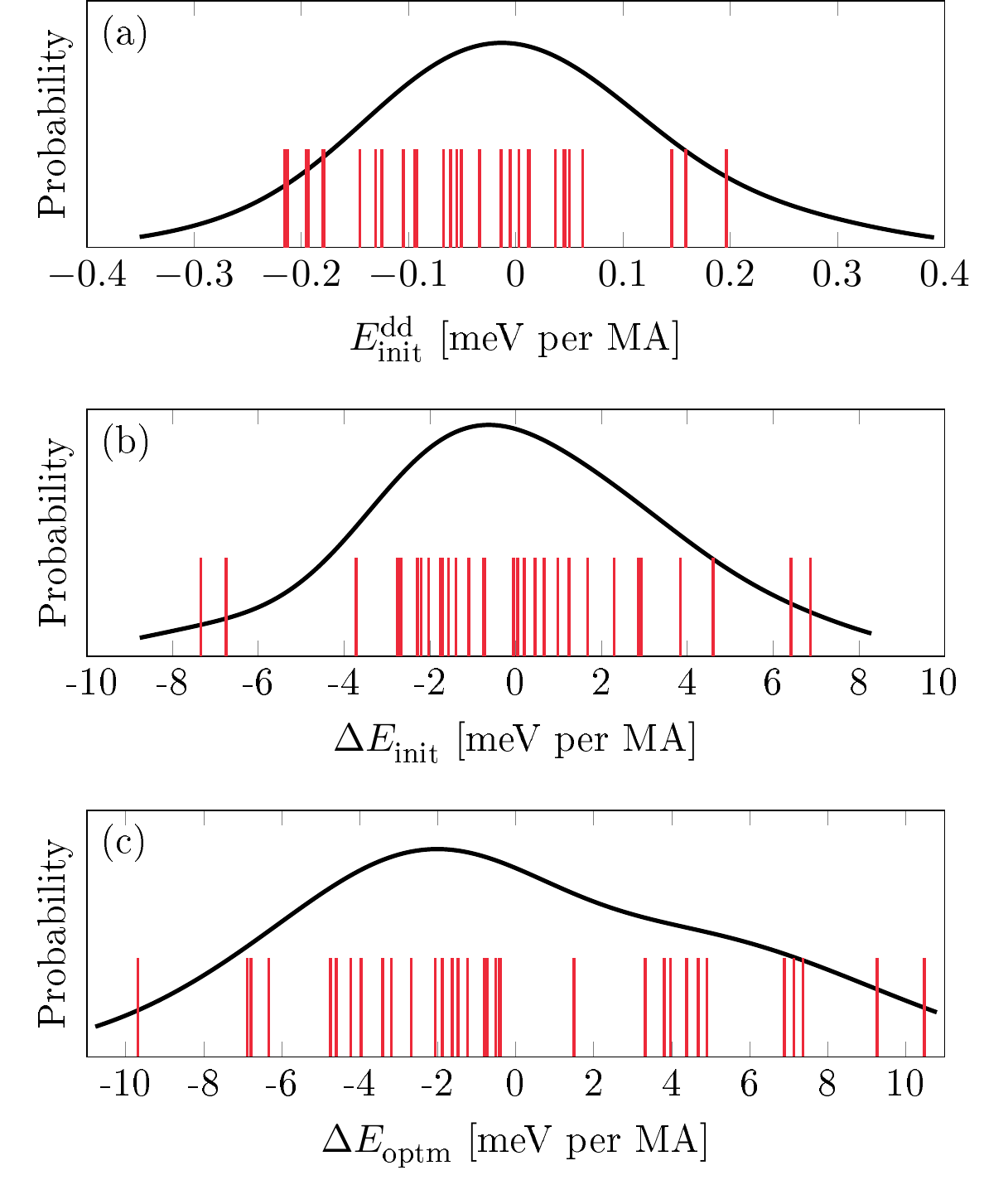}
\caption{Probability distribution of (a) electrostatic dipole-dipole interaction energy in the initial structures, (b) total energy in the initial structures, and (c) total energy in the DFT (PBE+vdW) optimized structures of $4\!\times\!4\!\times\!4$ supercell models (black curves). The red vertical lines indicate the energies of the samples discussed in this paper. All energies are given in meV per $\text{MAPbI}_3^{}$ unit.}\label{dd444}
\end{figure}

The atomic positions in the $4\!\times\!4\!\times\!4$ supercell models were randomly initialized using the following protocol:
\begin{enumerate}[(i)]
  \item The  $\text{C--N}$ bond midpoint of each $\text{MA}^+$ ion was located at the center of each single cell.
  \item For each $\text{MA}^+$, the direction of $\text{C--N}$ bond and the axial rotational angle of the ion around the $\text{C--N}$ bond was randomly set.
  \item The initial $\text{PbI}_3^-$ framework is undeformed.
\end{enumerate}
For each initial structure, we calculated the total electrostatic dipole-dipole interaction energy $E_{\text{init}}^{\text{dd}}$ by summing the interaction energy of each dipole pair. We used $p_0^{}=2.2~\text{D}$ and a dielectric constant of $25.7$ \cite{Brivio13} in this paper. The electrostatic energy is fast to compute and allows us to sample many thousands of $4\!\times\!4\!\times\!4$ models to generate an energy distribution as shown in Fig.~\ref{dd444}(a). The distribution is of Gaussian character centered at $0$. Due to the small dipole moment of each $\text{MA}$-dipole and the large dielectric constant, the distribution width is quite small ($\sim\!\!0.4~\text{meV}$ per $\text{MA}$-dipole).

We randomly selected 33 initial structures from Fig.~\ref{dd444}(a) in this paper. The distribution of their single-point total energies (centered at their mean value), as plotted in Fig.~\ref{dd444}(b), exhibits a quasi Gaussian character with a much larger width ($\sim\!\!10~\text{meV}$ per $\text{MA}^+$) than Fig.~\ref{dd444}(a). In addition, there is no clear correlation between the electrostatic-energy and total-energy sequence in these model systems. These results tell us that the electrostatic dipole-dipole interaction energy has only a minor contribution to the total energy. This implies that previously proposed large-scale simulation models based on the dipole-dipole interaction energy, such as the classical Monte Carlo approaches in Refs.~\cite{Frost14b,Leguy15}, would erroneously overemphasize the electrostatic interaction energy.

\begin{figure*}[!ht]
\begin{tabular}{ccc}
\multicolumn{3}{c}{(a) Highest-energy structure} \\
\multirow{20}{*}{\includegraphics[clip=true,trim=1in 4.4in 1in 0.9in,page=1,scale=0.5]{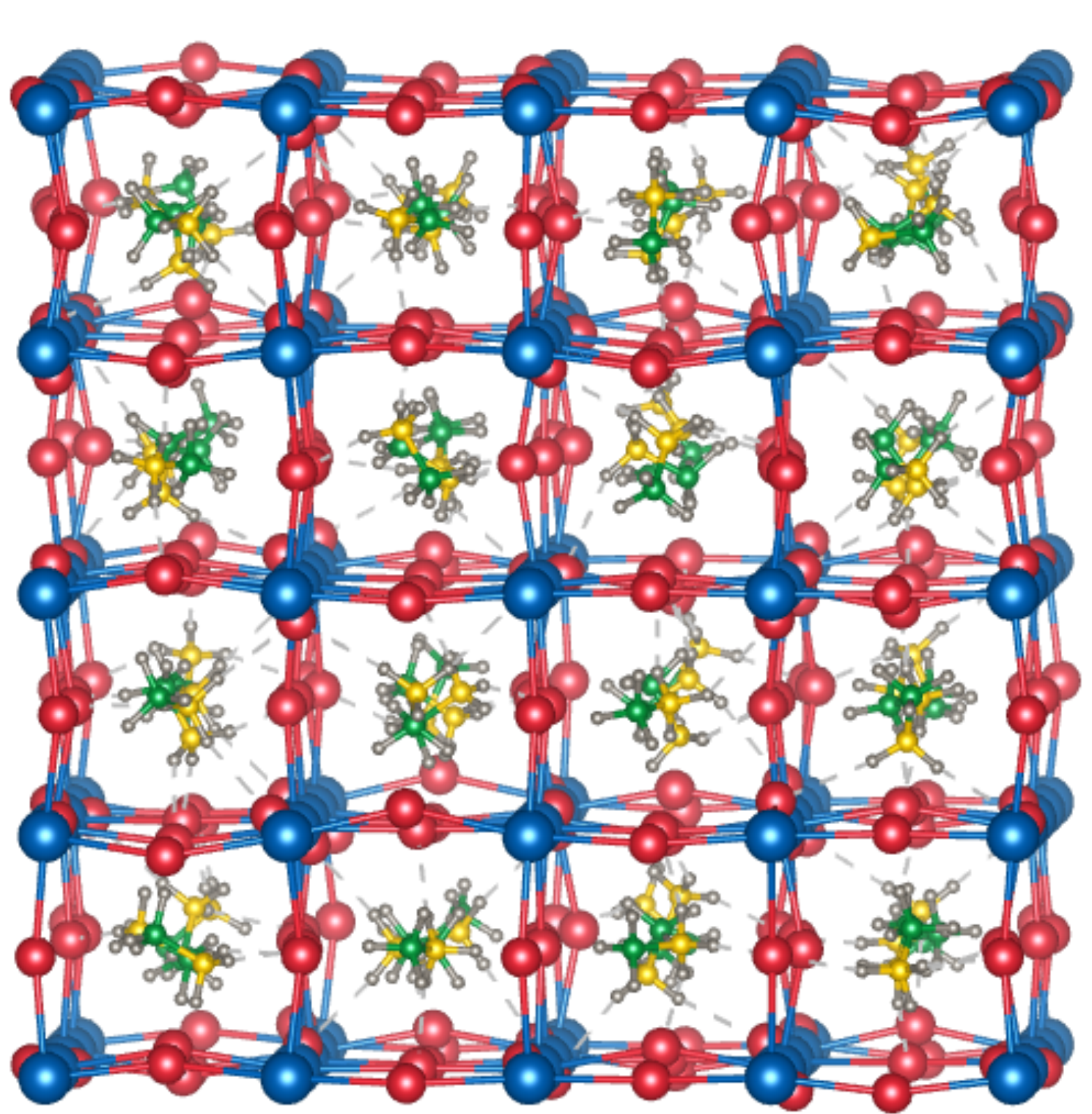}} &
\multicolumn{2}{c}{$\text{MA}$-dipole pattern} \\
&
\includegraphics[clip=true,trim=3.1in 8.8in 3.1in 1.0in,page=3,scale=0.8]{./Fig07.pdf} &
\includegraphics[clip=true,trim=3.1in 8.8in 3.1in 1.0in,page=4,scale=0.8]{./Fig07.pdf} \\
&
\includegraphics[clip=true,trim=3.1in 8.8in 3.1in 1.0in,page=5,scale=0.8]{./Fig07.pdf} &
\includegraphics[clip=true,trim=3.1in 8.8in 3.1in 1.0in,page=6,scale=0.8]{./Fig07.pdf} \\
\\
\multicolumn{3}{c}{(b) Lowest-energy structure} \\
\multirow{20}{*}{\includegraphics[clip=true,trim=1in 4.4in 1in 0.9in,page=2,scale=0.5]{./Fig07.pdf}} &
\multicolumn{2}{c}{$\text{MA}$-dipole pattern} \\
&
\includegraphics[clip=true,trim=3.1in 8.8in 3.1in 1.0in,page=7,scale=0.8]{./Fig07.pdf} &
\includegraphics[clip=true,trim=3.1in 8.8in 3.1in 1.0in,page=8,scale=0.8]{./Fig07.pdf} \\
&
\includegraphics[clip=true,trim=3.1in 8.8in 3.1in 1.0in,page=9,scale=0.8]{./Fig07.pdf} &
\includegraphics[clip=true,trim=3.1in 8.8in 3.1in 1.0in,page=10,scale=0.8]{./Fig07.pdf}
\end{tabular}
\caption{Two optimized structures of $4\!\times\!4\!\times\!4$ supercell models and their $\text{MA}$-dipole patterns within $xy$ layers of different $z$ ranges.}\label{444models}
\end{figure*}

For the DFT-optimized structures of these 33 samples, we plot the distribution of total energies in Fig.~\ref{dd444}(c). This distribution is somewhat broader than Fig.~\ref{dd444}(b) ($\sim\!\!15~\text{meV}$). Figures~\ref{444models}(a) and (b) show the optimized structures that have the highest and lowest total energies, respectively. Their difference is only $20~\text{meV}$ per $\text{MAPbI}_3^{}$ unit, which is very small compared to the total-energy difference among the $2\!\times\!2\!\times\!2$ supercell models discussed in the previous section. We can therefore use these 33 optimized structures to properly represent the many possibilities of the disordered structure of the cubic (high-temperature) $\text{MAPbI}_3^{}$ phase.

\begin{figure}[!ht]
\includegraphics[clip=true,trim=1.6in 8in 1.6in 1in,scale=.6]{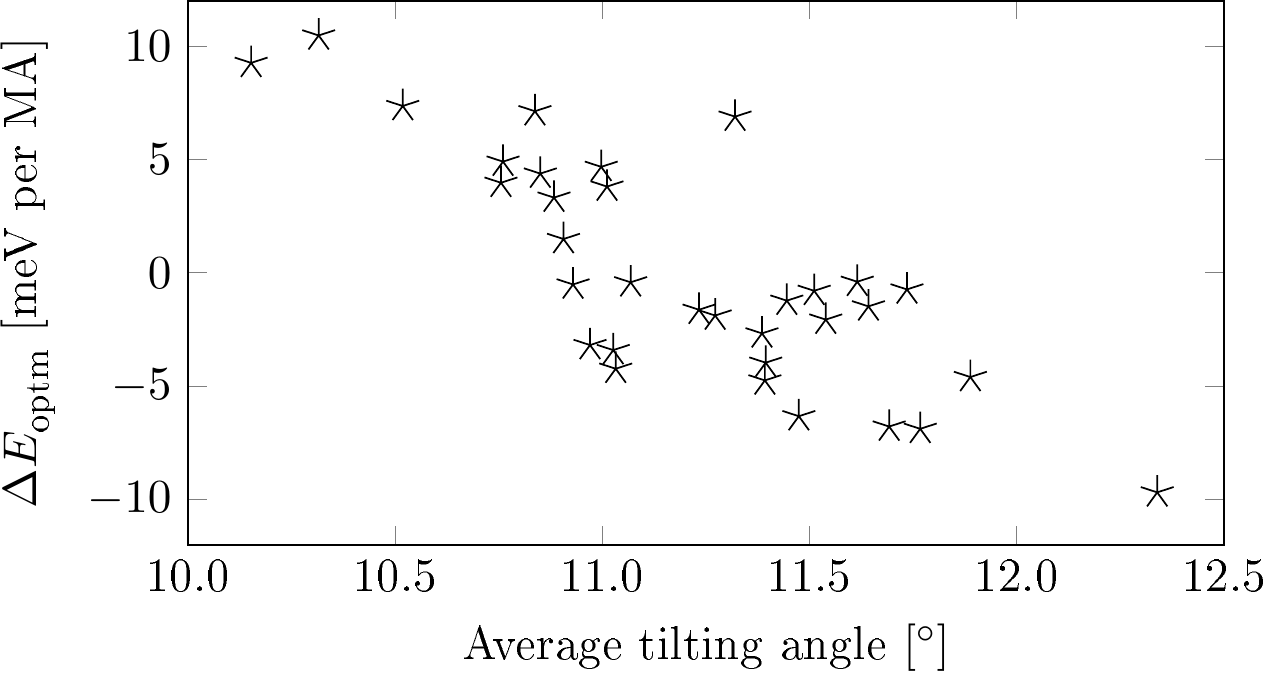}
\caption{Relative total energy (per unit $\text{MAPbI}_3^{}$) \textit{vs.} the average octahedron-tilting angle of each optimized $4\!\times\!4\!\times\!4$ supercell structure.}\label{tilting}
\end{figure}

We observe three common features in the optimized structures of these 33 model systems [Figures~\ref{444models}(a) and (b) as two examples]. First, the average dipole moments are small (data not shown), thus they can be considered approximately nonpolar. Second, most of the $\text{MA}$ dipoles, which were fully-randomly initialized, were reoriented into face-to-face directions in the DFT-optimized structures. Third, they exhibit noticeable inorganic-framework deformation, which is irregular and local, and occurs along all three lattice vectors. This is very different to the deformation patterns in the ordered $2\!\times\!2\!\times\!2$ structures shown in Figs.~\ref{f2f222}(a)--(c) (they are periodically extended within parallel lattice planes). The average octahedron-tilting angle is $11.2\text{\textdegree}$, clearly smaller than the (in-plane) tilting angles in Figs.~\ref{f2f222}(a)--(c) (ranging between $15.5$ and $16.9\text{\textdegree}$), as well as the average tilting angle in the irregular structure Fig.~\ref{f2f222}(d) ($14.5\text{\textdegree}$). Figure~\ref{tilting} shows that, there is a rough correlation between the larger average tilting angle and the lower total energy within these 33 optimized cubic supercell structures.

In these 33 model systems, the  total energy of the relaxed structures is on average $225~\text{meV}$ per $\text{MAPbI}_3^{}$ unit lower than that of the initial structures. The contribution to this total-energy minimization, as we understand, consists of two major components: the formation of hydrogen bonds ($40\text{--}50~\text{meV}$ per bond, thus $120\text{--}150~\text{meV}$ for three bonds \cite{Egger14,LiJ16,LiJ18a,LiJ18b}), and the inorganic-cage deformation. Estimated in this way, the latter contribution is much larger than the $\sim\!\!20~\text{meV}$ per unit $\text{MAPbI}_3^{}$ (with an average deformation angle of $5.4\text{\textdegree}$) reported in our previous primitive-cell study \cite{LiJ16}. This can be rationalized by the much larger average tilting angle ($11.2\text{\textdegree}$) in the optimized $4\!\times\!4\!\times\!4$ structures. Most of the $\text{MA}$-dipoles (a) are properly bound to the inorganic framework and (b) adopt the (quasi-)face-to-face direction in the optimized structures as alluded to earlier. We therefore argue that the difference in total energies among the 33 DFT-relaxed structures is mainly due to different alignments of $\text{MA}$-dipoles which result in different magnitudes and shapes of inorganic-framework deformation.

\subsubsection{Pair-mode analysis}

To understand the final dipole alignments we make use of our pair-mode concept. Here we first analyze the pair-mode distribution of the optimized structures with the highest [Fig.~\ref{444models}(a)] and lowest [Fig.~\ref{444models}(b)] total energies. Their dipole patterns in different $xy$ layers (the $z$ coordinates of $\text{MA}^+$ ions are similar within the same layer) are plotted in Fig.~\ref{444models}. Table~\ref{444modelsgeom} summarizes the distribution of altogether $\frac{1}{2}\cdot6\cdot4^3=192$ modes in each structure.

\begin{table}[!ht]
\caption{Pair-mode distribution of optimized $4\!\times\!4\!\times\!4$ supercell models shown in Figs.~\ref{444models}(a) and (b).}\label{444modelsgeom}
\begin{tabular}{ccc} \hline\hline
\multirow{2}{*}{Mode No.} & \hspace{2.em}Fig.~\ref{444models}(a)\hspace{2.em} & \hspace{2.em}Fig.~\ref{444models}(b)\hspace{2.em} \\
& (highest energy) & (lowest energy) \\ \hline
\multicolumn{3}{l}{Diagonal-dipole containing pair modes} \\
1--17 & $~~6$ & $~~0$ \\
\multicolumn{3}{l}{Linear modes} \\
18 & $13$  & $~~2$ \\
19 & $~~3$ & $~~1$ \\
21 & $~~2$ & $~~3$ \\
\multicolumn{3}{l}{Vertical modes} \\
20 & $42$ & $60$ \\
22 & $44$ & $56$ \\
25 & $40$ & $22$ \\
\multicolumn{3}{l}{Parallel modes} \\
23 & $21$ & $17$ \\
24 & $21$ & $31$ \\ \hline\hline
\end{tabular}
\end{table}

Four aspects characterize the pair-mode distributions:
\begin{enumerate}[(1)]
  \item Diagonal dipoles are very rare: only $1$ in Fig.~\ref{444models}(a), resulting in $6$ pair modes in the mode 1--17 category. All other dipoles are distributed over all six face-to-face directions $\pm x$, $\pm y$ and $\pm z$, similar to structure~IV of $2\!\times\!2\!\times\!2$ supercell model discussed previously.
  \item Vertical modes, especially the in-plane modes (No.~20 and 22), dominate in both structures. The out-of-plane mode (No.~25) is of nearly equal importance in Fig.~\ref{444models}(a), while in Fig.~\ref{444models}(b) it is less frequent.
  \item The only noticeable distribution of linear modes is No.~18 in Fig.~\ref{444models}(a) ($13/192=0.068$).
  \item The two parallel modes are equally distributed in Fig.~\ref{444models}(a), while in Fig.~\ref{444models}(b) the antiparallel mode (No.~24) is significantly more populated than No.~23.
\end{enumerate}
To summarize, the lowest-energy structure [Fig.~\ref{444models}(b)] exhibits much higher distribution in pair modes No.~20 and 22, nearly no population in linear modes, and obvious importance in mode No.~24 over No.~23.

\begin{figure}[!ht]
{\footnotesize (a) Distribution of each sample} \\
\includegraphics[clip=true,trim=1.55in 6.9in 1.55in 1.in,page=1,scale=.6]{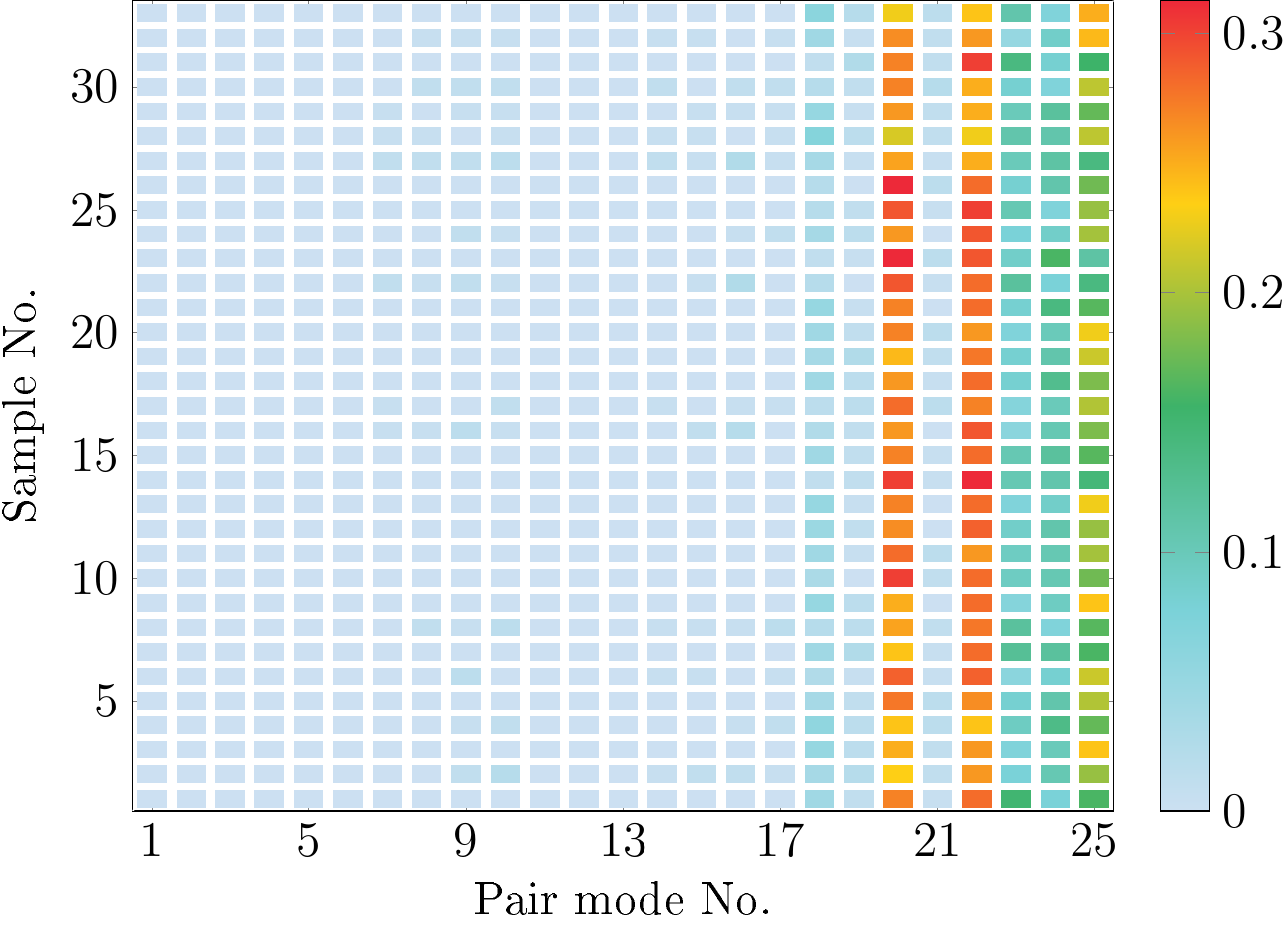}

{\footnotesize (b) Overall distribution} \\
\includegraphics[clip=true,trim=1.80in 7.3in 0.95in 1.in,page=2,scale=.6]{./Fig09.pdf}
\caption{Pair-mode distribution of optimized $4\!\times\!4\!\times\!4$ supercell models: (a) distribution of each individual sample, and (b) the overall distribution of all samples.}\label{pmall}
\end{figure}

To quantify the pair-mode pattern, we generate a pair-mode distribution from the 33 considered samples in this work. Figure~\ref{pmall}(a) shows the pair-mode distribution of each sample as a heat map, and Fig.~\ref{pmall}(b) shows the overall pair-mode distribution of all 33 samples (altogether $33\cdot\frac{1}{2}\cdot6\cdot4^3=6336$ modes). Figure~\ref{pmall}(a) indicates that the pair-mode distributions associated with all samples are similar. In general, the population of all pair modes that involve one or two diagonal dipoles, i.e., No.~1--17, is negligible. This indicates that the probability to find a diagonally oriented $\text{MA}^+$ cation in the relaxed disordered cubic structure is very low, in good agreement with the \textit{ab initio} MD results of polar-angle distribution of $\text{MA}^+$ ions \cite{Lahnsteiner16}. The vertical modes (No.~20, 22 and 25), especially the in-plane modes No.~20 and 22, are the most dominant. The parallel (No.~23) and antiparallel (No.~24) modes are also noticeably distributed and they have almost equal population. Finally, the population of the linear modes (No.~18, 19 and 21) is small. The occurrence of the head-to-head (No.~19) and tail-to-tail (No.~21) modes is almost negligible.

The pair-mode distribution shown in Fig.~\ref{pmall} enables us to construct large $\text{MAPbI}_3^{}$ models that are out of reach of DFT. Since such large models follow the $\text{MA}$-distribution in stable configurations and include $\text{MA}^+$ nearest-neighbor interactions, they will provide good models to study realistic $\text{MAPbI}_3^{}$ structures under realistic conditions. 
The construction of such large multi-scale $\text{MAPbI}_3^{}$ models will be the subject of future work \cite{Jaervi18}.

\begin{figure}[!ht]
\subfigure[~Pair mode No.~18]{\includegraphics[clip=true,trim=3.5in 9.5in 3.5in 1.in,page=1,scale=1]{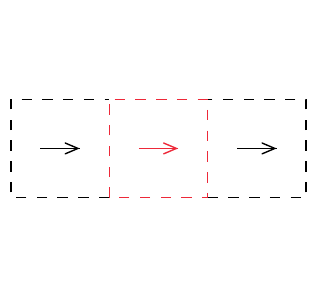}} $\quad$
\subfigure[~Pair mode No.~24]{\includegraphics[clip=true,trim=3.5in 9.5in 3.5in 1.in,page=2,scale=1]{./Fig10.pdf}}
\caption{Possibilities to construct pair mode (a) No.~18 and (b) No.~24 for a face-to-face dipole (colored in red) with its nearest neighbors (colored in black).}\label{possibilities}
\end{figure}

To better understand the pair-mode distribution in Fig.~\ref{pmall}, we focus on the modes No.~18--25. For an arbitrary face-to-face dipole, the possibilities to construct different pair modes with another face-to-face nearest neighbor are different. For example, there are 2 ways to construct mode No.~18 [Fig.~\ref{possibilities}(a)] and 4 ways for No.~24 [Fig.~\ref{possibilities}(b)]. Table~\ref{f2f} lists the number of possibilities and the corresponding probabilities (i.e., number of possibilities divided by $6\cdot6=36$). These \emph{probabilities} refer to \emph{fully-random} systems, in which the dipoles are (a) oriented along face-to-face directions and (b) do not interacting with each other (i.e., neither electrostatically nor via cage deformation). From the DFT results, we can then extract the probabilities in the relaxed structures associated with these modes that take the full electrostatic and structural response into account. We define these probabilities as the number of times a mode occurs in the 33 samples divided by the total number of modes in the $4\!\times\!4\!\times\!4$ supercell models ($6336$). The resulting probabilities are also listed in Table~\ref{f2f} and enable a direct comparison between non-interacting dipoles and the real systems.

\begin{table}[!ht]
\caption{Number of possibilities and probability to construct a pair mode (from No.~18 to No.~25) for an arbitrary dipole in a system in which all dipoles are fully randomly distributed in 6 face-to-face directions, as well as the probability of this mode in the 33 investigated relaxed $4\!\times\!4\!\times\!4$ supercell-models.}\label{f2f}
\begin{tabular}{ccccc} \hline\hline
\multirow{2}{*}{Mode No.} & Number of       & Probability in       & ~ & Probability in \\
                            & possibilities & fully-random systems & & relaxed systems \\ \hline
18 & 2 & $0.056$ & & $0.036$ \\
19 & 1 & $0.028$ & & $0.012$ \\
20 & 8 & $0.222$ & & $0.267$ \\
21 & 1 & $0.028$ & & $0.009$ \\
22 & 8 & $0.222$ & & $0.274$ \\
23 & 4 & $0.111$ & & $0.092$ \\
24 & 4 & $0.111$ & & $0.103$ \\
25 & 8 & $0.222$ & & $0.189$ \\ \hline\hline
\end{tabular}
\end{table}


We start from the vertical modes No.~20, 22 and 25. They are the highest populated modes in fully-random systems thus can be understood as ``intrinsically'' dominant modes. In relaxed structures, there are significant increases in the distribution in both in-plane modes, No.~20 ($0.045$, namely $20.3\%$) and 22 ($0.051$ or $23.1\%$). In contrast, a noticeable drop ($0.033$ or $14.9\%$) can be observed in the DFT results for the out-of-plane mode No.~25. For the least populated linear modes, the $\text{MA}$-pair interaction results in a $0.019$ ($34.4\%$) drop for mode No.~18, and a more significant decrease in the population in both modes No.~19 and 22 (the distribution in these two modes almost vanish in the optimized structures). Finally, the distribution in both parallel (No.~23) and antiparallel (No.~24) modes are nearly identical and slightly lower than the theoretical values.

\subsubsection{Pair-mode expansion of the total energy}

We can now use the dominant modes in Tab.~\ref{f2f}, that is modes No.~18, 20 and 22--25, to perform a mode expansion of the total energy of a supercell structure:
\begin{align}
E_{\text{optm}}^{} = \sum_n p_n^{} E_n^{} + \text{const}, &\quad
n\in\{18,20,22,23,24,25\}. \label{fit}
\end{align}
Here $n$ labels the pair modes, $p_n^{}$ is the probability of the $n$th mode (last column of Tab.~\ref{f2f}), and  $E_n^{}$ the associated ``pair-mode energy''. Using Eq.~(\ref{fit}) to fit the total-energy and pair-mode-distribution data, we obtained the set of $E_n^{}$ listed in Table~\ref{pmen}. We use the constant term in Eq.~\ref{fit} to shift  the smallest pair-mode energy value, i.e. $E_{20}^{}$, to $0$. The resulting pair-mode energis agree reasonably well with our analysis of the pair-mode distribution, and can give us an estimate of how the system will react to the change of an $\text{MA}$-dipole from one face-to-face direction to another. For example, from mode No.~20 to No.~23, the system total energy will increase by $\sim\!\!50~\text{meV}$.

\begin{table}[!ht]
\caption{Pair-mode energies (in $\text{meV}$) of modes No.~18, 20 and 22--25 calculated by fitting Eq.~(\ref{fit}).}\label{pmen}
\begin{tabular}{ccc} \hline\hline
Mode No. ($n$) & $\quad$ & Pair-mode energy ($E_n^{}$) \\ \hline
18 & & $52.9$ \\
20 & & $~~0~~~$ \\
22 & & $~~6.8$ \\
23 & & $49.9$ \\
24 & & $24.5$ \\
25 & & $31.9$ \\ \hline\hline
\end{tabular}
\end{table}

Our results indicate that, in $\text{MAPbI}_3^{}$, the interaction between the neighboring $\text{MA}^+$ ions favors the in-plane vertical modes No.~20 and 22. The prevalent population of these two modes (together with the out-of-plane vertical mode No.~25) causes a three-dimensional isotropy on a large length scale, which effectively makes the material cubic. Conversely, the vanishing population in the linear modes (No.~18, especially No.~19 and 21) strongly limits the formation of linearly aligned neighboring $\text{MA}^+$ ions. This is very different to the formation of long linear $\text{MA}$-chains predicted by the combination of \textit{ab initio} MD and classical Monte Carlo simulations based on purely electrostatic interactions \cite{Frost14b,Leguy15}.

\subsubsection{Angular distribution of dipoles}

\begin{figure}[!ht]
\includegraphics[clip=true,trim=1.6in 7.8in 1.6in 1in,scale=.6]{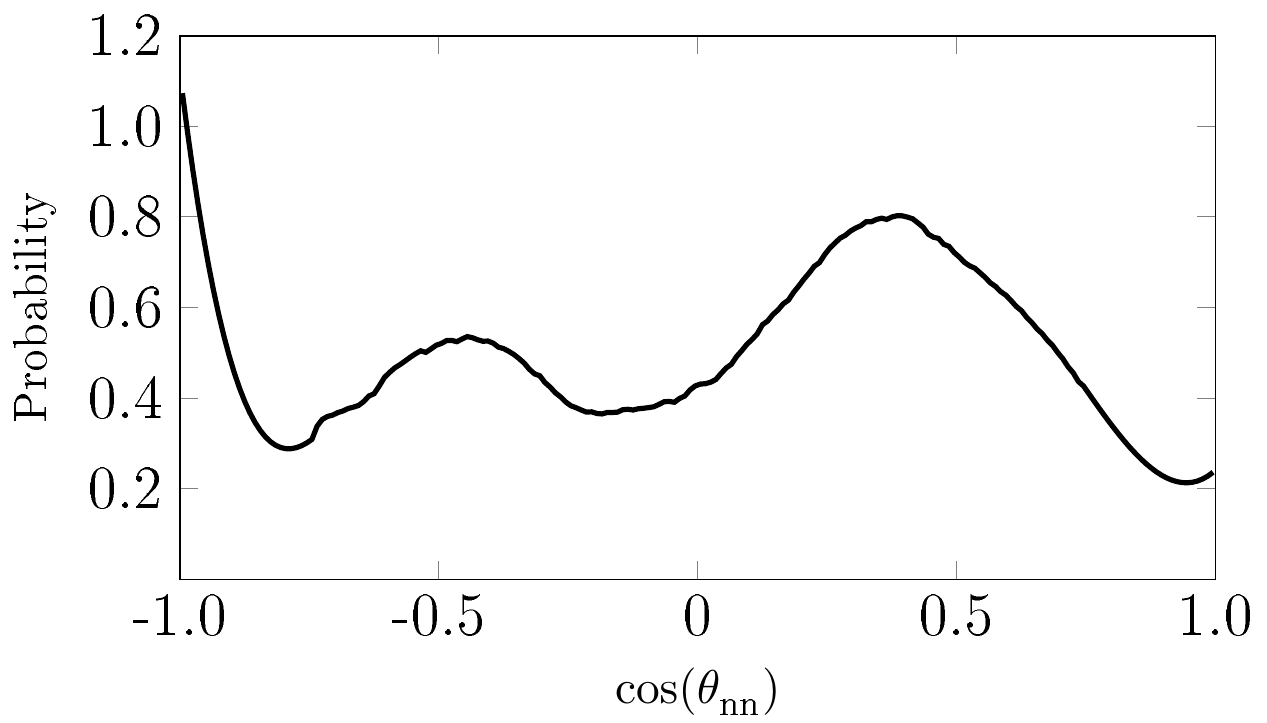}
\caption{Distribution of the cosine function of $\theta_{\text{nn}}^{}$ (the angle between two nearest-neighbor $\text{MA}$-dipoles) averaged over all 33 optimized $4\!\times\!4\!\times\!4$ supercell structures.}\label{angle}
\end{figure}

So far we have limited our discussion to discrete $\text{MA}$-dipole angles ($0$, $90$ and $180\text{\textdegree}$). In reality, the angle varies continuously between $0$ and $180\text{\textdegree}$. From the DFT-optimized $4\!\times\!4\!\times\!4$ structures we extracted the angle between each pair of nearest-neighbor $\text{MA}$-dipoles. The distribution of this angle (plotted in Fig.~\ref{angle}) is similar to the previous \textit{ab initio} MD results \cite{Lahnsteiner16}. Here we briefly discuss the three characteristic features of the distribution:
\begin{enumerate}[(1)]
  \item $180\text{\textdegree}$ (cosine $=-1$), corresponding to pair modes No.~19, 21 and 24: the distribution exhibits a sharp peak, since only population in No.~24 is noticeable and the two dipoles tend to take nearly perfect antiparallel alignment.
  \item $0\text{\textdegree}$ (cosine $=1$), modes No.~18 and 23: in many cases the two dipoles form an angle (up to $\sim\!\!45\text{\textdegree}$) instead of being aligned perfectly in one direction, thus resulting in a broad distribution.
 \item $90\text{\textdegree}$ (cosine $=0$): the most populated modes No.~20, 22 and 25 are all included in the large region of $(\sim\!\!45,\sim\!\!135)\text{\textdegree}$.
\end{enumerate}

\section{Conclusions}\label{concl}

We have studied a series of $\text{MAPbI}_3^{}$ supercell models using DFT. To establish a multi-scale model, we derived and analyzed the concept of ``pair modes'', i.e., the interaction of nearest $\text{MA}$-ion pairs. We first investigated several small $2\!\times\!2\!\times\!2$ supercell models, in which we can cancel out the overall dipole moment by hand using suitable dipole orientations. Our DFT results indicate that differences in pair modes have a significant effect on the atomic and electronic structure of these models. This finding motivated our exploration into larger $4\!\times\!4\!\times\!4$ $\text{MAPbI}_3^{}$ supercell models, which we used to simulate disordered $\text{MAPbI}_3^{}$ structures by randomly initializing the $\text{MA}^+$ orientations. Structural optimization using DFT and our pair mode analysis reveal that the final locations of $\text{MA}^+$ ions is not fully random, but follows certain preferred orientations that depend on the surrounding. Our results indicate that vertical geometries are preferred for nearest $\text{MA}$-pairs, which will lead to the formation of three-dimensionally isotropic network of $\text{MA}$ dipoles. In contrast, linearly extended $\text{MA}$-pairs in neighboring cells are largely suppressed.

The discussion of pair modes and their distribution in this paper is based on a series of (meta)stable configurations of disordered $\text{MAPbI}_3^{}$. DFT is an appropriate tool for this purpose as it can predict the geometries and give a reliable estimation of total energies of these configurations. We can use this dipole distribution to build a multi-scale model to generate the local structure in bulk $\text{MAPbI}_3^{}$ samples on large length scales, e.g., of a few tens of single cells \cite{Jaervi18}. The fitted pair-mode energies obtained in this work and the energy barriers for an $\text{MA}$ dipole to change its direction calculated in our other works using DFT \cite{LiJ18a,LiJ18b} allow us to include temperature in this multi-scale modeling scheme

\section*{Acknowledgment}

We thank H.~Levard as well as M.~Bokdam and J.~Lahnsteiner for fruitful discussions. The generous allocation of computing resources by the CSC-IT Center for Science (via the Project No.~ay6311) and the Aalto Science-IT project are gratefully acknowledged. An award of computer time was provided by the Innovative and Novel Computational Impact on Theory and Experiment (INCITE) program. This research used resources of the Argonne Leadership Computing Facility, which is a DOE Office of Science User Facility supported under Contract DE-AC02-06CH11357. This work was supported by the Academy of Finland through its Centres of Excellence Programme (2012-2014 and 2015-2017) under project numbers 251748 and 284621, as well as its Key Project Funding scheme under project number 305632.

\bibliographystyle{apsrev}

\begin{thebibliography}{53}
\expandafter\ifx\csname natexlab\endcsname\relax\def\natexlab#1{#1}\fi
\expandafter\ifx\csname bibnamefont\endcsname\relax
  \def\bibnamefont#1{#1}\fi
\expandafter\ifx\csname bibfnamefont\endcsname\relax
  \def\bibfnamefont#1{#1}\fi
\expandafter\ifx\csname citenamefont\endcsname\relax
  \def\citenamefont#1{#1}\fi
\expandafter\ifx\csname url\endcsname\relax
  \def\url#1{\texttt{#1}}\fi
\expandafter\ifx\csname urlprefix\endcsname\relax\def\urlprefix{URL }\fi
\providecommand{\bibinfo}[2]{#2}
\providecommand{\eprint}[2][]{\url{#2}}

\bibitem[{\citenamefont{Snaith}(2013)}]{Snaith13}
\bibinfo{author}{\bibfnamefont{H.~J.}~\bibnamefont{Snaith}},
  \bibinfo{journal}{J. Phys. Chem. Lett.} \textbf{\bibinfo{volume}{4}},
  \bibinfo{pages}{3623} (\bibinfo{year}{2013}).

\bibitem[{\citenamefont{Green et~al.}(2014)\citenamefont{Green, Ho-Baillie, and
  Snaith}}]{Green14}
\bibinfo{author}{\bibfnamefont{M.~A.} \bibnamefont{Green}},
  \bibinfo{author}{\bibfnamefont{A.}~\bibnamefont{Ho-Baillie}},
  \bibnamefont{and} \bibinfo{author}{\bibfnamefont{H.~J.}
  \bibnamefont{Snaith}}, \bibinfo{journal}{Nature Photon.}
  \textbf{\bibinfo{volume}{8}}, \bibinfo{pages}{506} (\bibinfo{year}{2014}).

\bibitem[{NRE()}]{NRELchart}
\bibinfo{note}{{\htmladdnormallink{\color{blue}{https://www.nrel.gov/pv/assets/images/efficiency-chart.png}}{https://www.nrel.gov/pv/assets/images/efficiency-chart.png}}
  \\ (National Renewable Energy Laboratory: Best research-cell efficiencies,
  2016).}

\bibitem[{\citenamefont{Kim et~al.}(2012)\citenamefont{Kim, Lee, Im, Lee,
  Moehl, Marchioro, Moon, Humphry-Baker, Yum, Moser et~al.}}]{KimHS12}
\bibinfo{author}{\bibfnamefont{H.-S.}~\bibnamefont{Kim}},
  \bibinfo{author}{\bibfnamefont{C.-R.}~\bibnamefont{Lee}},
  \bibinfo{author}{\bibfnamefont{J.-H.}~\bibnamefont{Im}},
  \bibinfo{author}{\bibfnamefont{K.-B.}~\bibnamefont{Lee}},
  \bibinfo{author}{\bibfnamefont{T.}~\bibnamefont{Moehl}},
  \bibinfo{author}{\bibfnamefont{A.}~\bibnamefont{Marchioro}},
  \bibinfo{author}{\bibfnamefont{S.-J.}~\bibnamefont{Moon}},
  \bibinfo{author}{\bibfnamefont{R.}~\bibnamefont{Humphry-Baker}},
  \bibinfo{author}{\bibfnamefont{J.-H.}~\bibnamefont{Yum}},
  \bibinfo{author}{\bibfnamefont{J.~E.}~\bibnamefont{Moser}},
  \bibnamefont{et~al.}, \bibinfo{journal}{Sci. Rep.}
  \textbf{\bibinfo{volume}{2}}, \bibinfo{pages}{591} (\bibinfo{year}{2012}).

\bibitem[{\citenamefont{Lee et~al.}(2012)\citenamefont{Lee, Teuscher, Miyasaka,
  Murakami, and Snaith}}]{LeeMM12}
\bibinfo{author}{\bibfnamefont{M.~M.}~\bibnamefont{Lee}},
  \bibinfo{author}{\bibfnamefont{J.}~\bibnamefont{Teuscher}},
  \bibinfo{author}{\bibfnamefont{T.}~\bibnamefont{Miyasaka}},
  \bibinfo{author}{\bibfnamefont{T.~N.}~\bibnamefont{Murakami}},
  \bibnamefont{and} \bibinfo{author}{\bibfnamefont{H.~J.}
  \bibnamefont{Snaith}}, \bibinfo{journal}{Science}
  \textbf{\bibinfo{volume}{338}}, \bibinfo{pages}{643} (\bibinfo{year}{2012}).

\bibitem[{\citenamefont{Jackson et~al.}(2016)\citenamefont{Jackson, Wuerz,
  Hariskos, Lotter, Witte, and Powalla}}]{Jackson16}
\bibinfo{author}{\bibfnamefont{P.}~\bibnamefont{Jackson}},
  \bibinfo{author}{\bibfnamefont{R.}~\bibnamefont{Wuerz}},
  \bibinfo{author}{\bibfnamefont{D.}~\bibnamefont{Hariskos}},
  \bibinfo{author}{\bibfnamefont{E.}~\bibnamefont{Lotter}},
  \bibinfo{author}{\bibfnamefont{W.}~\bibnamefont{Witte}}, \bibnamefont{and}
  \bibinfo{author}{\bibfnamefont{M.}~\bibnamefont{Powalla}},
  \bibinfo{journal}{Phys. Status Solidi Rapid Res. Lett.}
  \textbf{\bibinfo{volume}{10}}, \bibinfo{pages}{583} (\bibinfo{year}{2016}).

\bibitem[{\citenamefont{Stoumpos et~al.}(2013)\citenamefont{Stoumpos,
  Malliakas, and Kanatzidis}}]{Stoumpos13}
\bibinfo{author}{\bibfnamefont{C.~C.}~\bibnamefont{Stoumpos}},
  \bibinfo{author}{\bibfnamefont{C.~D.}~\bibnamefont{Malliakas}},
  \bibnamefont{and} \bibinfo{author}{\bibfnamefont{M.~G.}
  \bibnamefont{Kanatzidis}}, \bibinfo{journal}{Inorg. Chem.}
  \textbf{\bibinfo{volume}{52}}, \bibinfo{pages}{9019} (\bibinfo{year}{2013}).

\bibitem[{\citenamefont{de~Wolf et~al.}(2014)\citenamefont{de~Wolf, Holovsky,
  Moon, L\"oper, Niesen, Ledinsky, Haug, Yum, and Ballif}}]{deWolf14}
\bibinfo{author}{\bibfnamefont{S.}~\bibnamefont{de~Wolf}},
  \bibinfo{author}{\bibfnamefont{J.}~\bibnamefont{Holovsky}},
  \bibinfo{author}{\bibfnamefont{S.-J.}~\bibnamefont{Moon}},
  \bibinfo{author}{\bibfnamefont{P.}~\bibnamefont{L\"oper}},
  \bibinfo{author}{\bibfnamefont{B.}~\bibnamefont{Niesen}},
  \bibinfo{author}{\bibfnamefont{M.}~\bibnamefont{Ledinsky}},
  \bibinfo{author}{\bibfnamefont{F.-J.}~\bibnamefont{Haug}},
  \bibinfo{author}{\bibfnamefont{J.-H.}~\bibnamefont{Yum}}, \bibnamefont{and}
  \bibinfo{author}{\bibfnamefont{C.}~\bibnamefont{Ballif}},
  \bibinfo{journal}{J. Phys. Chem. Lett.} \textbf{\bibinfo{volume}{5}},
  \bibinfo{pages}{1035} (\bibinfo{year}{2014}).

\bibitem[{\citenamefont{Stranks et~al.}(2013)\citenamefont{Stranks, Eperon,
  Grancini, Menelaou, Alcocer, Leijtens, Herz, Petrozza, and
  Snaith}}]{Stranks13}
\bibinfo{author}{\bibfnamefont{S.~D.}~\bibnamefont{Stranks}},
  \bibinfo{author}{\bibfnamefont{G.~E.}~\bibnamefont{Eperon}},
  \bibinfo{author}{\bibfnamefont{G.}~\bibnamefont{Grancini}},
  \bibinfo{author}{\bibfnamefont{C.}~\bibnamefont{Menelaou}},
  \bibinfo{author}{\bibfnamefont{M.~J.~P.}~\bibnamefont{Alcocer}},
  \bibinfo{author}{\bibfnamefont{T.}~\bibnamefont{Leijtens}},
  \bibinfo{author}{\bibfnamefont{L.~M.}~\bibnamefont{Herz}},
  \bibinfo{author}{\bibfnamefont{A.}~\bibnamefont{Petrozza}}, \bibnamefont{and}
  \bibinfo{author}{\bibfnamefont{H.~J.}~\bibnamefont{Snaith}},
  \bibinfo{journal}{Science} \textbf{\bibinfo{volume}{342}},
  \bibinfo{pages}{341} (\bibinfo{year}{2013}).

\bibitem[{\citenamefont{Xing et~al.}(2013)\citenamefont{Xing, Mathews, Sun,
  Lim, Lam, Gr\"atzel, Mhaisalkar, and Sum}}]{Xing13}
\bibinfo{author}{\bibfnamefont{G.}~\bibnamefont{Xing}},
  \bibinfo{author}{\bibfnamefont{N.}~\bibnamefont{Mathews}},
  \bibinfo{author}{\bibfnamefont{S.}~\bibnamefont{Sun}},
  \bibinfo{author}{\bibfnamefont{S.~S.}~\bibnamefont{Lim}},
  \bibinfo{author}{\bibfnamefont{Y.~M.}~\bibnamefont{Lam}},
  \bibinfo{author}{\bibfnamefont{M.}~\bibnamefont{Gr\"atzel}},
  \bibinfo{author}{\bibfnamefont{S.}~\bibnamefont{Mhaisalkar}},
  \bibnamefont{and} \bibinfo{author}{\bibfnamefont{T.~C.}~\bibnamefont{Sum}},
  \bibinfo{journal}{Science} \textbf{\bibinfo{volume}{342}},
  \bibinfo{pages}{344} (\bibinfo{year}{2013}).

\bibitem[{\citenamefont{{Ponseca,~Jr.}
  et~al.}(2014)\citenamefont{{Ponseca,~Jr.}, Savenije, Abdellah, Zheng,
  Yartsev, Pascher, Harlang, Chabera, Pullerits, Stepanov
  et~al.}}]{PonsecaJr14}
\bibinfo{author}{\bibfnamefont{C.~S.}~\bibnamefont{{Ponseca,~Jr.}}},
  \bibinfo{author}{\bibfnamefont{T.~J.}~\bibnamefont{Savenije}},
  \bibinfo{author}{\bibfnamefont{M.}~\bibnamefont{Abdellah}},
  \bibinfo{author}{\bibfnamefont{K.}~\bibnamefont{Zheng}},
  \bibinfo{author}{\bibfnamefont{A.}~\bibnamefont{Yartsev}},
  \bibinfo{author}{\bibfnamefont{T.}~\bibnamefont{Pascher}},
  \bibinfo{author}{\bibfnamefont{T.}~\bibnamefont{Harlang}},
  \bibinfo{author}{\bibfnamefont{P.}~\bibnamefont{Chabera}},
  \bibinfo{author}{\bibfnamefont{T.}~\bibnamefont{Pullerits}},
  \bibinfo{author}{\bibfnamefont{A.}~\bibnamefont{Stepanov}},
  \bibnamefont{et~al.}, \bibinfo{journal}{J. Am. Chem. Soc.}
  \textbf{\bibinfo{volume}{136}}, \bibinfo{pages}{5189} (\bibinfo{year}{2014}).

\bibitem[{\citenamefont{Johnston and Herz}(2016)}]{Johnston16}
\bibinfo{author}{\bibfnamefont{M.~B.}~\bibnamefont{Johnston}} \bibnamefont{and}
  \bibinfo{author}{\bibfnamefont{L.~M.}~\bibnamefont{Herz}},
  \bibinfo{journal}{Acc. Chem. Res.} \textbf{\bibinfo{volume}{49}},
  \bibinfo{pages}{146} (\bibinfo{year}{2016}).

\bibitem[{\citenamefont{Snaith et~al.}(2014)\citenamefont{Snaith, Abate, Ball,
  Eperon, Leijtens, Noel, Stanks, Wang, Wojciechowski, and Zhang}}]{Snaith14a}
\bibinfo{author}{\bibfnamefont{H.~J.}~\bibnamefont{Snaith}},
  \bibinfo{author}{\bibfnamefont{A.}~\bibnamefont{Abate}},
  \bibinfo{author}{\bibfnamefont{J.~M.}~\bibnamefont{Ball}},
  \bibinfo{author}{\bibfnamefont{G.~E.}~\bibnamefont{Eperon}},
  \bibinfo{author}{\bibfnamefont{T.}~\bibnamefont{Leijtens}},
  \bibinfo{author}{\bibfnamefont{N.~K.}~\bibnamefont{Noel}},
  \bibinfo{author}{\bibfnamefont{S.~D.}~\bibnamefont{Stanks}},
  \bibinfo{author}{\bibfnamefont{J.~T.-W.}~\bibnamefont{Wang}},
  \bibinfo{author}{\bibfnamefont{K.}~\bibnamefont{Wojciechowski}},
  \bibnamefont{and} \bibinfo{author}{\bibfnamefont{W.}~\bibnamefont{Zhang}},
  \bibinfo{journal}{J. Phys. Chem. Lett.} \textbf{\bibinfo{volume}{5}},
  \bibinfo{pages}{1511} (\bibinfo{year}{2014}).

\bibitem[{\citenamefont{Kim et~al.}(2015)\citenamefont{Kim, Jand, Ahn, Choi,
  Guerrero, Bisquert, and Park}}]{KimHS15}
\bibinfo{author}{\bibfnamefont{H.-S.}~\bibnamefont{Kim}},
  \bibinfo{author}{\bibfnamefont{I.-H.}~\bibnamefont{Jand}},
  \bibinfo{author}{\bibfnamefont{N.-Y.}~\bibnamefont{Ahn}},
  \bibinfo{author}{\bibfnamefont{M.-S.}~\bibnamefont{Choi}},
  \bibinfo{author}{\bibfnamefont{A.}~\bibnamefont{Guerrero}},
  \bibinfo{author}{\bibfnamefont{J.}~\bibnamefont{Bisquert}}, \bibnamefont{and}
  \bibinfo{author}{\bibfnamefont{N.-G.}~\bibnamefont{Park}},
  \bibinfo{journal}{J. Phys. Chem. Lett.} \textbf{\bibinfo{volume}{6}},
  \bibinfo{pages}{4633} (\bibinfo{year}{2015}).

\bibitem[{\citenamefont{Chen et~al.}(2016)\citenamefont{Chen, Yang, Priya, and
  Zhu}}]{ChenB16}
\bibinfo{author}{\bibfnamefont{B.}~\bibnamefont{Chen}},
  \bibinfo{author}{\bibfnamefont{M.}~\bibnamefont{Yang}},
  \bibinfo{author}{\bibfnamefont{S.}~\bibnamefont{Priya}}, \bibnamefont{and}
  \bibinfo{author}{\bibfnamefont{K.}~\bibnamefont{Zhu}}, \bibinfo{journal}{J.
  Phys. Chem. Lett.} \textbf{\bibinfo{volume}{7}}, \bibinfo{pages}{905}
  (\bibinfo{year}{2016}).

\bibitem[{\citenamefont{Noh et~al.}(2013)\citenamefont{Noh, Im, Heo, Mandal,
  and Seok}}]{Noh13}
\bibinfo{author}{\bibfnamefont{J.-H.}~\bibnamefont{Noh}},
  \bibinfo{author}{\bibfnamefont{S.-H.}~\bibnamefont{Im}},
  \bibinfo{author}{\bibfnamefont{J.-H.}~\bibnamefont{Heo}},
  \bibinfo{author}{\bibfnamefont{T.~N.}~\bibnamefont{Mandal}},
  \bibnamefont{and} \bibinfo{author}{\bibfnamefont{S.-I.}~\bibnamefont{Seok}},
  \bibinfo{journal}{Nano Lett.} \textbf{\bibinfo{volume}{13}},
  \bibinfo{pages}{1764} (\bibinfo{year}{2013}).

\bibitem[{\citenamefont{Niu et~al.}(2014)\citenamefont{Niu, Li, Meng, Wang,
  Dong, and Qiu}}]{Niu14}
\bibinfo{author}{\bibfnamefont{G.}~\bibnamefont{Niu}},
  \bibinfo{author}{\bibfnamefont{W.}~\bibnamefont{Li}},
  \bibinfo{author}{\bibfnamefont{F.}~\bibnamefont{Meng}},
  \bibinfo{author}{\bibfnamefont{L.}~\bibnamefont{Wang}},
  \bibinfo{author}{\bibfnamefont{H.}~\bibnamefont{Dong}}, \bibnamefont{and}
  \bibinfo{author}{\bibfnamefont{Y.}~\bibnamefont{Qiu}}, \bibinfo{journal}{J.
  Mater. Chem. A} \textbf{\bibinfo{volume}{2}}, \bibinfo{pages}{705}
  (\bibinfo{year}{2014}).

\bibitem[{\citenamefont{Niu et~al.}(2015)\citenamefont{Niu, Guo, and
  Wang}}]{Niu15}
\bibinfo{author}{\bibfnamefont{G.}~\bibnamefont{Niu}},
  \bibinfo{author}{\bibfnamefont{X.}~\bibnamefont{Guo}}, \bibnamefont{and}
  \bibinfo{author}{\bibfnamefont{L.}~\bibnamefont{Wang}}, \bibinfo{journal}{J.
  Mater. Chem. A} \textbf{\bibinfo{volume}{3}}, \bibinfo{pages}{8970}
  (\bibinfo{year}{2015}).

\bibitem[{\citenamefont{Poglitsch and Weber}(1987)}]{Poglitsch87}
\bibinfo{author}{\bibfnamefont{A.}~\bibnamefont{Poglitsch}} \bibnamefont{and}
  \bibinfo{author}{\bibfnamefont{D.}~\bibnamefont{Weber}}, \bibinfo{journal}{J.
  Chem. Phys.} \textbf{\bibinfo{volume}{87}}, \bibinfo{pages}{6373}
  (\bibinfo{year}{1987}).

\bibitem[{\citenamefont{Weller et~al.}(2015)\citenamefont{Weller, Weber, Henry,
  di~Pumpo, and Hansen}}]{Weller15}
\bibinfo{author}{\bibfnamefont{M.~T.}~\bibnamefont{Weller}},
  \bibinfo{author}{\bibfnamefont{O.~J.}~\bibnamefont{Weber}},
  \bibinfo{author}{\bibfnamefont{P.~F.}~\bibnamefont{Henry}},
  \bibinfo{author}{\bibfnamefont{A.~M.}~\bibnamefont{di~Pumpo}},
  \bibnamefont{and} \bibinfo{author}{\bibfnamefont{T.~C.}
  \bibnamefont{Hansen}}, \bibinfo{journal}{Chem. Commun.}
  \textbf{\bibinfo{volume}{51}}, \bibinfo{pages}{4180} (\bibinfo{year}{2015}).

\bibitem[{\citenamefont{Egger et~al.}(2016)\citenamefont{Egger, Rappe, and
  Kronik}}]{Egger16}
\bibinfo{author}{\bibfnamefont{D.~A.}~\bibnamefont{Egger}},
  \bibinfo{author}{\bibfnamefont{A.~M.}~\bibnamefont{Rappe}}, \bibnamefont{and}
  \bibinfo{author}{\bibfnamefont{L.}~\bibnamefont{Kronik}},
  \bibinfo{journal}{Acc. Chem. Res.} \textbf{\bibinfo{volume}{49}},
  \bibinfo{pages}{573} (\bibinfo{year}{2016}).

\bibitem[{\citenamefont{Wasylishen et~al.}(1985)\citenamefont{Wasylishen, Knop, and
  MacDonald}}]{Wasylishen85}
\bibinfo{author}{\bibfnamefont{R.~E.}~\bibnamefont{Wasylishen}},
  \bibinfo{author}{\bibfnamefont{O.}~\bibnamefont{Knop}}, \bibnamefont{and}
  \bibinfo{author}{\bibfnamefont{J.~B.}~\bibnamefont{MacDonald}},
  \bibinfo{journal}{Solid State Commun.} \textbf{\bibinfo{volume}{56}},
  \bibinfo{pages}{581} (\bibinfo{year}{1985}).

\bibitem[{\citenamefont{Bakulin et~al.}(2015)\citenamefont{Bakulin, Selig, Bakker, Rezus, M\"uller, Glaser, Lovrincic, Sun, Chen, Walsh, Frost, and Jansen}}]{Bakulin15}
\bibinfo{author}{\bibfnamefont{A.~A.}~\bibnamefont{Bakulin}},
  \bibinfo{author}{\bibfnamefont{O.}~\bibnamefont{Selig}}, \bibinfo{author}{\bibfnamefont{H.~J.}~\bibnamefont{Bakker}},
  \bibinfo{author}{\bibfnamefont{Y.~L.~A.}~\bibnamefont{Rezus}}, \bibinfo{author}{\bibfnamefont{C.}~\bibnamefont{M\"uller}},
  \bibinfo{author}{\bibfnamefont{T.}~\bibnamefont{Glaser}}, \bibinfo{author}{\bibfnamefont{R.}~\bibnamefont{Lovrincic}},
  \bibinfo{author}{\bibfnamefont{Z.}~\bibnamefont{Sun}}, \bibinfo{author}{\bibfnamefont{Z.}~\bibnamefont{Chenn}}, \bibinfo{author}{\bibfnamefont{A.}~\bibnamefont{Walsh}},
  \bibinfo{author}{\bibfnamefont{J.~M.}~\bibnamefont{Frost}}, \bibnamefont{and}
  \bibinfo{author}{\bibfnamefont{T.~L.~C.}~\bibnamefont{Jansen}},
  \bibinfo{journal}{J. Phys. Chem. Lett.} \textbf{\bibinfo{volume}{6}},
  \bibinfo{pages}{3663} (\bibinfo{year}{2015}).

\bibitem[{\citenamefont{Chen et~al.}(2015)\citenamefont{Chen, Foley, Ipek,
  Tyagi, Copley, Brown, Choi, and Lee}}]{ChenT15}
\bibinfo{author}{\bibfnamefont{T.}~\bibnamefont{Chen}},
  \bibinfo{author}{\bibfnamefont{B.~J.}~\bibnamefont{Foley}},
  \bibinfo{author}{\bibfnamefont{B.}~\bibnamefont{Ipek}},
  \bibinfo{author}{\bibfnamefont{M.}~\bibnamefont{Tyagi}},
  \bibinfo{author}{\bibfnamefont{J.~R.~D.}~\bibnamefont{Copley}},
  \bibinfo{author}{\bibfnamefont{C.~M.}~\bibnamefont{Brown}},
  \bibinfo{author}{\bibfnamefont{J.~J.}~\bibnamefont{Choi}}, \bibnamefont{and}
  \bibinfo{author}{\bibfnamefont{S.-H.}~\bibnamefont{Lee}},
  \bibinfo{journal}{Phys. Chem. Chem. Phys.} \textbf{\bibinfo{volume}{17}},
  \bibinfo{pages}{31278} (\bibinfo{year}{2015}).

\bibitem[{\citenamefont{Meloni et~al.}(2016)\citenamefont{Meloni, Moehl, Tress, Franckevicius, Saliba, Lee, Gao, Nazeeruddin, Zakeeruddin, Rothlisberger, and Gr\"atzel}}]{Meloni16}
\bibinfo{author}{\bibfnamefont{S.}~\bibnamefont{Meloni}},
  \bibinfo{author}{\bibfnamefont{T.}~\bibnamefont{Moehl}},
  \bibinfo{author}{\bibfnamefont{W.}~\bibnamefont{Tress}},
  \bibinfo{author}{\bibfnamefont{M.}~\bibnamefont{Franckevi\v{c}ius}},
  \bibinfo{author}{\bibfnamefont{M.}~\bibnamefont{Saliba}},
  \bibinfo{author}{\bibfnamefont{Y.~H.}~\bibnamefont{Lee}},
  \bibinfo{author}{\bibfnamefont{P.}~\bibnamefont{Gao}}, \bibinfo{author}{\bibfnamefont{M.~K.}~\bibnamefont{Nazeeruddin}},
  \bibinfo{author}{\bibfnamefont{S.~M.}~\bibnamefont{Zakeeruddin}},
  \bibinfo{author}{\bibfnamefont{U.}~\bibnamefont{Rothlisberger}}, \bibnamefont{and}
  \bibinfo{author}{\bibfnamefont{M.}~\bibnamefont{Gr\"atzel}},
  \bibinfo{journal}{Nature Comm.} \textbf{\bibinfo{volume}{7}},
  \bibinfo{pages}{10334} (\bibinfo{year}{2016}).

\bibitem[{\citenamefont{Onoda-Yamamuro
  et~al.}(1992)\citenamefont{Onoda-Yamamuro, Matsuo, and
  Suga}}]{OnodaYamamuro92}
\bibinfo{author}{\bibfnamefont{N.}~\bibnamefont{Onoda-Yamamuro}},
  \bibinfo{author}{\bibfnamefont{T.}~\bibnamefont{Matsuo}}, \bibnamefont{and}
  \bibinfo{author}{\bibfnamefont{H.}~\bibnamefont{Suga}}, \bibinfo{journal}{J.
  Phys. Chem. Solids} \textbf{\bibinfo{volume}{53}}, \bibinfo{pages}{935}
  (\bibinfo{year}{1992}).

\bibitem[{\citenamefont{Mosconi et~al.}(2014)\citenamefont{Mosconi, Quarti,
  Ivanovska, Ruani, and de~Angelis}}]{Mosconi14b}
\bibinfo{author}{\bibfnamefont{E.}~\bibnamefont{Mosconi}},
  \bibinfo{author}{\bibfnamefont{C.}~\bibnamefont{Quarti}},
  \bibinfo{author}{\bibfnamefont{T.}~\bibnamefont{Ivanovska}},
  \bibinfo{author}{\bibfnamefont{G.}~\bibnamefont{Ruani}}, \bibnamefont{and}
  \bibinfo{author}{\bibfnamefont{F.}~\bibnamefont{de~Angelis}},
  \bibinfo{journal}{Phys. Chem. Chem. Phys.} \textbf{\bibinfo{volume}{16}},
  \bibinfo{pages}{16137} (\bibinfo{year}{2014}).

\bibitem[{\citenamefont{Lee et~al.}(2015)\citenamefont{Lee, Bristowe, Bristowe,
  and Cheetham}}]{LeeJH15}
\bibinfo{author}{\bibfnamefont{J.-H.}~\bibnamefont{Lee}},
  \bibinfo{author}{\bibfnamefont{N.~C.}~\bibnamefont{Bristowe}},
  \bibinfo{author}{\bibfnamefont{P.~D.}~\bibnamefont{Bristowe}},
  \bibnamefont{and} \bibinfo{author}{\bibfnamefont{A.~K.}
  \bibnamefont{Cheetham}}, \bibinfo{journal}{Chem. Commun.}
  \textbf{\bibinfo{volume}{51}}, \bibinfo{pages}{6434} (\bibinfo{year}{2015}).

\bibitem[{\citenamefont{Li et~al.}()}]{LiJ18a}
\bibinfo{author}{\bibfnamefont{J.}~\bibnamefont{Li}},
  \bibinfo{author}{\bibfnamefont{M.}~\bibnamefont{Bouchard}},
  \bibinfo{author}{\bibfnamefont{P.}~\bibnamefont{Reiss}},
  \bibinfo{author}{\bibfnamefont{D.}~\bibnamefont{Aldakov}},
  \bibinfo{author}{\bibfnamefont{S.}~\bibnamefont{Pouget}},
  \bibinfo{author}{\bibfnamefont{R.}~\bibnamefont{Demadrille}},
  \bibinfo{author}{\bibfnamefont{C.}~\bibnamefont{Aumaitre}},
  \bibinfo{author}{\bibfnamefont{B.}~\bibnamefont{Frick}},
  \bibinfo{author}{\bibfnamefont{D.}~\bibnamefont{Djurado}},
  \bibinfo{author}{\bibfnamefont{M.}~\bibnamefont{Rossi}},
  \bibnamefont{and} \bibinfo{author}{\bibfnamefont{P.}~\bibnamefont{Rinke}}, \bibinfo{journal}{J. Phys. Chem. Lett.} \bibinfo{note}{submitted}.

\bibitem[{\citenamefont{Li and Rinke}()}]{LiJ18b}
\bibinfo{author}{\bibfnamefont{J.}~\bibnamefont{Li}} \bibnamefont{and}
  \bibinfo{author}{\bibfnamefont{P.}~\bibnamefont{Rinke}}, \bibinfo{note}{in
  preparation}.

\bibitem[{\citenamefont{Egger and Kronik}(2014)}]{Egger14}
\bibinfo{author}{\bibfnamefont{D.~A.}~\bibnamefont{Egger}} \bibnamefont{and}
  \bibinfo{author}{\bibfnamefont{L.}~\bibnamefont{Kronik}},
  \bibinfo{journal}{J. Phys. Chem. Lett.} \textbf{\bibinfo{volume}{5}},
  \bibinfo{pages}{2728} (\bibinfo{year}{2014}).

\bibitem[{\citenamefont{Li and Rinke}(2016)}]{LiJ16}
\bibinfo{author}{\bibfnamefont{J.}~\bibnamefont{Li}} \bibnamefont{and}
  \bibinfo{author}{\bibfnamefont{P.}~\bibnamefont{Rinke}},
  \bibinfo{journal}{Phys. Rev. B} \textbf{\bibinfo{volume}{94}},
  \bibinfo{pages}{045201} (\bibinfo{year}{2016}).

\bibitem[{\citenamefont{Frost et~al.}(2014{\natexlab{a}})\citenamefont{Frost,
  Butler, Brivio, Hendon, van Schilfgaarde, and Walsh}}]{Frost14a}
\bibinfo{author}{\bibfnamefont{J.~M.}~\bibnamefont{Frost}},
  \bibinfo{author}{\bibfnamefont{K.~T.}~\bibnamefont{Butler}},
  \bibinfo{author}{\bibfnamefont{F.}~\bibnamefont{Brivio}},
  \bibinfo{author}{\bibfnamefont{C.~H.}~\bibnamefont{Hendon}},
  \bibinfo{author}{\bibfnamefont{M.}~\bibnamefont{van Schilfgaarde}},
  \bibnamefont{and} \bibinfo{author}{\bibfnamefont{A.}~\bibnamefont{Walsh}},
  \bibinfo{journal}{Nano Lett.} \textbf{\bibinfo{volume}{14}},
  \bibinfo{pages}{2584} (\bibinfo{year}{2014}{\natexlab{a}}).

\bibitem[{\citenamefont{Frost et~al.}(2014{\natexlab{b}})\citenamefont{Frost,
  Butler, and Walsh}}]{Frost14b}
\bibinfo{author}{\bibfnamefont{J.~M.}~\bibnamefont{Frost}},
  \bibinfo{author}{\bibfnamefont{K.~T.}~\bibnamefont{Butler}},
  \bibnamefont{and} \bibinfo{author}{\bibfnamefont{A.}~\bibnamefont{Walsh}},
  \bibinfo{journal}{APL Mater.} \textbf{\bibinfo{volume}{2}},
  \bibinfo{pages}{081506} (\bibinfo{year}{2014}{\natexlab{b}}).

\bibitem[{\citenamefont{Leguy et~al.}(2015)\citenamefont{Leguy, Frost, McMahon,
  {Garcia~Sakai}, Kochelmann, Law, Li, Foglia, Walsh, O'Regan
  et~al.}}]{Leguy15}
\bibinfo{author}{\bibfnamefont{A.~M.~A.}~\bibnamefont{Leguy}},
  \bibinfo{author}{\bibfnamefont{J.~M.}~\bibnamefont{Frost}},
  \bibinfo{author}{\bibfnamefont{A.~P.}~\bibnamefont{McMahon}},
  \bibinfo{author}{\bibfnamefont{V.}~\bibnamefont{{Garcia~Sakai}}},
  \bibinfo{author}{\bibfnamefont{W.}~\bibnamefont{Kochelmann}},
  \bibinfo{author}{\bibfnamefont{C.~H.}~\bibnamefont{Law}},
  \bibinfo{author}{\bibfnamefont{X.}~\bibnamefont{Li}},
  \bibinfo{author}{\bibfnamefont{F.}~\bibnamefont{Foglia}},
  \bibinfo{author}{\bibfnamefont{A.}~\bibnamefont{Walsh}},
  \bibinfo{author}{\bibfnamefont{B.~C.}~\bibnamefont{O'Regan}},
  \bibnamefont{et~al.}, \bibinfo{journal}{Nature Comm.}
  \textbf{\bibinfo{volume}{6}}, \bibinfo{pages}{7124} (\bibinfo{year}{2015}).

\bibitem[{\citenamefont{Umari et~al.}(2014)\citenamefont{Umari, Mosconi, and
  de~Angelis}}]{Umari14}
\bibinfo{author}{\bibfnamefont{P.}~\bibnamefont{Umari}},
  \bibinfo{author}{\bibfnamefont{E.}~\bibnamefont{Mosconi}}, \bibnamefont{and}
  \bibinfo{author}{\bibfnamefont{F.}~\bibnamefont{de~Angelis}},
  \bibinfo{journal}{Sci. Rep.} \textbf{\bibinfo{volume}{4}},
  \bibinfo{pages}{4467} (\bibinfo{year}{2014}).

\bibitem[{\citenamefont{Men\'endez-Proupin
  et~al.}(2014)\citenamefont{Men\'endez-Proupin, Palacios, Wahn\'on, and
  Conesa}}]{MenendezProupin14}
\bibinfo{author}{\bibfnamefont{E.}~\bibnamefont{Men\'endez-Proupin}},
  \bibinfo{author}{\bibfnamefont{P.}~\bibnamefont{Palacios}},
  \bibinfo{author}{\bibfnamefont{P.}~\bibnamefont{Wahn\'on}}, \bibnamefont{and}
  \bibinfo{author}{\bibfnamefont{J.~C.}~\bibnamefont{Conesa}},
  \bibinfo{journal}{Phys. Rev. B} \textbf{\bibinfo{volume}{90}},
  \bibinfo{pages}{045207} (\bibinfo{year}{2014}).

\bibitem[{\citenamefont{Yin et~al.}(2015)\citenamefont{Yin, Yang, Kang, Yan,
  and Wei}}]{Yin15}
\bibinfo{author}{\bibfnamefont{W.}~\bibnamefont{Yin}},
  \bibinfo{author}{\bibfnamefont{J.}~\bibnamefont{Yang}},
  \bibinfo{author}{\bibfnamefont{J.-G.}~\bibnamefont{Kang}},
  \bibinfo{author}{\bibfnamefont{Y.}~\bibnamefont{Yan}}, \bibnamefont{and}
  \bibinfo{author}{\bibfnamefont{S.}~\bibnamefont{Wei}}, \bibinfo{journal}{J.
  Mater. Chem. A} \textbf{\bibinfo{volume}{3}}, \bibinfo{pages}{8926}
  (\bibinfo{year}{2015}).

\bibitem[{\citenamefont{Brivio et~al.}(2015)\citenamefont{Brivio, Frost,
  Skelton, Jackson, Weber, Weller, {n}i, Leguy, Barnes, and Walsh}}]{Brivio15}
\bibinfo{author}{\bibfnamefont{F.}~\bibnamefont{Brivio}},
  \bibinfo{author}{\bibfnamefont{J.~M.}~\bibnamefont{Frost}},
  \bibinfo{author}{\bibfnamefont{J.~M.}~\bibnamefont{Skelton}},
  \bibinfo{author}{\bibfnamefont{A.~J.}~\bibnamefont{Jackson}},
  \bibinfo{author}{\bibfnamefont{O.~J.}~\bibnamefont{Weber}},
  \bibinfo{author}{\bibfnamefont{M.~T.}~\bibnamefont{Weller}},
  \bibinfo{author}{\bibfnamefont{A.~R.}~\bibnamefont{Go\~{n}i}},
  \bibinfo{author}{\bibfnamefont{A.~M.~A.}~\bibnamefont{Leguy}},
  \bibinfo{author}{\bibfnamefont{P.~R.~F.}~\bibnamefont{Barnes}},
  \bibnamefont{and} \bibinfo{author}{\bibfnamefont{A.}~\bibnamefont{Walsh}},
  \bibinfo{journal}{Phys. Rev. B} \textbf{\bibinfo{volume}{92}},
  \bibinfo{pages}{144308} (\bibinfo{year}{2015}).

\bibitem[{\citenamefont{Deretzis et~al.}(2016)\citenamefont{Deretzis,
  {di~Mauro}, Alberti, Pellegrino, Smecca, and {la~Magna}}}]{Deretzis16}
\bibinfo{author}{\bibfnamefont{I.}~\bibnamefont{Deretzis}},
  \bibinfo{author}{\bibfnamefont{B.~N.}~\bibnamefont{{di~Mauro}}},
  \bibinfo{author}{\bibfnamefont{A.}~\bibnamefont{Alberti}},
  \bibinfo{author}{\bibfnamefont{G.}~\bibnamefont{Pellegrino}},
  \bibinfo{author}{\bibfnamefont{E.}~\bibnamefont{Smecca}}, \bibnamefont{and}
  \bibinfo{author}{\bibfnamefont{A.}~\bibnamefont{{la~Magna}}},
  \bibinfo{journal}{Sci. Rep.} \textbf{\bibinfo{volume}{6}},
  \bibinfo{pages}{24443} (\bibinfo{year}{2016}).

\bibitem[{\citenamefont{Yin et~al.}(2014)\citenamefont{Yin, Shi, and
  Yan}}]{Yin14}
\bibinfo{author}{\bibfnamefont{W.}~\bibnamefont{Yin}},
  \bibinfo{author}{\bibfnamefont{T.}~\bibnamefont{Shi}}, \bibnamefont{and}
  \bibinfo{author}{\bibfnamefont{Y.}~\bibnamefont{Yan}},
  \bibinfo{journal}{Appl. Phys. Lett.} \textbf{\bibinfo{volume}{104}},
  \bibinfo{pages}{063903} (\bibinfo{year}{2014}).

\bibitem[{\citenamefont{Kim et~al.}(2014)\citenamefont{Kim, Lee, Lee, and
  Hong}}]{KimJS14}
\bibinfo{author}{\bibfnamefont{J.-S.}~\bibnamefont{Kim}},
  \bibinfo{author}{\bibfnamefont{S.-H.}~\bibnamefont{Lee}},
  \bibinfo{author}{\bibfnamefont{J.-H.}~\bibnamefont{Lee}}, \bibnamefont{and}
  \bibinfo{author}{\bibfnamefont{K.-H.}~\bibnamefont{Hong}},
  \bibinfo{journal}{J. Phys. Chem. Lett.} \textbf{\bibinfo{volume}{5}},
  \bibinfo{pages}{1312} (\bibinfo{year}{2014}).

\bibitem[{\citenamefont{Ma and Wang}(2015)}]{Ma15}
\bibinfo{author}{\bibfnamefont{J.}~\bibnamefont{Ma}} \bibnamefont{and}
  \bibinfo{author}{\bibfnamefont{L.-W.}~\bibnamefont{Wang}},
  \bibinfo{journal}{Nano Lett.} \textbf{\bibinfo{volume}{15}},
  \bibinfo{pages}{248} (\bibinfo{year}{2015}).

\bibitem[{\citenamefont{Lahnsteiner et~al.}(2016)\citenamefont{Lahnsteiner,
  Kresse, Kumar, Sarma, Franchini, and Bokdam}}]{Lahnsteiner16}
\bibinfo{author}{\bibfnamefont{J.}~\bibnamefont{Lahnsteiner}},
  \bibinfo{author}{\bibfnamefont{G.}~\bibnamefont{Kresse}},
  \bibinfo{author}{\bibfnamefont{A.}~\bibnamefont{Kumar}},
  \bibinfo{author}{\bibfnamefont{D.~D.} \bibnamefont{Sarma}},
  \bibinfo{author}{\bibfnamefont{C.}~\bibnamefont{Franchini}},
  \bibnamefont{and} \bibinfo{author}{\bibfnamefont{M.}~\bibnamefont{Bokdam}},
  \bibinfo{journal}{Phys. Rev. B} \textbf{\bibinfo{volume}{94}},
  \bibinfo{pages}{214114} (\bibinfo{year}{2016}).

\bibitem[{NOT()}]{NOTEsupercell}
\bibinfo{note}{This is a short notation for simplicity and does not mean that
  we treat an $\text{MA}^+$ ion only as a dipole.}

\bibitem[{\citenamefont{Even et~al.}(1996)\citenamefont{Even, Carignano, and
  Katan}}]{Even16}
\bibinfo{author}{\bibfnamefont{J.}~\bibnamefont{Even}},
  \bibinfo{author}{\bibfnamefont{M.}~\bibnamefont{Carignano}}, \bibnamefont{and}
  \bibinfo{author}{\bibfnamefont{C.}~\bibnamefont{Katan}},
  \bibinfo{journal}{Nanoscale} \textbf{\bibinfo{volume}{8}},
  \bibinfo{pages}{6222} (\bibinfo{year}{2016}).

\bibitem[{\citenamefont{Perdew et~al.}(1996)\citenamefont{Perdew, Burke, and
  Ernzerhof}}]{Perdew96}
\bibinfo{author}{\bibfnamefont{J.~P.}~\bibnamefont{Perdew}},
  \bibinfo{author}{\bibfnamefont{K.}~\bibnamefont{Burke}}, \bibnamefont{and}
  \bibinfo{author}{\bibfnamefont{M.}~\bibnamefont{Ernzerhof}},
  \bibinfo{journal}{Phys. Rev. Lett.} \textbf{\bibinfo{volume}{77}},
  \bibinfo{pages}{3865} (\bibinfo{year}{1996}).

\bibitem[{\citenamefont{Tkatchenko and Scheffler}(2009)}]{Tkatchenko09}
\bibinfo{author}{\bibfnamefont{A.}~\bibnamefont{Tkatchenko}} \bibnamefont{and}
  \bibinfo{author}{\bibfnamefont{M.}~\bibnamefont{Scheffler}},
  \bibinfo{journal}{Phys. Rev. Lett.} \textbf{\bibinfo{volume}{102}},
  \bibinfo{pages}{073005} (\bibinfo{year}{2009}).

\bibitem[{\citenamefont{van Lenthe et~al.}(1993)\citenamefont{van Lenthe,
  Baerends, and Sneijders}}]{vanLenthe93}
\bibinfo{author}{\bibfnamefont{E.}~\bibnamefont{van Lenthe}},
  \bibinfo{author}{\bibfnamefont{E.~J.}~\bibnamefont{Baerends}},
  \bibnamefont{and} \bibinfo{author}{\bibfnamefont{J.~G.}
  \bibnamefont{Sneijders}}, \bibinfo{journal}{J. Chem. Phys.}
  \textbf{\bibinfo{volume}{99}}, \bibinfo{pages}{4597} (\bibinfo{year}{1993}).

\bibitem[{\citenamefont{Even et~al.}(2012)\citenamefont{Even, Pedesseau,
  Dupertuis, Jancu, and Katan}}]{Even12}
\bibinfo{author}{\bibfnamefont{J.}~\bibnamefont{Even}},
  \bibinfo{author}{\bibfnamefont{L.}~\bibnamefont{Pedesseau}},
  \bibinfo{author}{\bibfnamefont{M.-A.}~\bibnamefont{Dupertuis}},
  \bibinfo{author}{\bibfnamefont{J.-M.}~\bibnamefont{Jancu}}, \bibnamefont{and}
  \bibinfo{author}{\bibfnamefont{C.}~\bibnamefont{Katan}},
  \bibinfo{journal}{Phys. Rev. B} \textbf{\bibinfo{volume}{86}},
  \bibinfo{pages}{205301} (\bibinfo{year}{2012}).

\bibitem[{\citenamefont{Even et~al.}(2013)\citenamefont{Even, Pedesseau, Jancu,
  and Katan}}]{Even13}
\bibinfo{author}{\bibfnamefont{J.}~\bibnamefont{Even}},
  \bibinfo{author}{\bibfnamefont{L.}~\bibnamefont{Pedesseau}},
  \bibinfo{author}{\bibfnamefont{J.-M.}~\bibnamefont{Jancu}}, \bibnamefont{and}
  \bibinfo{author}{\bibfnamefont{C.}~\bibnamefont{Katan}}, \bibinfo{journal}{J.
  Phys. Chem. Lett.} \textbf{\bibinfo{volume}{4}}, \bibinfo{pages}{2999}
  (\bibinfo{year}{2013}).

\bibitem[{\citenamefont{Brivio et~al.}(2014)\citenamefont{Brivio, Butler,
  Walsh, and van Schilfgaarde}}]{Brivio14}
\bibinfo{author}{\bibfnamefont{F.}~\bibnamefont{Brivio}},
  \bibinfo{author}{\bibfnamefont{K.~T.}~\bibnamefont{Butler}},
  \bibinfo{author}{\bibfnamefont{A.}~\bibnamefont{Walsh}}, \bibnamefont{and}
  \bibinfo{author}{\bibfnamefont{M.}~\bibnamefont{van Schilfgaarde}},
  \bibinfo{journal}{Phys. Rev. B} \textbf{\bibinfo{volume}{89}},
  \bibinfo{pages}{155204} (\bibinfo{year}{2014}).

\bibitem[{\citenamefont{Katan et~al.}(2015)\citenamefont{Katan, Pedesseau,
  Kepenekian, Rolland, and Even}}]{Katan15}
\bibinfo{author}{\bibfnamefont{C.}~\bibnamefont{Katan}},
  \bibinfo{author}{\bibfnamefont{L.}~\bibnamefont{Pedesseau}},
  \bibinfo{author}{\bibfnamefont{M.}~\bibnamefont{Kepenekian}},
  \bibinfo{author}{\bibfnamefont{A.}~\bibnamefont{Rolland}}, \bibnamefont{and}
  \bibinfo{author}{\bibfnamefont{J.}~\bibnamefont{Even}}, \bibinfo{journal}{J.
  Mater. Chem. A} \textbf{\bibinfo{volume}{3}}, \bibinfo{pages}{9232}
  (\bibinfo{year}{2015}).

\bibitem[{\citenamefont{Chiarella et~al.}(2008)\citenamefont{Chiarella,
  Zappettini, and Licci}}]{Chiarella08}
\bibinfo{author}{\bibfnamefont{F.}~\bibnamefont{Chiarella}},
  \bibinfo{author}{\bibfnamefont{A.}~\bibnamefont{Zappettini}},
  \bibinfo{author}{\bibfnamefont{F.}~\bibnamefont{Licci}},
  \bibinfo{author}{\bibfnamefont{I.}~\bibnamefont{Borriello}},
  \bibinfo{author}{\bibfnamefont{G.}~\bibnamefont{Cantele}},
  \bibinfo{author}{\bibfnamefont{D.}~\bibnamefont{Ninno}},
  \bibinfo{author}{\bibfnamefont{A.}~\bibnamefont{Cassinese}}, \bibnamefont{and}
  \bibinfo{author}{\bibfnamefont{R.}~\bibnamefont{Vaglio}},
  \bibinfo{journal}{Phys. Rev. B} \textbf{\bibinfo{volume}{77}},
  \bibinfo{pages}{045129} (\bibinfo{year}{2008}).

\bibitem[{\citenamefont{Brivio et~al.}(2013)\citenamefont{Brivio, Walker, and
  Walsh}}]{Brivio13}
\bibinfo{author}{\bibfnamefont{F.}~\bibnamefont{Brivio}},
  \bibinfo{author}{\bibfnamefont{A.~B.}~\bibnamefont{Walker}},
  \bibnamefont{and} \bibinfo{author}{\bibfnamefont{A.}~\bibnamefont{Walsh}},
  \bibinfo{journal}{APL Mater.} \textbf{\bibinfo{volume}{1}},
  \bibinfo{pages}{042111} (\bibinfo{year}{2013}).

\bibitem[{\citenamefont{Blum et~al.}(2009)\citenamefont{Blum, Gehrke, Hanke,
  Havu, Havu, Ren, Reuter, and Scheffler}}]{Blum09}
\bibinfo{author}{\bibfnamefont{V.}~\bibnamefont{Blum}},
  \bibinfo{author}{\bibfnamefont{R.}~\bibnamefont{Gehrke}},
  \bibinfo{author}{\bibfnamefont{F.}~\bibnamefont{Hanke}},
  \bibinfo{author}{\bibfnamefont{P.}~\bibnamefont{Havu}},
  \bibinfo{author}{\bibfnamefont{V.}~\bibnamefont{Havu}},
  \bibinfo{author}{\bibfnamefont{X.}~\bibnamefont{Ren}},
  \bibinfo{author}{\bibfnamefont{K.}~\bibnamefont{Reuter}}, \bibnamefont{and}
  \bibinfo{author}{\bibfnamefont{M.}~\bibnamefont{Scheffler}},
  \bibinfo{journal}{Comput. Phys. Comm.} \textbf{\bibinfo{volume}{180}},
  \bibinfo{pages}{2175} (\bibinfo{year}{2009}).

\bibitem[{\citenamefont{Havu et~al.}(2009)\citenamefont{Havu, Blum, Havu, and
  Scheffler}}]{HavuV09}
\bibinfo{author}{\bibfnamefont{V.}~\bibnamefont{Havu}},
  \bibinfo{author}{\bibfnamefont{V.}~\bibnamefont{Blum}},
  \bibinfo{author}{\bibfnamefont{P.}~\bibnamefont{Havu}}, \bibnamefont{and}
  \bibinfo{author}{\bibfnamefont{M.}~\bibnamefont{Scheffler}},
  \bibinfo{journal}{J. Comput. Phys.} \textbf{\bibinfo{volume}{228}},
  \bibinfo{pages}{8367} (\bibinfo{year}{2009}).

\bibitem[{\citenamefont{Levchenko et~al.}(2015)\citenamefont{Levchenko, Ren,
  Wieferink, Johanni, Rinke, Blum, and Scheffler}}]{Levchenko15}
\bibinfo{author}{\bibfnamefont{S.~V.}~\bibnamefont{Levchenko}},
  \bibinfo{author}{\bibfnamefont{X.}~\bibnamefont{Ren}},
  \bibinfo{author}{\bibfnamefont{J.}~\bibnamefont{Wieferink}},
  \bibinfo{author}{\bibfnamefont{R.}~\bibnamefont{Johanni}},
  \bibinfo{author}{\bibfnamefont{P.}~\bibnamefont{Rinke}},
  \bibinfo{author}{\bibfnamefont{V.}~\bibnamefont{Blum}}, \bibnamefont{and}
  \bibinfo{author}{\bibfnamefont{M.}~\bibnamefont{Scheffler}},
  \bibinfo{journal}{Comput. Phys. Comm.} \textbf{\bibinfo{volume}{192}},
  \bibinfo{pages}{60} (\bibinfo{year}{2015}), ISSN \bibinfo{issn}{0010-4655}.

\bibitem[{\citenamefont{Knuth et~al.}(2015)\citenamefont{Knuth, Carbogno,
  Atalla, Blum, and Scheffler}}]{Knuth15}
\bibinfo{author}{\bibfnamefont{F.}~\bibnamefont{Knuth}},
  \bibinfo{author}{\bibfnamefont{C.}~\bibnamefont{Carbogno}},
  \bibinfo{author}{\bibfnamefont{V.}~\bibnamefont{Atalla}},
  \bibinfo{author}{\bibfnamefont{V.}~\bibnamefont{Blum}}, \bibnamefont{and}
  \bibinfo{author}{\bibfnamefont{M.}~\bibnamefont{Scheffler}},
  \bibinfo{journal}{Comput. Phys. Comm.} \textbf{\bibinfo{volume}{190}},
  \bibinfo{pages}{33} (\bibinfo{year}{2015}).

\bibitem[{NoM()}]{NoMaDsupercell}
\bibinfo{note}{See
  {\htmladdnormallink{\color{blue}{http://dx.doi.org/10.17172/NOMAD/2018.05.05-1}}{http://dx.doi.org/10.17172/NOMAD/2018.05.05-1}}}.

\bibitem[{SMs()}]{SMsupercell}
\bibinfo{note}{See Supplemental Material at
  {\htmladdnormallink{\color{blue}{http://link.aps.org/}}{http://link.aps.org/}} for individual $\text{MA}$-dipoles and pair modes including the deviation from face-to-face directions; structural parameters and band structures of several $2\!\times\!2\!\times\!2$ supercells; pair-mode distribution from DFT results of  $3\!\times\!3\!\times\!3$ supercells.}

\bibitem[{\citenamefont{Baikie et~al.}(2013)\citenamefont{Baikie, Fang, Kadro,
  Schreyer, Wei, Mhaisalkar, Gr\"atzel, and White}}]{Baikie13}
\bibinfo{author}{\bibfnamefont{T.}~\bibnamefont{Baikie}},
  \bibinfo{author}{\bibfnamefont{Y.}~\bibnamefont{Fang}},
  \bibinfo{author}{\bibfnamefont{J.~M.}~\bibnamefont{Kadro}},
  \bibinfo{author}{\bibfnamefont{M.}~\bibnamefont{Schreyer}},
  \bibinfo{author}{\bibfnamefont{F.}~\bibnamefont{Wei}},
  \bibinfo{author}{\bibfnamefont{S.~G.}~\bibnamefont{Mhaisalkar}},
  \bibinfo{author}{\bibfnamefont{M.}~\bibnamefont{Gr\"atzel}},
  \bibnamefont{and} \bibinfo{author}{\bibfnamefont{T.~J.}~\bibnamefont{White}},
  \bibinfo{journal}{J. Mater. Chem. A} \textbf{\bibinfo{volume}{1}},
  \bibinfo{pages}{5628} (\bibinfo{year}{2013}).

\bibitem[{\citenamefont{Jaervi et~al.}(2018)\citenamefont{Jaervi, Li and Rinke}}]{Jaervi18}
\bibinfo{author}{\bibfnamefont{J.}~\bibfnamefont{J\"arvi}},
  \bibinfo{author}{\bibfnamefont{J.}~\bibfnamefont{Li}}, \bibnamefont{and}
  \bibinfo{author}{\bibfnamefont{P.}~\bibfnamefont{Rinke}},
  \bibinfo{note}{in preparation}.

\end{thebibliography}

\onecolumngrid

\clearpage

\thispagestyle{empty}

\onecolumngrid

\def\thepage{S\arabic{page}}
\setcounter{page}{0}

\def\thesection{S\arabic{section}}
\setcounter{section}{0}

\def\thefigure{S\arabic{figure}}
\setcounter{figure}{0}

\parskip 1.5em
\linespread{1.1}

\begin{center}
Supplemental Material for

\textbf{\large Multi-scale model for disordered hybrid perovskites: the concept of organic cation pair modes}

Jingrui Li$^{1,\ast}$, Jari J\"arvi$^{1,2}$ and Patrick Rinke$^1$

$^1$\textit{Centre of Excellence in Computational Nanoscience (COMP) and Department of Applied Physics, \\ Aalto University, P.O.Box 11100, FI-00076 AALTO, Finland}\\
$^2$\textit{Department of Physics, University of Helsinki, P.O.Box 64, FI-00014 University of Helsinki, Finland}

\end{center}

\vspace{48.2em}

\begin{flushleft}
\rule{.2\textwidth}{0.8bp} \\
$^{\ast}$: \href{mailto:jingrui.li@aalto.fi}{\footnotesize jingrui.li@aalto.fi}
\end{flushleft}

\clearpage

\section{Pair-mode definition when considering the deviation of \texorpdfstring{$\text{MA}^+$}{}-dipoles from the face-to-face direction}

In our previous work \hyperlink{LiJ16}{[1]} we found two stable orientations for $\text{MA}^+$ cations in the cubic primitive-cell model. In the ``diagonal'' structure [the left structure of Fig.~1(a) in the manuscript], the $\text{C}$--$\text{N}$ bond is oriented along the diagonal direction of the single unit cell. We thus denote the direction of the $\text{MA}^+$-dipole $[111]$, $[11\bar{1}]$, and so forth. In the ``face-to-face'' structure [the right structure of Fig.~1(b) in the manuscript], the $\text{C}$--$\text{N}$ bond is oriented along the face-to-face direction of the single unit cell with a small deviation. This deviation is system- and model-dependent. For example, it is $23.2\text{\textdegree}$ in the primitive-cell model of $\text{MAPbI}_3^{}$ \hyperlink{LiJ16}{[1]}, $22.3\text{\textdegree}$ on average in structure~II, and $8.4\text{\textdegree}$ on average in Structure~III. According to the primitive-cell results \hyperlink{LiJ16}{[1]}, the $\text{C}$--$\text{N}$ bond along the $[100]$ direction is located in the symmetry planes $[002]$ or $[020]$. The resulting $\text{C}$--$\text{N}$ bond directions including the deviation are then $[1\epsilon0]$, $[1\bar{\epsilon}0]$, $[10\epsilon]$ or $[10\bar{\epsilon}]$.

The six face-to-face $\text{MA}^+$ dipoles in Fig.~1(c) in the manuscript can be subdivided into twenty-four dipole directions as follows (thick solid and dashed arrows point out of and into the plane of the paper, respectively):

\begin{center}
\begin{tabular}{ccc} \hline\hline
Without & & \multirow{2}{*}{With deviation} \\
deviation & & \\
\hline\hline
\end{tabular}
\end{center}

Together with the eight diagonal directions, there are 32 possible directions for an $\text{MA}^+$ dipole. This results in 86 symmetry inequivalent modes as follows:

\clearpage
\begin{itemize}
  \item Pair modes No.~1--6:

  \begin{tabular}{ccl}
  \begin{tikzpicture}
  \node at (0,0) [left,color=white] {\small$z$};
  \draw[color=white,midarrow=1.] (.1,0) -- (2.2,0) node [right,xshift=-.1cm,color=white] {\small$x$};
  \end{tikzpicture} & \begin{tikzpicture} \draw[color=white] (0,-0.1) -- (0,1.2); \node at (0,0.5) [color=white] {$\Rightarrow$}; \end{tikzpicture} &
  \begin{tikzpicture}
  \draw (0,0) circle (.1);
  \fill (0,0) circle (.03);
  \node at (0,0) [left] {\small$z$};
  \draw[midarrow=1.] (.1,0) -- (2.2,0) node [right,xshift=-.1cm] {\small$x$};
  \draw[midarrow=1.] (0,.1) -- (0,1.2) node [above,yshift=-.1cm] {\small$y$};
  \draw[dashed] (1,0) -- (1,1);
  \draw[dashed] (0,1) -- (2,1) -- (2,0);
  \fill (.41,.37) -- (.63,.59) -- (.59,.63) -- (.37,.41) -- cycle;
  \draw[thick,midarrow=1.] (.39,.39) -- (.61,.61);
  \fill (1.41,.37) -- (1.63,.59) -- (1.59,.63) -- (1.37,.41) -- cycle;
  \draw[thick,midarrow=1.] (1.39,.39) -- (1.61,.61);
  \node at (1,1.5) [above,yshift=-.1cm] {\small $_{}^{}$No.~1$_{}^{}$};
  \end{tikzpicture} ~
  \begin{tikzpicture}
  \draw (0,0) circle (.1);
  \fill (0,0) circle (.03);
  \node at (0,0) [left] {\small$z$};
  \draw[midarrow=1.] (.1,0) -- (2.2,0) node [right,xshift=-.1cm] {\small$x$};
  \draw[midarrow=1.] (0,.1) -- (0,1.2) node [above,yshift=-.1cm] {\small$y$};
  \draw[dashed] (1,0) -- (1,1);
  \draw[dashed] (0,1) -- (2,1) -- (2,0);
  \fill (.41,.37) -- (.63,.59) -- (.59,.63) -- (.37,.41) -- cycle;
  \draw[thick,midarrow=1.] (.39,.39) -- (.61,.61);
  \draw[very thick,dotted] (1.39,.39) -- (1.61,.61);
  \draw[thick,dotted,midarrow=1.] (1.39,.39) -- (1.61,.61);
  \node at (1,1.5) [above,yshift=-.1cm] {\small $_{}^{}$No.~2$_{}^{}$};
  \end{tikzpicture} ~
  \begin{tikzpicture}
  \draw (0,0) circle (.1);
  \fill (0,0) circle (.03);
  \node at (0,0) [left] {\small$z$};
  \draw[midarrow=1.] (.1,0) -- (2.2,0) node [right,xshift=-.1cm] {\small$x$};
  \draw[midarrow=1.] (0,.1) -- (0,1.2) node [above,yshift=-.1cm] {\small$y$};
  \draw[dashed] (1,0) -- (1,1);
  \draw[dashed] (0,1) -- (2,1) -- (2,0);
  \fill (.41,.37) -- (.63,.59) -- (.59,.63) -- (.37,.41) -- cycle;
  \draw[thick,midarrow=1.] (.39,.39) -- (.61,.61);
  \draw[very thick,dotted] (1.39,.61) -- (1.61,.39);
  \draw[thick,dotted,midarrow=1.] (1.39,.61) -- (1.61,.39);
  \node at (1,1.5) [above,yshift=-.1cm] {\small $_{}^{}$No.~3$_{}^{}$};
  \end{tikzpicture} \\
\end{tabular}

  \item Pair mode No.~7 can be splits into 2 variations:

  \begin{tabular}{ccl}
  \begin{tikzpicture}
  \draw (0,0) circle (.1);
  \fill (0,0) circle (.03);
  \node at (0,0) [left] {\small$z$};
  \draw[midarrow=1.] (.1,0) -- (2.2,0) node [right,xshift=-.1cm] {\small$x$};
  \draw[midarrow=1.] (0,.1) -- (0,1.2) node [above,yshift=-.1cm] {\small$y$};
  \draw[dashed] (1,0) -- (1,1);
  \draw[dashed] (0,1) -- (2,1) -- (2,0);
  \fill (.41,.37) -- (.63,.59) -- (.59,.63) -- (.37,.41) -- cycle;
  \draw[thick,midarrow=1.] (.39,.39) -- (.61,.61);
  \draw[midarrow=1.] (1.3,.5) -- (1.7,.5);
  \node at (1,1.5) [above,yshift=-.1cm] {\small $_{}^{}$No.~7$_{}^{}$};
  \end{tikzpicture} & \begin{tikzpicture} \draw[color=white] (0,-0.1) -- (0,1.2); \node at (0,0.5) {$\Rightarrow$}; \end{tikzpicture} &
  \begin{tikzpicture}
  \draw (0,0) circle (.1);
  \fill (0,0) circle (.03);
  \node at (0,0) [left] {\small$z$};
  \draw[midarrow=1.] (.1,0) -- (2.2,0) node [right,xshift=-.1cm] {\small$x$};
  \draw[midarrow=1.] (0,.1) -- (0,1.2) node [above,yshift=-.1cm] {\small$y$};
  \draw[dashed] (1,0) -- (1,1);
  \draw[dashed] (0,1) -- (2,1) -- (2,0);
  \fill (.41,.37) -- (.63,.59) -- (.59,.63) -- (.37,.41) -- cycle;
  \draw[thick,midarrow=1.] (.39,.39) -- (.61,.61);
  \draw[midarrow=1.] (1.31,.44) -- (1.69,.56);
  \node at (1,1.5) [above,yshift=-.1cm] {\small No.~7$_1^{}$};
  \end{tikzpicture} ~
  \begin{tikzpicture}
  \draw (0,0) circle (.1);
  \fill (0,0) circle (.03);
  \node at (0,0) [left] {\small$z$};
  \draw[midarrow=1.] (.1,0) -- (2.2,0) node [right,xshift=-.1cm] {\small$x$};
  \draw[midarrow=1.] (0,.1) -- (0,1.2) node [above,yshift=-.1cm] {\small$y$};
  \draw[dashed] (1,0) -- (1,1);
  \draw[dashed] (0,1) -- (2,1) -- (2,0);
  \fill (.41,.37) -- (.63,.59) -- (.59,.63) -- (.37,.41) -- cycle;
  \draw[thick,midarrow=1.] (.39,.39) -- (.61,.61);
  \draw[midarrow=1.] (1.31,.56) -- (1.69,.44);
  \node at (1,1.5) [above,yshift=-.1cm] {\small No.~7$_2^{}$};
  \end{tikzpicture}
  \end{tabular}

  \item Pair mode No.~8 splits into 2 variations:

  \begin{tabular}{ccl}
  \begin{tikzpicture}
  \draw (0,0) circle (.1);
  \fill (0,0) circle (.03);
  \node at (0,0) [left] {\small$z$};
  \draw[midarrow=1.] (.1,0) -- (2.2,0) node [right,xshift=-.1cm] {\small$x$};
  \draw[midarrow=1.] (0,.1) -- (0,1.2) node [above,yshift=-.1cm] {\small$y$};
  \draw[dashed] (1,0) -- (1,1);
  \draw[dashed] (0,1) -- (2,1) -- (2,0);
  \fill (.41,.37) -- (.63,.59) -- (.59,.63) -- (.37,.41) -- cycle;
  \draw[thick,midarrow=1.] (.39,.39) -- (.61,.61);
  \draw[midarrow=1.] (1.7,.5) -- (1.3,.5);
  \node at (1,1.5) [above,yshift=-.1cm] {\small $_{}^{}$No.~8$_{}^{}$};
  \end{tikzpicture} & \begin{tikzpicture} \draw[color=white] (0,-0.1) -- (0,1.2); \node at (0,0.5) {$\Rightarrow$}; \end{tikzpicture} &
  \begin{tikzpicture}
  \draw (0,0) circle (.1);
  \fill (0,0) circle (.03);
  \node at (0,0) [left] {\small$z$};
  \draw[midarrow=1.] (.1,0) -- (2.2,0) node [right,xshift=-.1cm] {\small$x$};
  \draw[midarrow=1.] (0,.1) -- (0,1.2) node [above,yshift=-.1cm] {\small$y$};
  \draw[dashed] (1,0) -- (1,1);
  \draw[dashed] (0,1) -- (2,1) -- (2,0);
  \fill (.41,.37) -- (.63,.59) -- (.59,.63) -- (.37,.41) -- cycle;
  \draw[thick,midarrow=1.] (.39,.39) -- (.61,.61);
  \draw[midarrow=1.] (1.69,.44) -- (1.31,.56);
  \node at (1,1.5) [above,yshift=-.1cm] {\small No.~8$_1^{}$};
  \end{tikzpicture} ~
  \begin{tikzpicture}
  \draw (0,0) circle (.1);
  \fill (0,0) circle (.03);
  \node at (0,0) [left] {\small$z$};
  \draw[midarrow=1.] (.1,0) -- (2.2,0) node [right,xshift=-.1cm] {\small$x$};
  \draw[midarrow=1.] (0,.1) -- (0,1.2) node [above,yshift=-.1cm] {\small$y$};
  \draw[dashed] (1,0) -- (1,1);
  \draw[dashed] (0,1) -- (2,1) -- (2,0);
  \fill (.41,.37) -- (.63,.59) -- (.59,.63) -- (.37,.41) -- cycle;
  \draw[thick,midarrow=1.] (.39,.39) -- (.61,.61);
  \draw[midarrow=1.] (1.69,.56) -- (1.31,.44);
  \node at (1,1.5) [above,yshift=-.1cm] {\small No.~8$_2^{}$};
  \end{tikzpicture}
  \end{tabular}

  \item Pair mode No.~9 splits into 4 variations:

  %

  \item Pair mode No.~10 splits into 4 variations:

  %

  \item Pair modes No.~11--13:

  

  \item Pair mode No.~14 splits into 2 variations:

  \begin{tabular}{ccl}
  \begin{tikzpicture}
  \draw (0,0) circle (.1);
  \fill (0,0) circle (.03);
  \node at (0,0) [left] {\small$z$};
  \draw[midarrow=1.] (.1,0) -- (2.2,0) node [right,xshift=-.1cm] {\small$x$};
  \draw[midarrow=1.] (0,.1) -- (0,1.2) node [above,yshift=-.1cm] {\small$y$};
  \draw[dashed] (1,0) -- (1,1);
  \draw[dashed] (0,1) -- (2,1) -- (2,0);
  \fill (.59,.37) -- (.37,.59) -- (.41,.63) -- (.63,.41) -- cycle;
  \draw[thick,midarrow=1.] (.61,.39) -- (.39,.61);
  \draw[midarrow=1.] (1.3,.5) -- (1.7,.5);
  \node at (1,1.5) [above,yshift=-.1cm] {\small $_{}^{}$No.~14$_{}^{}$};
  \end{tikzpicture} & \begin{tikzpicture} \draw[color=white] (0,-0.1) -- (0,1.2); \node at (0,0.5) {$\Rightarrow$}; \end{tikzpicture} &
  \begin{tikzpicture}
  \draw (0,0) circle (.1);
  \fill (0,0) circle (.03);
  \node at (0,0) [left] {\small$z$};
  \draw[midarrow=1.] (.1,0) -- (2.2,0) node [right,xshift=-.1cm] {\small$x$};
  \draw[midarrow=1.] (0,.1) -- (0,1.2) node [above,yshift=-.1cm] {\small$y$};
  \draw[dashed] (1,0) -- (1,1);
  \draw[dashed] (0,1) -- (2,1) -- (2,0);
  \fill (.59,.37) -- (.37,.59) -- (.41,.63) -- (.63,.41) -- cycle;
  \draw[thick,midarrow=1.] (.61,.39) -- (.39,.61);
  \draw[midarrow=1.] (1.31,.44) -- (1.69,.56);
  \node at (1,1.5) [above,yshift=-.1cm] {\small No.~14$_1^{}$};
  \end{tikzpicture} ~
  \begin{tikzpicture}
  \draw (0,0) circle (.1);
  \fill (0,0) circle (.03);
  \node at (0,0) [left] {\small$z$};
  \draw[midarrow=1.] (.1,0) -- (2.2,0) node [right,xshift=-.1cm] {\small$x$};
  \draw[midarrow=1.] (0,.1) -- (0,1.2) node [above,yshift=-.1cm] {\small$y$};
  \draw[dashed] (1,0) -- (1,1);
  \draw[dashed] (0,1) -- (2,1) -- (2,0);
  \fill (.59,.37) -- (.37,.59) -- (.41,.63) -- (.63,.41) -- cycle;
  \draw[thick,midarrow=1.] (.61,.39) -- (.39,.61);
  \draw[midarrow=1.] (1.31,.56) -- (1.69,.44);
  \node at (1,1.5) [above,yshift=-.1cm] {\small No.~14$_2^{}$};
  \end{tikzpicture}
  \end{tabular}

  \clearpage

  \item Pair mode No.~15 splits into 2 variations:

  \begin{tabular}{ccl}
  \begin{tikzpicture}
  \draw (0,0) circle (.1);
  \fill (0,0) circle (.03);
  \node at (0,0) [left] {\small$z$};
  \draw[midarrow=1.] (.1,0) -- (2.2,0) node [right,xshift=-.1cm] {\small$x$};
  \draw[midarrow=1.] (0,.1) -- (0,1.2) node [above,yshift=-.1cm] {\small$y$};
  \draw[dashed] (1,0) -- (1,1);
  \draw[dashed] (0,1) -- (2,1) -- (2,0);
  \fill (.59,.37) -- (.37,.59) -- (.41,.63) -- (.63,.41) -- cycle;
  \draw[thick,midarrow=1.] (.61,.39) -- (.39,.61);
  \draw[midarrow=1.] (1.7,.5) -- (1.3,.5);
  \node at (1,1.5) [above,yshift=-.1cm] {\small $_{}^{}$No.~15$_{}^{}$};
  \end{tikzpicture} & \begin{tikzpicture} \draw[color=white] (0,-0.1) -- (0,1.2); \node at (0,0.5) {$\Rightarrow$}; \end{tikzpicture} &
  \begin{tikzpicture}
  \draw (0,0) circle (.1);
  \fill (0,0) circle (.03);
  \node at (0,0) [left] {\small$z$};
  \draw[midarrow=1.] (.1,0) -- (2.2,0) node [right,xshift=-.1cm] {\small$x$};
  \draw[midarrow=1.] (0,.1) -- (0,1.2) node [above,yshift=-.1cm] {\small$y$};
  \draw[dashed] (1,0) -- (1,1);
  \draw[dashed] (0,1) -- (2,1) -- (2,0);
  \fill (.59,.37) -- (.37,.59) -- (.41,.63) -- (.63,.41) -- cycle;
  \draw[thick,midarrow=1.] (.61,.39) -- (.39,.61);
  \draw[midarrow=1.] (1.69,.44) -- (1.31,.56);
  \node at (1,1.5) [above,yshift=-.1cm] {\small No.~15$_1^{}$};
  \end{tikzpicture} ~
  \begin{tikzpicture}
  \draw (0,0) circle (.1);
  \fill (0,0) circle (.03);
  \node at (0,0) [left] {\small$z$};
  \draw[midarrow=1.] (.1,0) -- (2.2,0) node [right,xshift=-.1cm] {\small$x$};
  \draw[midarrow=1.] (0,.1) -- (0,1.2) node [above,yshift=-.1cm] {\small$y$};
  \draw[dashed] (1,0) -- (1,1);
  \draw[dashed] (0,1) -- (2,1) -- (2,0);
  \fill (.59,.37) -- (.37,.59) -- (.41,.63) -- (.63,.41) -- cycle;
  \draw[thick,midarrow=1.] (.61,.39) -- (.39,.61);
  \draw[midarrow=1.] (1.69,.56) -- (1.31,.44);
  \node at (1,1.5) [above,yshift=-.1cm] {\small No.~15$_2^{}$};
  \end{tikzpicture}
  \end{tabular}

  \item Pair mode No.~20 splits into 10 variations:

  \begin{tabular}{ccl}
  \begin{tikzpicture}
  \draw (0,0) circle (.1);
  \fill (0,0) circle (.03);
  \node at (0,0) [left] {\small$z$};
  \draw[midarrow=1.] (.1,0) -- (2.2,0) node [right,xshift=-.1cm] {\small$x$};
  \draw[midarrow=1.] (0,.1) -- (0,1.2) node [above,yshift=-.1cm] {\small$y$};
  \draw[dashed] (1,0) -- (1,1);
  \draw[dashed] (0,1) -- (2,1) -- (2,0);
  \draw[midarrow=1.] (.3,.5) -- (.7,.5);
  \draw[midarrow=1.] (1.5,.3) -- (1.5,.7);
  \node at (1,1.5) [above,yshift=-.1cm] {\small $_{}^{}$No.~20$_{}^{}$};
  \end{tikzpicture} & \begin{tikzpicture} \draw[color=white] (0,-0.1) -- (0,1.2); \node at (0,0.5) {$\Rightarrow$}; \end{tikzpicture} &
  \begin{tikzpicture}
  \draw (0,0) circle (.1);
  \fill (0,0) circle (.03);
  \node at (0,0) [left] {\small$z$};
  \draw[midarrow=1.] (.1,0) -- (2.2,0) node [right,xshift=-.1cm] {\small$x$};
  \draw[midarrow=1.] (0,.1) -- (0,1.2) node [above,yshift=-.1cm] {\small$y$};
  \draw[dashed] (1,0) -- (1,1);
  \draw[dashed] (0,1) -- (2,1) -- (2,0);
  \draw[midarrow=1.] (.31,.44) -- (.69,.56);
  \draw[midarrow=1.] (1.44,.31) -- (1.56,.69);
  \node at (1,1.5) [above,yshift=-.1cm] {\small No.~20$_1^1$};
  \end{tikzpicture} ~
  \begin{tikzpicture}
  \draw (0,0) circle (.1);
  \fill (0,0) circle (.03);
  \node at (0,0) [left] {\small$z$};
  \draw[midarrow=1.] (.1,0) -- (2.2,0) node [right,xshift=-.1cm] {\small$x$};
  \draw[midarrow=1.] (0,.1) -- (0,1.2) node [above,yshift=-.1cm] {\small$y$};
  \draw[dashed] (1,0) -- (1,1);
  \draw[dashed] (0,1) -- (2,1) -- (2,0);
  \draw[midarrow=1.] (.31,.44) -- (.69,.56);
  \draw[midarrow=1.] (1.56,.31) -- (1.44,.69);
  \node at (1,1.5) [above,yshift=-.1cm] {\small No.~20$_1^2$};
  \end{tikzpicture} ~
  \begin{tikzpicture}
  \draw (0,0) circle (.1);
  \fill (0,0) circle (.03);
  \node at (0,0) [left] {\small$z$};
  \draw[midarrow=1.] (.1,0) -- (2.2,0) node [right,xshift=-.1cm] {\small$x$};
  \draw[midarrow=1.] (0,.1) -- (0,1.2) node [above,yshift=-.1cm] {\small$y$};
  \draw[dashed] (1,0) -- (1,1);
  \draw[dashed] (0,1) -- (2,1) -- (2,0);
  \draw[midarrow=1.] (.31,.44) -- (.69,.56);
  \fill (1.47,.31) rectangle (1.53,.69);
  \draw[thick,midarrow=1.] (1.5,.31) -- (1.5,.69);
  \node at (1,1.5) [above,yshift=-.1cm] {\small No.~20$_1^3$};
  \end{tikzpicture} \\
\end{tabular}

  \clearpage

  \item Pair mode No.~21 splits into 3 variations:

  

  \item Pair mode No.~22 splits into 10 variations:

  \begin{tabular}{ccl}
  \begin{tikzpicture}
  \draw (0,0) circle (.1);
  \fill (0,0) circle (.03);
  \node at (0,0) [left] {\small$z$};
  \draw[midarrow=1.] (.1,0) -- (2.2,0) node [right,xshift=-.1cm] {\small$x$};
  \draw[midarrow=1.] (0,.1) -- (0,1.2) node [above,yshift=-.1cm] {\small$y$};
  \draw[dashed] (1,0) -- (1,1);
  \draw[dashed] (0,1) -- (2,1) -- (2,0);
  \draw[midarrow=1.] (.7,.5) -- (.3,.5);
  \draw[midarrow=1.] (1.5,.3) -- (1.5,.7);
  \node at (1,1.5) [above,yshift=-.1cm] {\small $_{}^{}$No.~22$_{}^{}$};
  \end{tikzpicture} & \begin{tikzpicture} \draw[color=white] (0,-0.1) -- (0,1.2); \node at (0,0.5) {$\Rightarrow$}; \end{tikzpicture} &
  \begin{tikzpicture}
  \draw (0,0) circle (.1);
  \fill (0,0) circle (.03);
  \node at (0,0) [left] {\small$z$};
  \draw[midarrow=1.] (.1,0) -- (2.2,0) node [right,xshift=-.1cm] {\small$x$};
  \draw[midarrow=1.] (0,.1) -- (0,1.2) node [above,yshift=-.1cm] {\small$y$};
  \draw[dashed] (1,0) -- (1,1);
  \draw[dashed] (0,1) -- (2,1) -- (2,0);
  \draw[midarrow=1.] (.69,.44) -- (.31,.56);
  \draw[midarrow=1.] (1.44,.31) -- (1.56,.69);
  \node at (1,1.5) [above,yshift=-.1cm] {\small No.~22$_1^1$};
  \end{tikzpicture} ~
  \begin{tikzpicture}
  \draw (0,0) circle (.1);
  \fill (0,0) circle (.03);
  \node at (0,0) [left] {\small$z$};
  \draw[midarrow=1.] (.1,0) -- (2.2,0) node [right,xshift=-.1cm] {\small$x$};
  \draw[midarrow=1.] (0,.1) -- (0,1.2) node [above,yshift=-.1cm] {\small$y$};
  \draw[dashed] (1,0) -- (1,1);
  \draw[dashed] (0,1) -- (2,1) -- (2,0);
  \draw[midarrow=1.] (.69,.44) -- (.31,.56);
  \draw[midarrow=1.] (1.56,.31) -- (1.44,.69);
  \node at (1,1.5) [above,yshift=-.1cm] {\small No.~22$_1^2$};
  \end{tikzpicture} ~
  \begin{tikzpicture}
  \draw (0,0) circle (.1);
  \fill (0,0) circle (.03);
  \node at (0,0) [left] {\small$z$};
  \draw[midarrow=1.] (.1,0) -- (2.2,0) node [right,xshift=-.1cm] {\small$x$};
  \draw[midarrow=1.] (0,.1) -- (0,1.2) node [above,yshift=-.1cm] {\small$y$};
  \draw[dashed] (1,0) -- (1,1);
  \draw[dashed] (0,1) -- (2,1) -- (2,0);
  \draw[midarrow=1.] (.69,.44) -- (.31,.56);
  \fill (1.47,.31) rectangle (1.53,.69);
  \draw[thick,midarrow=1.] (1.5,.31) -- (1.5,.69);
  \node at (1,1.5) [above,yshift=-.1cm] {\small No.~22$_1^3$};
  \end{tikzpicture} \\
\end{tabular}

  \item Pair mode No.~23 splits into 7 variations:

  \begin{tabular}{ccl}
  \begin{tikzpicture}
  \draw (0,0) circle (.1);
  \fill (0,0) circle (.03);
  \node at (0,0) [left] {\small$z$};
  \draw[midarrow=1.] (.1,0) -- (2.2,0) node [right,xshift=-.1cm] {\small$x$};
  \draw[midarrow=1.] (0,.1) -- (0,1.2) node [above,yshift=-.1cm] {\small$y$};
  \draw[dashed] (1,0) -- (1,1);
  \draw[dashed] (0,1) -- (2,1) -- (2,0);
  \draw[midarrow=1.] (.5,.3) -- (.5,.7);
  \draw[midarrow=1.] (1.5,.3) -- (1.5,.7);
  \node at (1,1.5) [above,yshift=-.1cm] {\small $_{}^{}$No.~23$_{}^{}$};
  \end{tikzpicture} & \begin{tikzpicture} \draw[color=white] (0,-0.1) -- (0,1.2); \node at (0,0.5) {$\Rightarrow$}; \end{tikzpicture} &
  \begin{tikzpicture}
  \draw (0,0) circle (.1);
  \fill (0,0) circle (.03);
  \node at (0,0) [left] {\small$z$};
  \draw[midarrow=1.] (.1,0) -- (2.2,0) node [right,xshift=-.1cm] {\small$x$};
  \draw[midarrow=1.] (0,.1) -- (0,1.2) node [above,yshift=-.1cm] {\small$y$};
  \draw[dashed] (1,0) -- (1,1);
  \draw[dashed] (0,1) -- (2,1) -- (2,0);
  \draw[midarrow=1.] (.44,.31) -- (.56,.69);
  \draw[midarrow=1.] (1.44,.31) -- (1.56,.69);
  \node at (1,1.5) [above,yshift=-.1cm] {\small No.~23$_1^1$};
  \end{tikzpicture} ~
  \begin{tikzpicture}
  \draw (0,0) circle (.1);
  \fill (0,0) circle (.03);
  \node at (0,0) [left] {\small$z$};
  \draw[midarrow=1.] (.1,0) -- (2.2,0) node [right,xshift=-.1cm] {\small$x$};
  \draw[midarrow=1.] (0,.1) -- (0,1.2) node [above,yshift=-.1cm] {\small$y$};
  \draw[dashed] (1,0) -- (1,1);
  \draw[dashed] (0,1) -- (2,1) -- (2,0);
  \draw[midarrow=1.] (.44,.31) -- (.56,.69);
  \draw[midarrow=1.] (1.56,.31) -- (1.44,.69);
  \node at (1,1.5) [above,yshift=-.1cm] {\small No.~23$_1^2$};
  \end{tikzpicture} ~
  \begin{tikzpicture}
  \draw (0,0) circle (.1);
  \fill (0,0) circle (.03);
  \node at (0,0) [left] {\small$z$};
  \draw[midarrow=1.] (.1,0) -- (2.2,0) node [right,xshift=-.1cm] {\small$x$};
  \draw[midarrow=1.] (0,.1) -- (0,1.2) node [above,yshift=-.1cm] {\small$y$};
  \draw[dashed] (1,0) -- (1,1);
  \draw[dashed] (0,1) -- (2,1) -- (2,0);
  \draw[midarrow=1.] (.44,.31) -- (.56,.69);
  \fill (1.47,.31) rectangle (1.53,.69);
  \draw[thick,midarrow=1.] (1.5,.31) -- (1.5,.69);
  \node at (1,1.5) [above,yshift=-.1cm] {\small No.~23$_1^3$};
  \end{tikzpicture} \\
\end{tabular}

  \item Pair mode No.~24 splits into 7 variations:

  \begin{tabular}{ccl}
  \begin{tikzpicture}
  \draw (0,0) circle (.1);
  \fill (0,0) circle (.03);
  \node at (0,0) [left] {\small$z$};
  \draw[midarrow=1.] (.1,0) -- (2.2,0) node [right,xshift=-.1cm] {\small$x$};
  \draw[midarrow=1.] (0,.1) -- (0,1.2) node [above,yshift=-.1cm] {\small$y$};
  \draw[dashed] (1,0) -- (1,1);
  \draw[dashed] (0,1) -- (2,1) -- (2,0);
  \draw[midarrow=1.] (.5,.3) -- (.5,.7);
  \draw[midarrow=1.] (1.5,.7) -- (1.5,.3);
  \node at (1,1.5) [above,yshift=-.1cm] {\small $_{}^{}$No.~24$_{}^{}$};
  \end{tikzpicture} & \begin{tikzpicture} \draw[color=white] (0,-0.1) -- (0,1.2); \node at (0,0.5) {$\Rightarrow$}; \end{tikzpicture} &
  \begin{tikzpicture}
  \draw (0,0) circle (.1);
  \fill (0,0) circle (.03);
  \node at (0,0) [left] {\small$z$};
  \draw[midarrow=1.] (.1,0) -- (2.2,0) node [right,xshift=-.1cm] {\small$x$};
  \draw[midarrow=1.] (0,.1) -- (0,1.2) node [above,yshift=-.1cm] {\small$y$};
  \draw[dashed] (1,0) -- (1,1);
  \draw[dashed] (0,1) -- (2,1) -- (2,0);
  \draw[midarrow=1.] (.44,.31) -- (.56,.69);
  \draw[midarrow=1.] (1.44,.69) -- (1.56,.31);
  \node at (1,1.5) [above,yshift=-.1cm] {\small No.~24$_1^1$};
  \end{tikzpicture} ~
  \begin{tikzpicture}
  \draw (0,0) circle (.1);
  \fill (0,0) circle (.03);
  \node at (0,0) [left] {\small$z$};
  \draw[midarrow=1.] (.1,0) -- (2.2,0) node [right,xshift=-.1cm] {\small$x$};
  \draw[midarrow=1.] (0,.1) -- (0,1.2) node [above,yshift=-.1cm] {\small$y$};
  \draw[dashed] (1,0) -- (1,1);
  \draw[dashed] (0,1) -- (2,1) -- (2,0);
  \draw[midarrow=1.] (.44,.31) -- (.56,.69);
  \draw[midarrow=1.] (1.56,.69) -- (1.44,.31);
  \node at (1,1.5) [above,yshift=-.1cm] {\small No.~24$_1^2$};
  \end{tikzpicture} ~
  \begin{tikzpicture}
  \draw (0,0) circle (.1);
  \fill (0,0) circle (.03);
  \node at (0,0) [left] {\small$z$};
  \draw[midarrow=1.] (.1,0) -- (2.2,0) node [right,xshift=-.1cm] {\small$x$};
  \draw[midarrow=1.] (0,.1) -- (0,1.2) node [above,yshift=-.1cm] {\small$y$};
  \draw[dashed] (1,0) -- (1,1);
  \draw[dashed] (0,1) -- (2,1) -- (2,0);
  \draw[midarrow=1.] (.44,.31) -- (.56,.69);
  \fill (1.47,.69) rectangle (1.53,.31);
  \draw[thick,midarrow=1.] (1.5,.69) -- (1.5,.31);
  \node at (1,1.5) [above,yshift=-.1cm] {\small No.~24$_1^3$};
  \end{tikzpicture} \\
\end{tabular}

  \clearpage

  \item Pair mode No.~25 splits into 10 variations:

  \begin{tabular}{ccl}
  \begin{tikzpicture}
  \draw (0,0) circle (.1);
  \fill (0,0) circle (.03);
  \node at (0,0) [left] {\small$z$};
  \draw[midarrow=1.] (.1,0) -- (2.2,0) node [right,xshift=-.1cm] {\small$x$};
  \draw[midarrow=1.] (0,.1) -- (0,1.2) node [above,yshift=-.1cm] {\small$y$};
  \draw[dashed] (1,0) -- (1,1);
  \draw[dashed] (0,1) -- (2,1) -- (2,0);
  \draw[midarrow=1.] (.5,.3) -- (.5,.7);
  \draw[thick] (1.5,.5) circle (.15);
  \fill (1.5,.5) circle (.05);
  \node at (1,1.5) [above,yshift=-.1cm] {\small $_{}^{}$No.~25$_{}^{}$};
  \end{tikzpicture} & \begin{tikzpicture} \draw[color=white] (0,-0.1) -- (0,1.2); \node at (0,0.5) {$\Rightarrow$}; \end{tikzpicture} &
  \begin{tikzpicture}
  \draw (0,0) circle (.1);
  \fill (0,0) circle (.03);
  \node at (0,0) [left] {\small$z$};
  \draw[midarrow=1.] (.1,0) -- (2.2,0) node [right,xshift=-.1cm] {\small$x$};
  \draw[midarrow=1.] (0,.1) -- (0,1.2) node [above,yshift=-.1cm] {\small$y$};
  \draw[dashed] (1,0) -- (1,1);
  \draw[dashed] (0,1) -- (2,1) -- (2,0);
  \draw[midarrow=1.] (.44,.31) -- (.56,.69);
  \draw[thick] (1.5,.5) circle (.15);
  \draw[thick,->] (1.44,.5) -- (1.56,.5);
  \node at (1,1.5) [above,yshift=-.1cm] {\small No.~25$_1^1$};
  \end{tikzpicture} ~
  \begin{tikzpicture}
  \draw (0,0) circle (.1);
  \fill (0,0) circle (.03);
  \node at (0,0) [left] {\small$z$};
  \draw[midarrow=1.] (.1,0) -- (2.2,0) node [right,xshift=-.1cm] {\small$x$};
  \draw[midarrow=1.] (0,.1) -- (0,1.2) node [above,yshift=-.1cm] {\small$y$};
  \draw[dashed] (1,0) -- (1,1);
  \draw[dashed] (0,1) -- (2,1) -- (2,0);
  \draw[midarrow=1.] (.44,.31) -- (.56,.69);
  \draw[thick] (1.5,.5) circle (.15);
  \draw[thick,->] (1.56,.5) -- (1.44,.5);
  \node at (1,1.5) [above,yshift=-.1cm] {\small No.~25$_1^2$};
  \end{tikzpicture} ~
  \begin{tikzpicture}
  \draw (0,0) circle (.1);
  \fill (0,0) circle (.03);
  \node at (0,0) [left] {\small$z$};
  \draw[midarrow=1.] (.1,0) -- (2.2,0) node [right,xshift=-.1cm] {\small$x$};
  \draw[midarrow=1.] (0,.1) -- (0,1.2) node [above,yshift=-.1cm] {\small$y$};
  \draw[dashed] (1,0) -- (1,1);
  \draw[dashed] (0,1) -- (2,1) -- (2,0);
  \draw[midarrow=1.] (.44,.31) -- (.56,.69);
  \draw[thick] (1.5,.5) circle (.15);
  \draw[thick,->] (1.5,.44) -- (1.5,.56);
  \node at (1,1.5) [above,yshift=-.1cm] {\small No.~25$_1^3$};
  \end{tikzpicture} ~
  \begin{tikzpicture}
  \draw (0,0) circle (.1);
  \fill (0,0) circle (.03);
  \node at (0,0) [left] {\small$z$};
  \draw[midarrow=1.] (.1,0) -- (2.2,0) node [right,xshift=-.1cm] {\small$x$};
  \draw[midarrow=1.] (0,.1) -- (0,1.2) node [above,yshift=-.1cm] {\small$y$};
  \draw[dashed] (1,0) -- (1,1);
  \draw[dashed] (0,1) -- (2,1) -- (2,0);
  \draw[midarrow=1.] (.44,.31) -- (.56,.69);
  \draw[thick] (1.5,.5) circle (.15);
  \draw[thick,->] (1.5,.56) -- (1.5,.44);
  \node at (1,1.5) [above,yshift=-.1cm] {\small No.~25$_1^4$};
  \end{tikzpicture} \\
\end{tabular}

\end{itemize}

\clearpage

\section{Band structure of \texorpdfstring{$2\!\times\!2\!\times\!2$}{} supercell structures~I, II, III and IV}

\begin{figure}[!ht]
(a) Structure~I \\
\includegraphics[clip=true,trim=1.in 6.in 1.in 1.in,page=1,scale=.9]{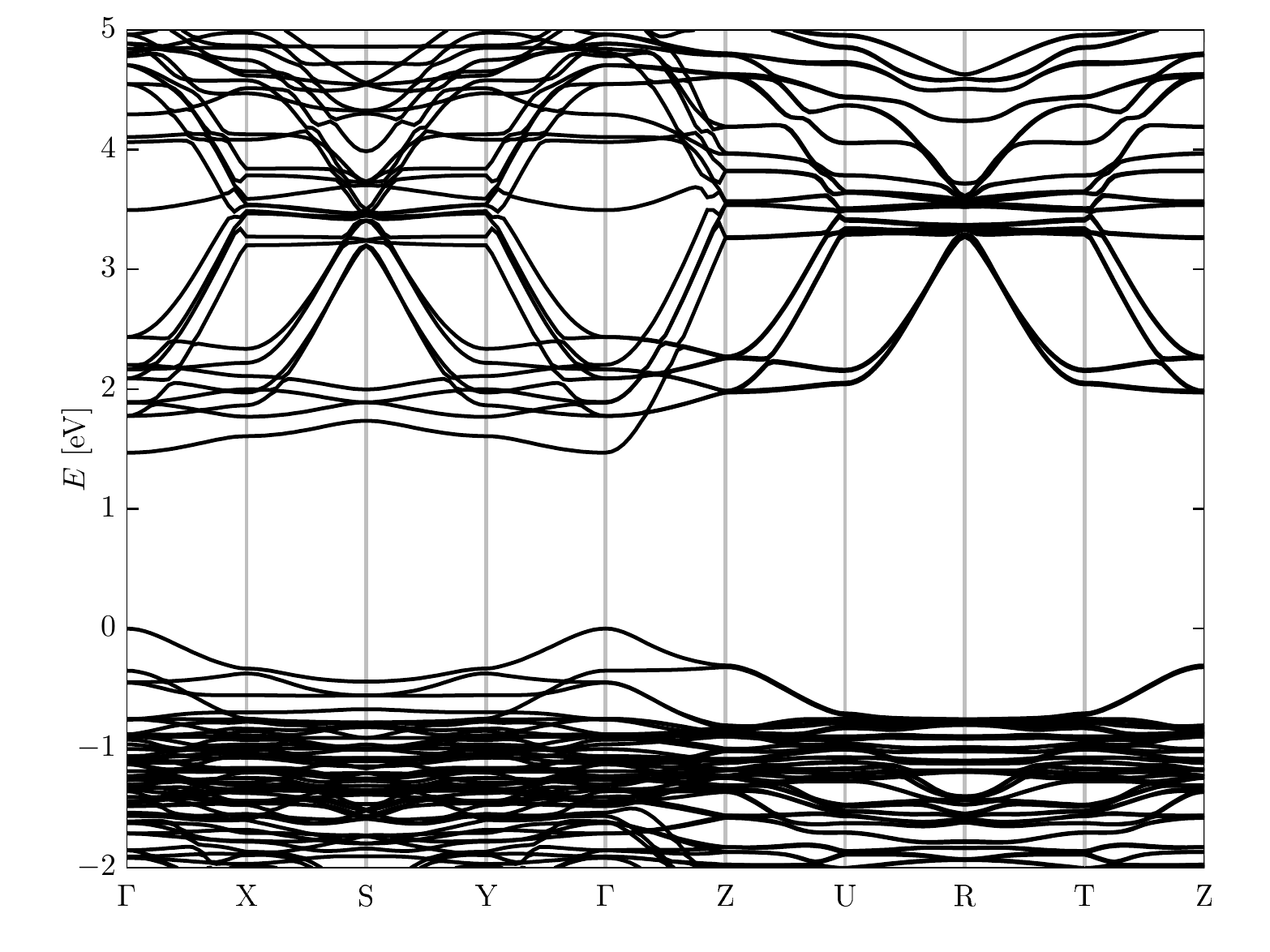}

(b) Structure~II \\
\includegraphics[clip=true,trim=1.in 6.in 1.in 1.in,page=2,scale=.9]{./FigS01.pdf}
\end{figure}
\begin{figure}[!ht]
(c) Structure~III \\
\includegraphics[clip=true,trim=1.in 6.in 1.in 1.in,page=3,scale=.9]{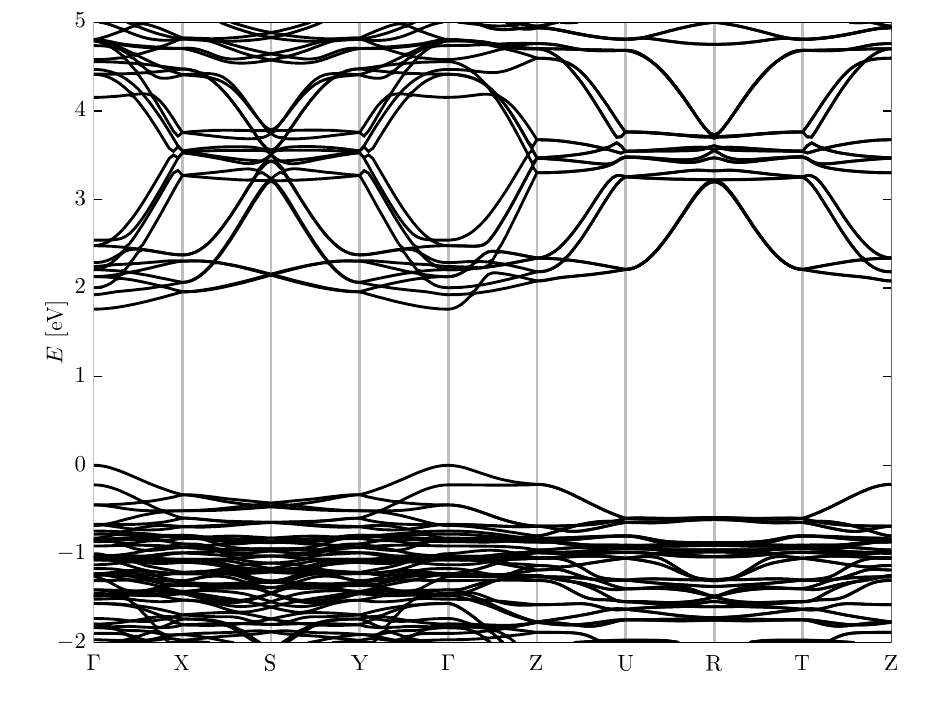}

(d) Structure~IV \\
\includegraphics[clip=true,trim=1.in 6.in 1.in 1.in,page=4,scale=.9]{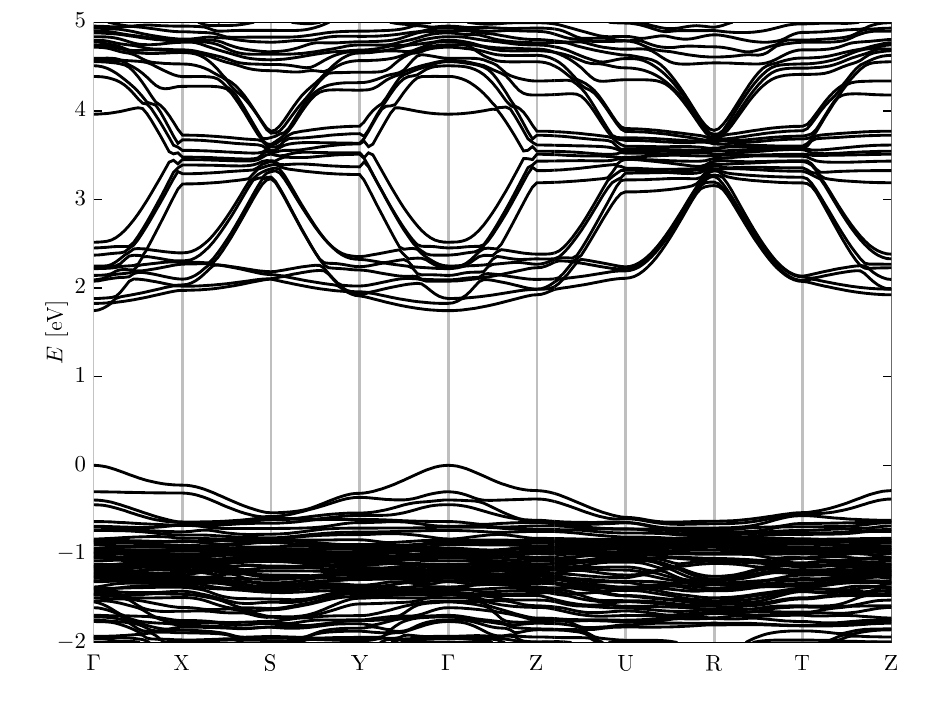}
\caption{Band structures of the four optimized $2\!\times\!2\!\times\!2$ supercell structures~I, II, III and IV.}
\end{figure}

\clearpage

\section{Optimized \texorpdfstring{$2\!\times\!2\!\times\!2$}{} supercell structure with all diagonally-oriented \texorpdfstring{$\text{MA}^+$}{} ions}

\begin{itemize}
  \item Atomic structure: cf. Fig.~4 in the manuscript.

  \item Lattice parameters: $a\approx b\approx c=12.60~\text{\AA}$.

  \item Band structure (band gap $=1.347~\text{eV}$):

  \begin{figure}[!ht]
  \includegraphics[clip=true,trim=1.in 6.in 1.in 1.in,page=1,scale=.9]{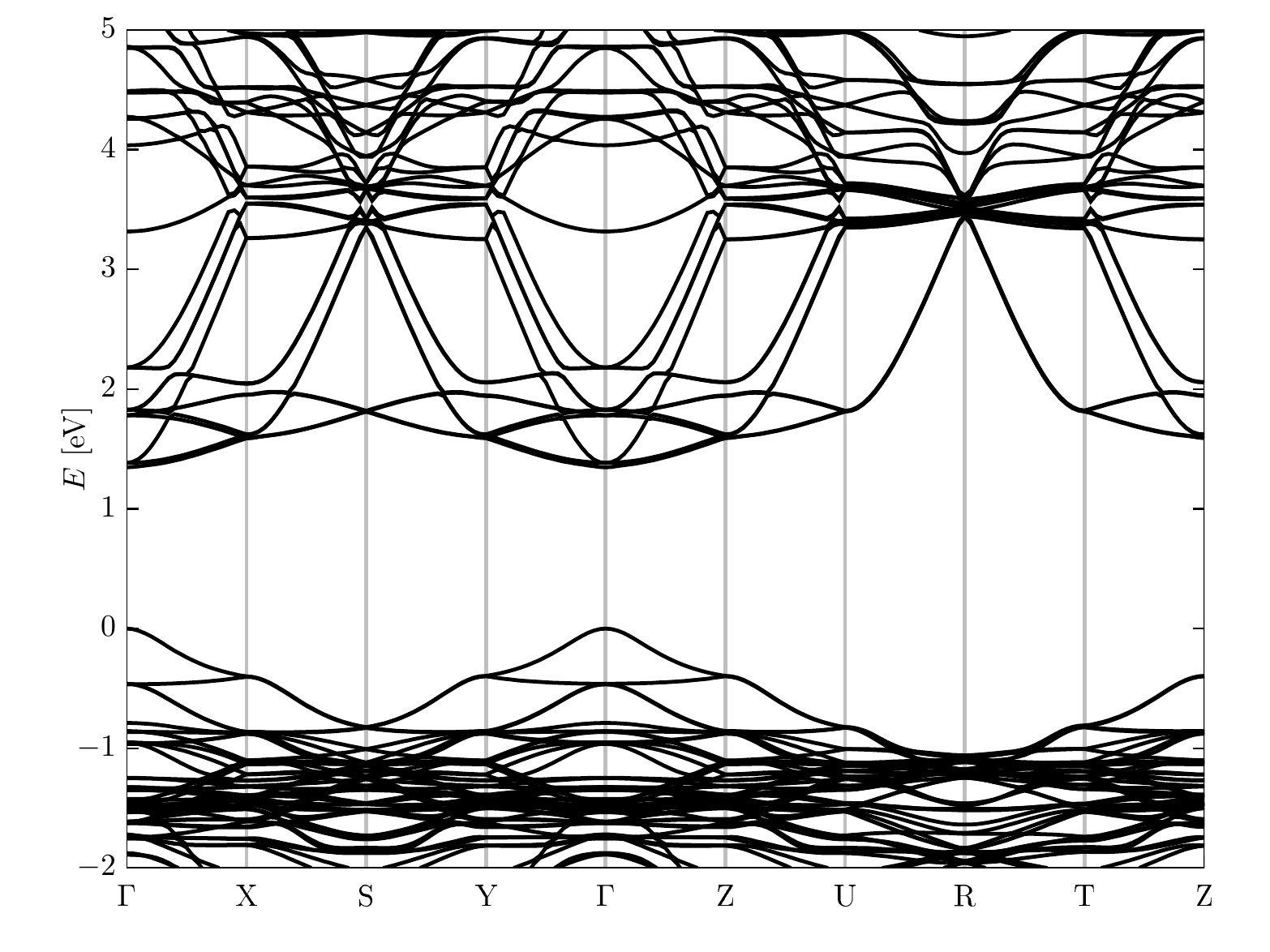}
  \caption{Band structure of the optimized $2\!\times\!2\!\times\!2$ supercell structure in which all $\text{MA}^+$ cations are oriented diagonally.}
  \end{figure}

\end{itemize}


\section{Pair-mode distribution in optimized \texorpdfstring{$3\!\times\!3\!\times\!3$}{} supercell structures}

\begin{figure}[!ht]
\begin{tabular}{cccc}
\multicolumn{4}{c}{(a) High-energy structure} \\
\multirow{14}{*}{\includegraphics[clip=true,trim=1in 4.5in 1in .9in,page=1,scale=.3]{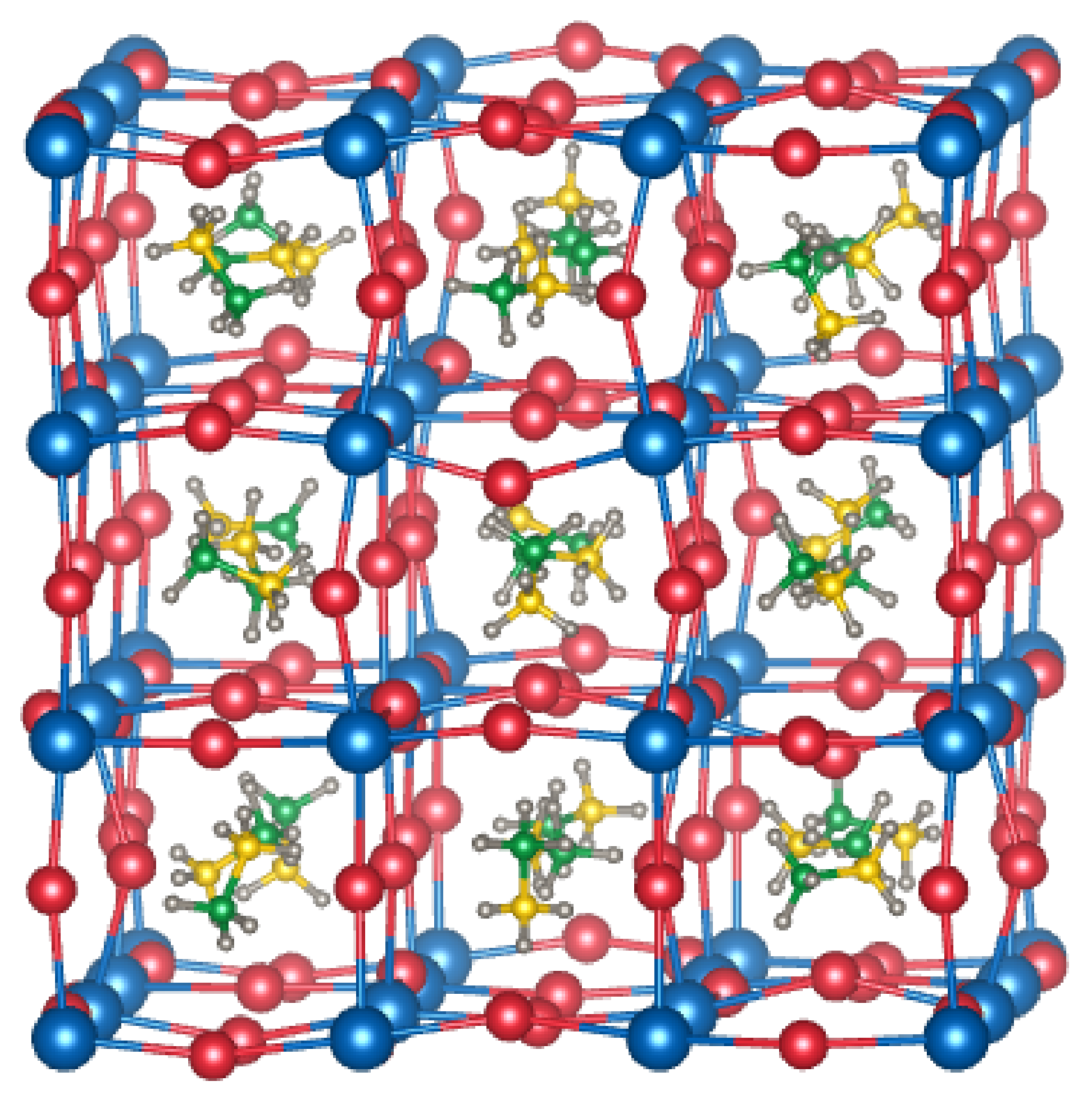}} & \multicolumn{3}{c}{$\text{MA}^+$-dipole pattern} \\
&
\includegraphics[clip=true,trim=3.37in 9.2in 3.37in 1.in,page=7,scale=1.]{./Fig07and08_2.pdf} &
\includegraphics[clip=true,trim=3.37in 9.2in 3.37in 1.in,page=8,scale=1.]{./Fig07and08_2.pdf} &
\includegraphics[clip=true,trim=3.37in 9.2in 3.37in 1.in,page=9,scale=1.]{./Fig07and08_2.pdf} \\
& Bottom layer & Middle layer & Top layer \\
\\
\multicolumn{4}{c}{(b) Low-energy structure} \\
\multirow{14}{*}{\includegraphics[clip=true,trim=1in 4.5in 1in .9in,page=2,scale=.3]{./Fig08_1.pdf}} & \multicolumn{3}{c}{$\text{MA}^+$-dipole pattern} \\
&
\includegraphics[clip=true,trim=3.37in 9.2in 3.37in 1.in,page=4,scale=1.]{./Fig07and08_2.pdf} &
\includegraphics[clip=true,trim=3.37in 9.2in 3.37in 1.in,page=5,scale=1.]{./Fig07and08_2.pdf} &
\includegraphics[clip=true,trim=3.37in 9.2in 3.37in 1.in,page=6,scale=1.]{./Fig07and08_2.pdf} \\
& Bottom layer & Middle layer & Top layer \\ \\
\end{tabular}
\caption{Two optimized structures of $3\!\times\!3\!\times\!3$ supercell models as well as their dipole patterns of the bottom, middle and top layers.}\label{333models}
\end{figure}

\begin{figure}[!ht]
\begin{tabular}{cc}
{\footnotesize (a) Distribution of each sample} & {\footnotesize (b) Overall distribution}\\
\includegraphics[clip=true,trim=1.55in 6.9in 1.55in 1.in,page=1,scale=.6]{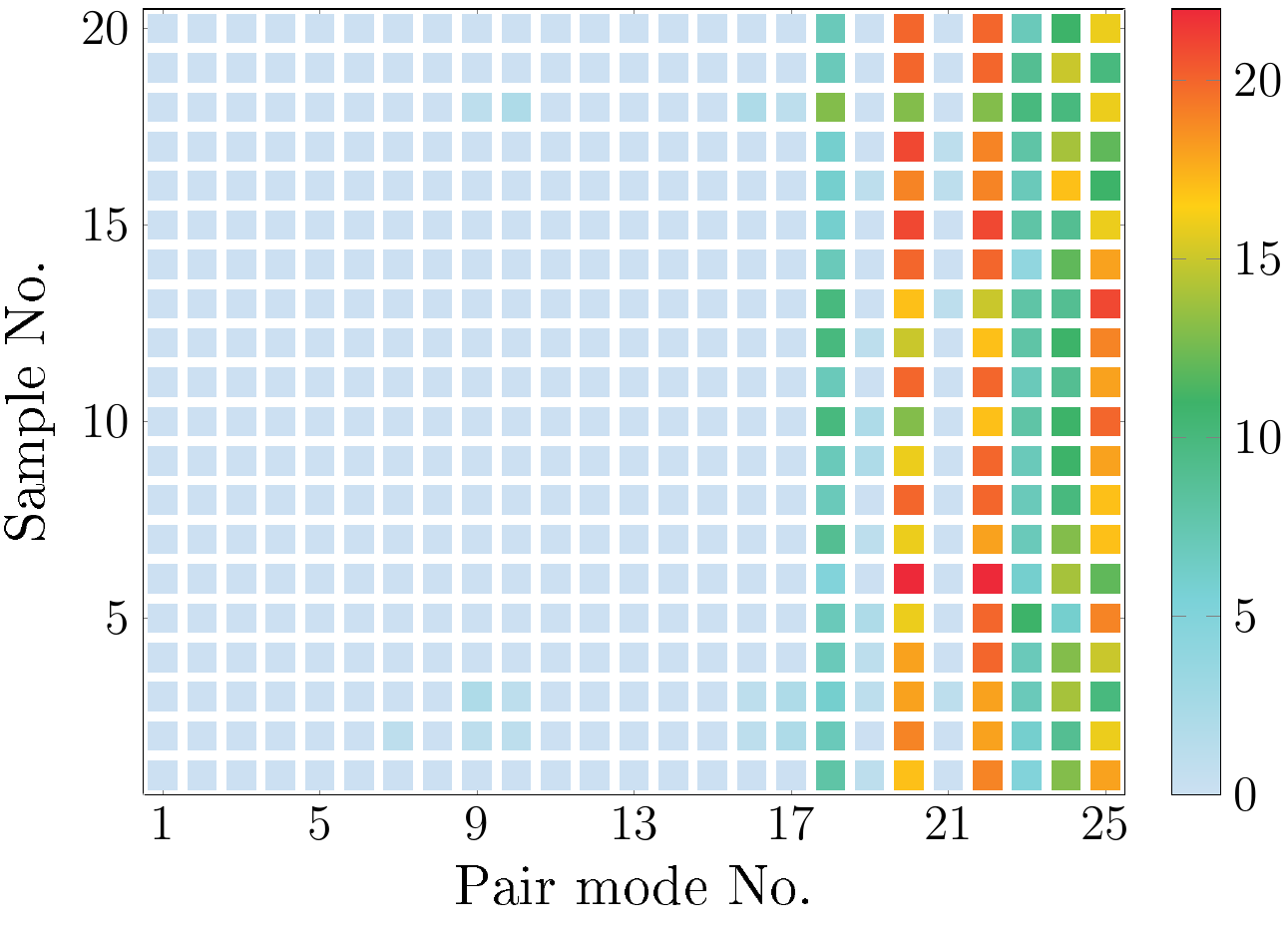} &
\includegraphics[clip=true,trim=1.75in 6.9in 1.55in 1.in,page=2,scale=.6]{./Fig09mod.pdf}
\end{tabular}
\caption{Pair-mode distribution of optimized $3\!\times\!3\!\times\!3$ supercell models: (a) distribution of each individual sample, and (b) the overall distribution of all samples.}\label{pm333all}
\end{figure}

We have also optimized 20 $3\!\times\!3\!\times\!3$ supercell structures of $\text{MAPbI}_3^{}$ using DFT (data available in Ref.~\hyperlink{NOMAD333}{[2]}). Figure~\ref{333models}(a) and (b) show the optimized structured that have the highest and lowest total energies, respectively. There difference is $15~\text{meV}$ per $\text{MAPbI}_3^{}$ unit. The pair-mode distribution of these 20 optimized structures is depicted in Fig.~\ref{pm333all}. Compared with the pair-mode distribution calculated with the more proper $4\!\times\!4\!\times\!4$ supercell models (Fig.~9 of the main text), the $3\!\times\!3\!\times\!3$ models show much more significant population ($0.094$ vs. $0.036$) in the linear-extending mode No.~18. We consider that this is an artefact due to the odd number of single cells along each lattice vector of the $3\!\times\!3\!\times\!3$ supercell models \hyperlink{Even16}{[3]}.

\clearpage

\vspace{2.em}

{\small
\begin{enumerate}[{[}1{]}]
  \item J.~Li and P.~Rinke, Phys.~Rev.~B \textbf{94}, 045210 (2016). \hypertarget{LiJ16}{}
  \item See {\htmladdnormallink{\color{blue}{http://dx.doi.org/10.17172/NOMAD/2018.05.15-1}}{http://dx.doi.org/10.17172/NOMAD/2018.05.15-1}}. \hypertarget{NOMAD333}{}
  \item J.~Even, M.~Carignano, and C.~Katan, Nanoscale \textbf{8}, 6222 (2016). \hypertarget{Even16}{}
\end{enumerate}
}

\twocolumngrid

\end{document}